\documentclass{aa}  
\usepackage{graphicx}
\usepackage[varg]{txfonts}
\usepackage{color}
\usepackage{natbib}
\bibpunct{(}{)}{;}{a}{}{,} 

\def\am{\mbox{\AA\,}}
\def\degr{\hbox{$^\circ$}}
\def\kms{km s$^{-1}$}
\def\kmsmpc{km s$^{-1}$ Mpc$^{-1}$}
\def\ergscmarcsec{erg\,s$^{-1}$\,cm$^{-2}$\,arcsec$^{-2}$}
\def\msun{M$_{\odot}$}
\def\msunpyr{M$_{\odot}$ yr$^{-1}$}
\def\mstar{$M_{\star}$}
\def\mstarsmsun{$M_{\star}/M_{\odot}$}

\def\hi{H\,{\sc i}}
\def\hii{H\,{\sc ii}}
\def\vmax{$V_{\rm max}$}
\def\gpk{{\sc GalPaK}$^{\rm 3D}$}
\def\DPA{$\Delta$PA}
\def\voversig{$V/\sigma$}

\def\oii{[O\,{\sc ii}]$\lambda\lambda 3727,3729$}

\def\oiib{[O\,{\sc ii}]$\lambda 3729$}
\def\oiii{[O\,{\sc iii}]$\lambda\lambda 4959,5007$}
\def\oiiib{[O\,{\sc iii}]$\lambda 5007$}
\def\hbeta{H$\beta$}
\def\halpha{H$\alpha$}
\def\nii{[N\,{\sc ii}]$\lambda\lambda 6548,6584$}
\def\neiii{[Ne\,{\sc iii}]}
\def\nev{[Ne\,{\sc v}]}
\def\mgii{Mg\,{\sc ii}}
\def\feii{Fe\,{\sc ii}}

\begin{document}

   \title{Deep MUSE observations in the HDFS
    \thanks{Based on observations made with ESO/VLT telescopes at the Paranal Observatory under program ID 60.A-9100(C). 
    Based on observations made with the NASA/ESA Hubble Space Telescope, obtained from the data archive at the Space Telescope Science Institute. STScI is operated by the Association of Universities for Research in Astronomy, Inc. under NASA contract NAS 5-26555.}
    }

   \titlerunning{Morpho-kinematics of low-mass distant galaxies in MUSE-HDFS}

   \subtitle{Morpho-kinematics of distant star-forming galaxies down to $10^8$ \msun}

   \author{
              T. Contini \inst{1,2} 
        \and B. Epinat \inst{1,2,3}
        \and N. Bouch\'e \inst{4} 
        \and J. Brinchmann \inst{5}      
        \and L. A. Boogaard \inst{5}     
        \and E. Ventou \inst{1,2}    
        \and R. Bacon \inst{6}
        \and J. Richard  \inst{6}
        \and P. M. Weilbacher \inst{7} 
        \and L. Wisotzki  \inst{7}
        \and D. Krajnovi\'c \inst{7}
        \and J-B. Vielfaure \inst{1,2} 
        \and E. Emsellem \inst{8,6}
        \and H. Finley \inst{1,2} 
        \and H. Inami  \inst{6}
                \and J. Schaye  \inst{5}
        \and M. Swinbank \inst{9}  
        \and A. Gu{\' e}rou \inst{8,1,2}
         \and T. Martinsson \inst{5,10,11}
         \and L. Michel-Dansac  \inst{6} 
         \and I. Schroetter \inst{1,2} 
         \and M. Shirazi \inst{12}       
         \and G. Soucail \inst{1,2}
    }

   \institute{
   IRAP, Institut de Recherche en Astrophysique et Plan\'etologie, CNRS, 14, avenue Edouard Belin, F-31400 Toulouse, France 
   \and Universit\'e de Toulouse, UPS-OMP, Toulouse, France
   \and Aix Marseille Universit\'e, CNRS, LAM, Laboratoire d'Astrophysique de Marseille, UMR 7326, 13388, Marseille, France
   \and IRAP/CNRS, 9, avenue Colonel Roche, F-31400 Toulouse, France 
   \and Leiden Observatory, Leiden University, P.O. Box 9513, 2300 RA Leiden, The Netherlands 
   \and Univ Lyon, Univ Lyon1, Ens de Lyon, CNRS, Centre de Recherche Astrophysique de Lyon UMR5574, F-69230, Saint-Genis-Laval, France
   \and Leibniz-Institut f{\"u}r Astrophysik Potsdam (AIP), An der Sternwarte 16, D-14482 Potsdam, Germany
   \and ESO, European Southern Observatory, Karl-Schwarzschild Str. 2, 85748 Garching bei Muenchen, Germany
   \and Institute for Computational Cosmology, Durham University, South Road, Durham DH1 3LE, UK   
   \and Instituto de Astrof{\` i}sica de Canarias (IAC), E-38205 La Laguna, Tenerife, Spain
   \and Departamento de Astrof{\` i}sica, Universidad de La Laguna, E-38206 La Laguna, Tenerife, Spain
   \and ETH Zurich, Institute of Astronomy, Wolfgang-Pauli-Str. 27, CH-8093 Zurich, Switzerland
   \\
   \\
    }


 
  \abstract
   {}
   {
   Whereas the evolution of gas kinematics of massive galaxies is now relatively well established up to redshift $z\sim 3$, little is known about the kinematics of lower mass (\mstar\ $\leq 10^{10}$\msun) galaxies. We use MUSE, a powerful wide-field, optical integral-field spectrograph (IFS) recently mounted on the VLT, to characterize this galaxy population at intermediate redshift.}
   {We made use of the deepest MUSE observations performed so far on the Hubble Deep Field South (HDFS). 
   This data cube, resulting from 27 hours of integration time, covers a one arcmin$^2$ field of view at an unprecedented depth (with a $1\sigma$ emission-line surface brightness limit of $1 \times 10^{-19}$ \ergscmarcsec)
   and a final spatial resolution of $\approx 0.7$\arcsec. We identified a sample of 28 resolved emission-line galaxies, extending over an area that is at least twice the seeing disk, spread over a redshift interval of $0.2 < z < 1.4$. More than half of the galaxies are at $z\sim 0.3-0.7$, which is a redshift range poorly studied so far with IFS kinematics.
   We used the public HST images and multiband photometry over the HDFS to constrain the stellar mass and star formation rate (SFR) of the galaxies and to perform a morphological analysis using {\sc Galfit}, providing estimates of the disk inclination, disk scale length, and position angle of the major axis. We derived the resolved ionized gas properties of these galaxies from the MUSE data and model the disk (both in 2D and in 3D with \gpk) to retrieve their intrinsic gas kinematics, including the maximum rotation velocity and velocity dispersion.}
   {
   We build a sample of resolved emission-line galaxies of much lower stellar mass and SFR  (by $\sim 1-2$ orders of magnitude) than previous IFS surveys.
   The gas kinematics of most of the spatially resolved MUSE-HDFS galaxies is consistent with disk-like rotation, but about 20\% have velocity dispersions that are larger than the rotation velocities  
   and 30\% are part of a close pair and/or show clear signs of recent gravitational interactions. These fractions are similar to what has been found in previous IFS surveys of more massive galaxies, indicating that the dynamical state of the ionized gas and the level of gravitational interactions of star-forming galaxies is not a strong function of their stellar mass. In the high-mass regime, the MUSE-HDFS galaxies follow the Tully-Fisher relation defined from previous IFS surveys in a similar redshift range. This scaling relation also extends to lower masses/velocities but with a higher dispersion. We find that 90\% of the MUSE-HDFS galaxies with stellar masses below $10^{9.5}$ \msun\ have settled gas disks.
   The MUSE-HDFS galaxies follow the scaling relations defined in the local universe between the specific angular momentum and stellar mass. However, we find that intermediate-redshift, star-forming galaxies 
fill a continuum transition from the spiral to elliptical local scaling relations, according to the dynamical state (i.e., rotation- or dispersion-dominated) of the gas. This indicates that some galaxies may lose 
their angular momentum and become dispersion-dominated prior to becoming passive.}
   {}


   \maketitle
%

\section{Introduction}

The dynamical properties of galaxies at different cosmic epochs provide key constraints on models of galaxy formation and evolution \citep[e.g.,][]{Steinmetz+Navarro1999,Governato+2007,Bouche+2010,Dutton+2011,Schaye+2015}. These properties are well known in the local universe over a large range of galaxy masses. 

So far, studies of the Tully-Fisher relation (hereafter TFR), which relates the maximum rotation velocity of disks to their luminosity \citep{Tully+Fisher1977}, their stellar mass \citep[e.g.,][]{Bell+deJong2001}, or even their total baryonic (i.e.,\, stars plus gas) mass \citep[e.g.,][]{McGaugh+2000} have been restricted to massive (\mstar\ $> 10^{9.5}$ \msun) galaxies \citep[e.g.,][]{Verheijen2001,Pizagno+2007,Courteau+2007,Bershady+2010,Reyes+2011}. In this mass range, most of star-forming galaxies in the local universe are rotation-dominated disks, which follow the TFR with a moderate scatter. Interacting galaxies or mergers \citep[e.g.,][]{Hibbard+2001}, such as the extreme (ultra-)luminous infrared galaxy population \citep[e.g.,][]{Arribas+2014}, are the only star-forming systems which are dispersion-dominated and thus depart significantly from the TFR. However, an {\sc Hi}-based TFR relation was recently investigated for passive early-type galaxies as part of the {\sc Atlas$^{\rm 3D}$} survey \citep{denHeijer+2015}.

In the low-mass regime (\mstar\ $< 10^{9.5}$ \msun), the Tully-Fisher relation of local galaxies has been much less studied, focusing mainly on gas-rich dwarf galaxies which scatter around the relation defined at higher masses and extrapolated to the low-mass domain. However, the TFR tightens when the gas mass is considered in the baryon budget, and is now well calibrated for local gas-dominated dwarfs \citep[e.g.,][]{Stark+2009}. Spanning a wide range of stellar masses ($10^{8.0} <$ \mstar\ $< 10^{11.7}$ \msun) and morphologies (from irregular to spirals), the GHASP survey \citep{Epinat+2008a,Epinat+2008b} is so far the largest and morphologically unbiased survey probing the ionized gas kinematics of local star-forming galaxies with Fabry-Perot integral-field spectroscopy (IFS).  Over the whole \mstar\ $=10^{8.0-11.7}$ \msun\ mass range, a well-selected sample of rotation-dominated GHASP disks follows a relatively tight TFR relation \citep{Torres-Flores+2011}. 
However, \cite{Simons+2015} argued for the existence of a transition stellar mass in the TFR at \mstar\ $= 10^{9.5}$ \msun, also referred to as the ``mass of disk formation''. Above this mass, nearly all galaxies, except for massive ellipticals, a small number of major mergers, and high star-forming systems, are rotation-dominated disks and lie on a relatively tight TFR. Below this mass, the TFR shows significant scatter to low rotation velocity and galaxies can either be rotation-dominated disks on the TFR or asymmetric or compact galaxies that scatter off. Thus, below this transition mass, a galaxy may or may not form a disk.

At high redshift ($z \sim 0.7-4$), our understanding of gas kinematics in emission-line galaxies has made enormous progress thanks to near-infrared, integral-field spectroscopic surveys with VLT/SINFONI and Keck/OSIRIS, such as MASSIV \citep{Contini+2012, Vergani+2012}, SINS \citep{Cresci+2009}, and AMAZE/LSD \citep{Gnerucci+2011}; see also \cite{Glazebrook2013} for a review. These surveys, ranging from a handful to about 100 star-forming galaxies, have been critical in revealing the variety and complexity of the dynamical state of the ionized gas among luminous and massive galaxies at high redshift. Although a large fraction of galaxies are rotation-dominated disks, a significant population (at least 20\%) is found in merging/interacting systems. Furthermore, high-redshift disks are often found to be  unstable and clumpy as revealed by their high (gaseous) velocity dispersion \citep[e.g.,][]{Forster-Schreiber+2009,Epinat+2009,Epinat+2012,Swinbank+2012a,Swinbank+2012b,Wisnioski+2011}. The new generation of near-infrared multi-IFU (KMOS/VLT) and multislit spectrographs (MOSFIRE/Keck) are now used to significantly increase the samples (up to several hundreds of galaxies) through ongoing surveys such as KROSS \citep{Stott+2016}, MOSDEF \citep{Price+2016}, KMOS$^{\rm 3D}$ \citep{Wisnioski+2015}, or KMOS-HIZELS \citep{Sobral+2013}. 

So far, galaxy kinematics at intermediate redshift ($z\sim 0.2-0.7$) has been addressed mainly with multislit spectroscopic surveys \citep[e.g.,\, DEEP2;][]{Kassin+2007,Kassin+2012,Miller+2011,Miller+2014}, due to the limited number of optical IFUs well suited for such analysis. The only exceptions are the sample of 19 Lyman-break analogs at $z\sim 0.2$ observed with OSIRIS/Keck \citep{Goncalves+2010} and the IMAGES survey of $\sim 60$ galaxies at $z\sim 0.6$ with GIRAFFE/VLT \citep{Puech+2010}. Integral-field spectroscopy nonetheless offers a key opportunity to investigate the complex structures of intermediate/high-redshift galaxies. With this technique it is possible to obtain a spectrum at each location in the image of a galaxy. In contrast, the classical technique of long(multi)-slit spectroscopy collects spectra along a 1D slice through a galaxy with the orientation chosen in advance. However, it is often very difficult to determine the major kinematic axis photometrically from the clumpy or barely resolved morphologies of galaxies at these redshifts. With a large field-of-view, high sensitivity, and a wavelength range in the optical domain, MUSE is an excellent instrument for gathering larger samples of IFU-based galaxy kinematics over a broad range of stellar masses.

In $\Lambda$CDM and other hierarchical models of structure formation, galaxies acquire their angular momenta $J$ by tidal torques \citep{Peebles1969,Doroshkevich1970,Barnes+Efstathiou1987}. With IFS surveys, it is now becoming possible to study the angular momentum properties of galaxies from accurate size and gas kinematics measurements \citep[e.g.,][]{Bouche+2007,Burkert+2015,Obreschkow+2015}.
However, because of the preselection of galaxies (magnitude-limited or color-selected samples) in previous IFS surveys and the limited sensitivity of previous instruments, such studies were limited to the upper end (\mstar\ $\geq 10^{10}$\msun) of the galaxy mass function. Indeed, low-mass galaxies are typically smaller, fainter, and have lower SFRs than massive galaxies. They are thus difficult to select, and obtaining high-quality resolved data for such galaxies has previously been very expensive in terms of telescope time, except for a handful of amplified galaxies in lensing clusters \citep[e.g.,][]{Swinbank+2006,Jones+2010,Livermore+2015,Leethochawalit+2016}. Thanks to MUSE \citep{Bacon+2010}, the new very sensitive and wide-field IFS on the VLT, it is now possible to perform blind and deep surveys to probe the mass regime below $10^{10}$ \msun\ down to $10^8$ \msun\ at various redshifts and thus cosmic epochs. 

We show in this paper a first illustration of the power of MUSE in this research area, making use of the deepest observation performed so far in the region of the Hubble Deep Field South \citep[HDFS;][]{Bacon+2015}. We identify a sample of 28 spatially resolved emission-line galaxies in this 27 hour exposure time data cube, spread over a large redshift interval of $0.2 < z < 1.4$ and covering three orders of magnitude in stellar mass. More than half of the galaxies are at $z\sim 0.3-0.7$, which is a redshift range poorly studied so far with IFS kinematics. Owing to the $\approx 0.7$\arcsec\ spatial resolution of the MUSE data, we are able to derive the spatially resolved gas kinematic properties of these galaxies; we shed light, for the first time, on the dynamical properties of low-mass, emission-line galaxies over the last 8 Gyr.  We emphasize that this analysis focuses on the kinematics of the ionized gas, which is traced by the bright emission lines. Stellar kinematics, which is much more difficult to derive in intermediate/high-redshift galaxies, will be the focus of further analyses. 

This paper is structured as follows. In section~\ref{sec:data} we present the MUSE-HDFS dataset, the selection criteria used to define a sample of 28 spatially resolved galaxies, and how this sample compares to previous IFS surveys of intermediate/high-redshift galaxies in terms of stellar masses and star formation rates. The morphological (based on HST images) and ionized gas kinematic (based on MUSE data) analysis of these 28 galaxies is described in sections~\ref{sec:morpho} and \ref{sec:kine}, respectively. Results concerning the dynamical state of galaxies, their location on the TFR and their angular momentum are presented in Sect.~\ref{sec:results}. Conclusions are given in Sect.~\ref{sec:conclu}. 
Throughout the paper, we assume a $\Lambda$CDM cosmology with $\Omega_m = 0.3$, $\Omega_{\Lambda} = 0.7$ and $H_0 = 70$ \kmsmpc. 

\section{Spatially resolved, emission-line galaxies in MUSE-HDFS}
\label{sec:data}

\begin{figure} \resizebox{\hsize}{!}{\includegraphics{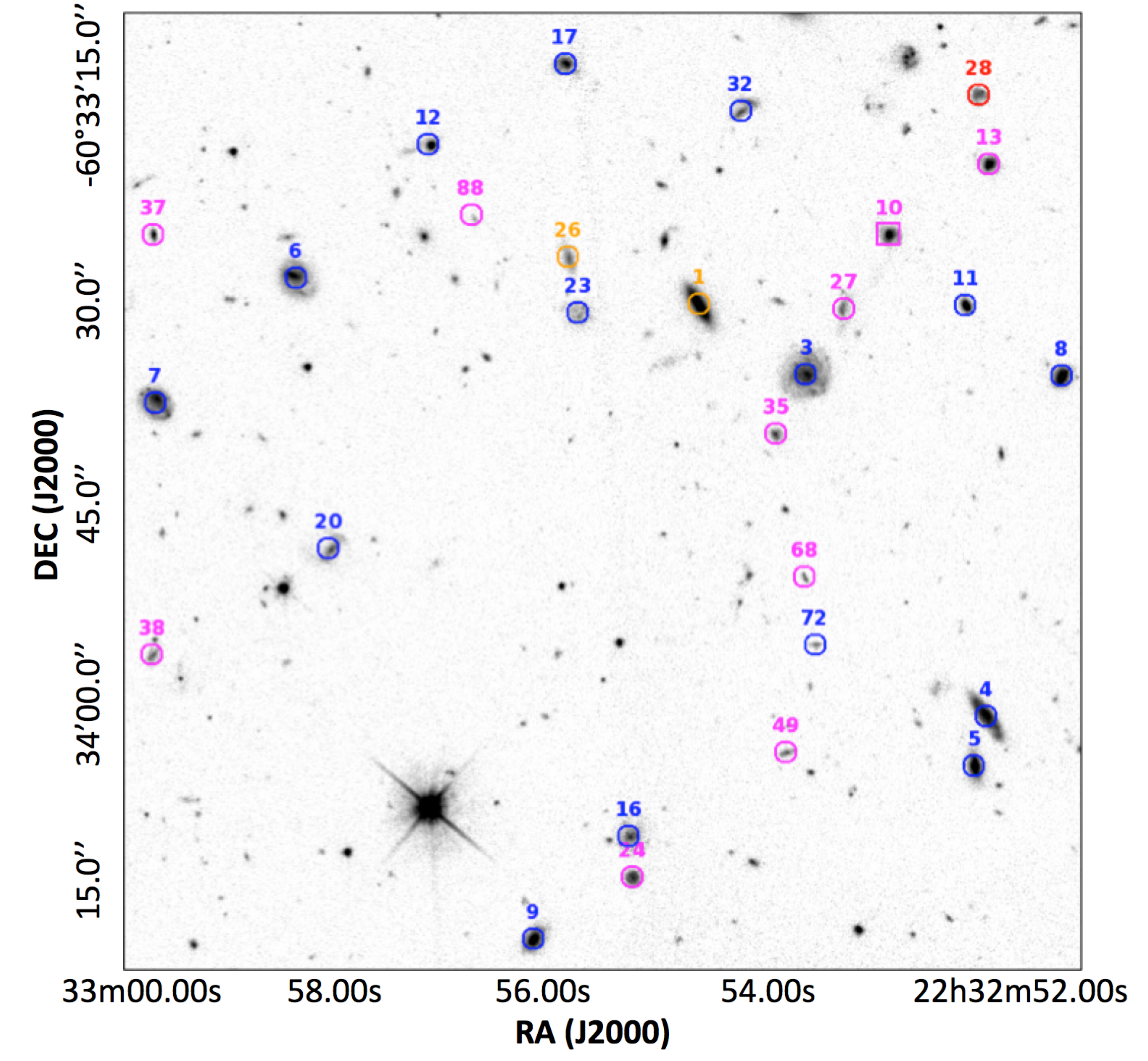}}
\caption{Location of the spatially resolved galaxies selected in the MUSE-HDFS data cube overlaid on the HST/WFPC2 F814W image. The diameter of the circles is equal to the effective spatial resolution ($\approx 0.7\arcsec$ FWHM) of the MUSE data cube. Identification numbers are those from the source catalog of \cite{Bacon+2015}. The AGN ID\#10 is indicated with a square. Colors correspond to different redshift intervals and thus to different set of bright emission lines accessible within the spectral range of MUSE (orange $= z < 0.27$, from \hbeta\ to \halpha; red $= 0.27 < z < 0.42$, from \oii\ to \halpha; blue $= 0.42 < z < 0.86$, from \oii\ to \oiiib; magenta $= 0.86 < z < 1.49$, \oii\ only).}
\label{fig:field}
\end{figure}

\subsection{The MUSE-HDFS dataset}

The Hubble Deep Field South (HDFS) was observed with MUSE, mounted at the ESO-VLT UT4, during six nights of the last commissioning run of the instrument between July 25 and August 3, 2014. A single $1\arcmin \times 1\arcmin$ MUSE field centered on $\alpha = 22\rm{h}32$\arcmin$55.64$\arcsec, $\delta = -60$\degr33\arcmin47\arcsec\ in the HDFS was observed, with a total combined on-target exposure time of 27 hours, reaching a $1\sigma$ emission-line surface brightness limit of $1 \times 10^{-19}$ \ergscmarcsec. The observing strategy and data reduction procedure are fully described in \citet{Bacon+2015}. In this paper, we simply summarize the main steps of the data acquisition/processing. 

The final MUSE-HDFS data cube is the result of a combination of 54 individual exposures of 30 minutes exposure time each. The instrument position angle (PA) was rotated by $90$\degr\ between two subsequent individual exposures together with random dithering offsets within a box of $3$\arcsec.  This strategy ensures that each position in the field (except for the center) falls onto four completely different locations on the detectors. The individual reduced and registered data cubes were coadded into a final data cube of $326 \times 331$ spatial pixels (spaxels), 
each with $3641$ spectral pixels ranging from $4750$\,\am\ to $9300$\,\am. As a result of the observing strategy (rotations and dithering), regions at the very border of the MUSE field of view (FoV) received less than the full exposure time. The spectral resolution of the final data cube is $\approx 2.3$\,\am\ FWHM with a sampling of $1.25$\,\am/pixel. The effective spatial resolution in the combined data cube is $0.66$\arcsec\ FWHM at $7000$\,\am and is about 10\% better (worse) at the red (blue) end of the spectral range. The spatial sampling is equal to $0.2$\arcsec/pixel. The flux calibration is consistent with HST photometry of the stars in the HDFS-MUSE FoV to within $\pm 0.05$ mag. The final wavelength solution is calibrated on air and is corrected for the heliocentric velocity.

This fully calibrated MUSE-HDFS data cube comes with a cube of the propagated variance estimate, which is used in the subsequent analysis especially during the line fitting process (see sect.~\ref{sec:linefitting}), as well as an exposure cube giving the number of exposures used for each spaxel. The reduced data cube, the associated source catalog \citep{Bacon+2015}, and an interactive source browser have all been made available to the public\footnote{\tt http://muse-vlt.eu/science/hdfs-v1-0}.

\begin{figure*}
\includegraphics[width=6cm]{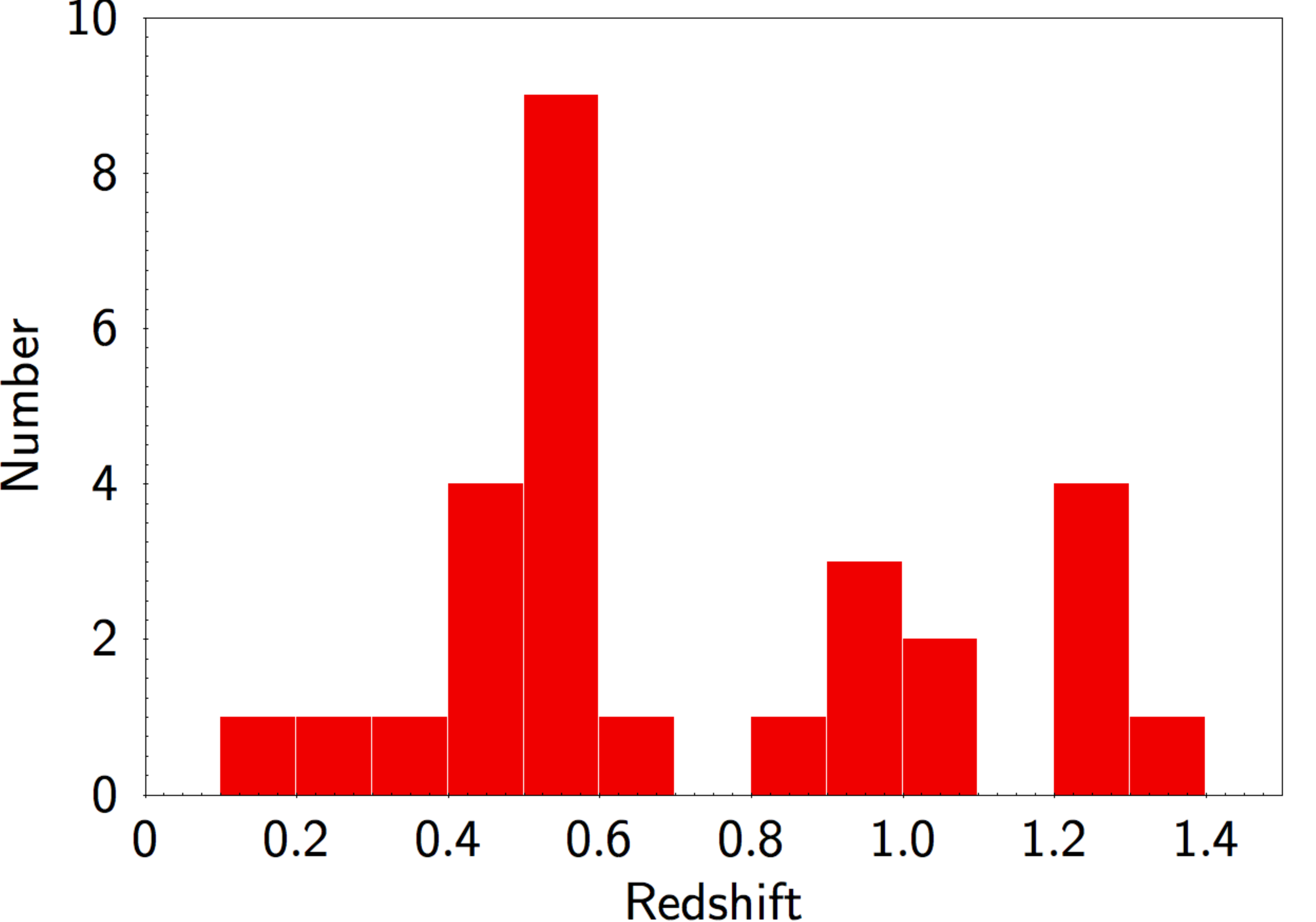}
\includegraphics[width=6cm]{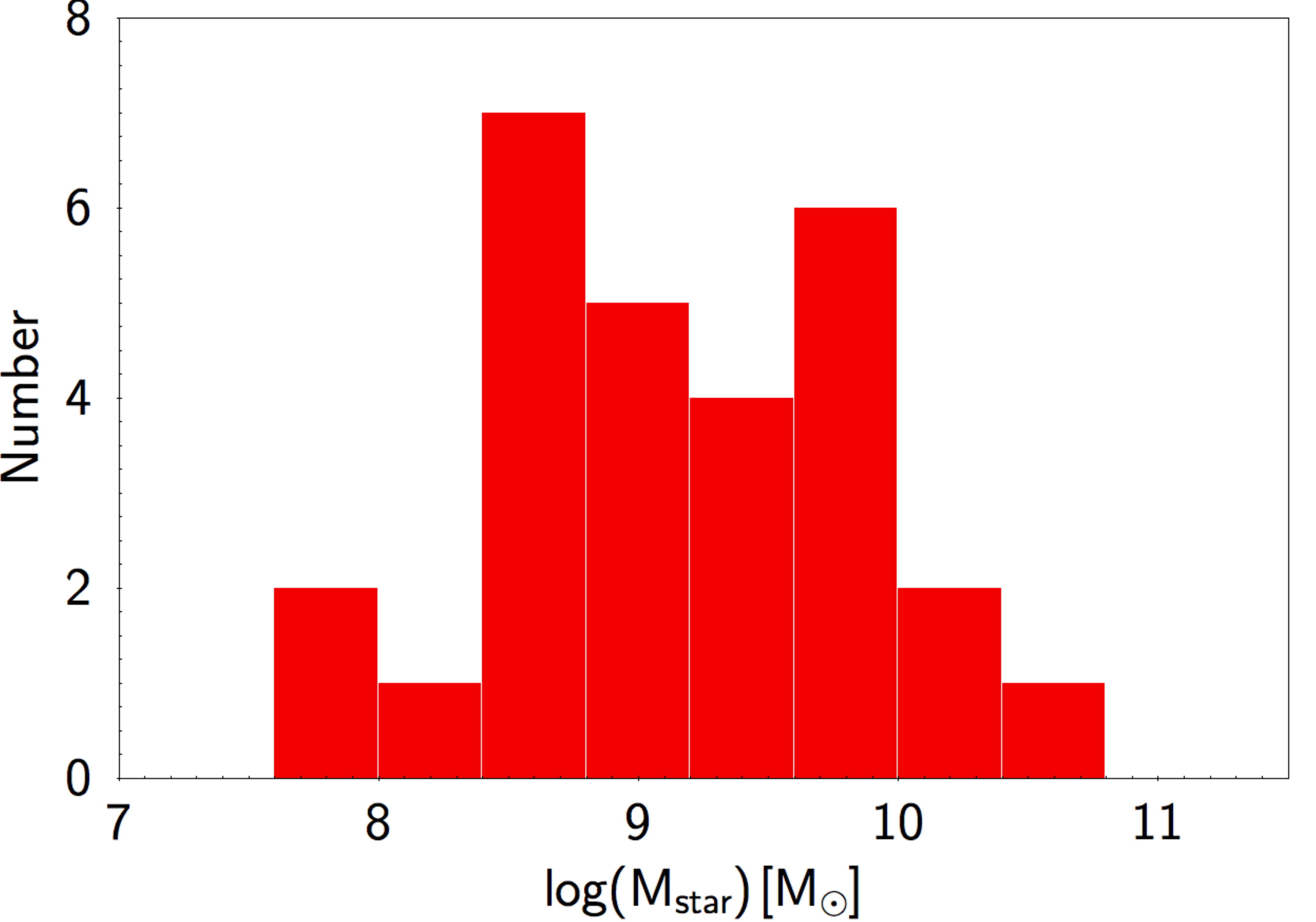}
\includegraphics[width=6cm]{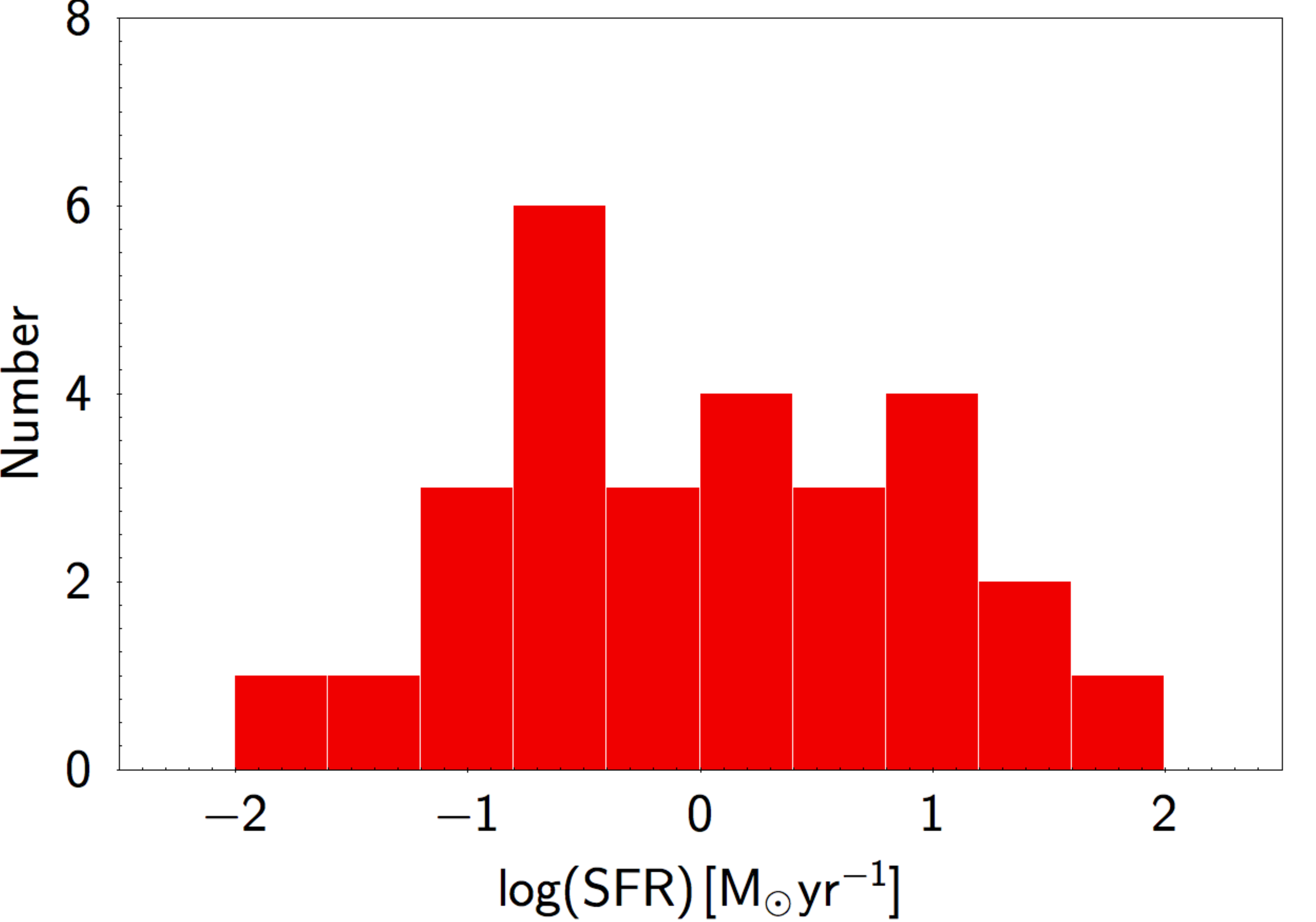}
\caption{Histograms of redshift ({\it left panel}), stellar mass ({\it middle panel}), and star formation rate ({\it right panel}) for the spatially resolved galaxies selected in the MUSE-HDFS.}
\label{fig:zmstarsfr}
\end{figure*}

\subsection{A sample of spatially resolved emission-line galaxies}

\cite{Bacon+2015} published a catalog of 189 sources from the MUSE-HDFS data cube with extracted spectra and securely determined redshifts. Most of these objects come from the HST photometric catalog of \cite{Casertano+2000}, but 27 of the Lyman$-\alpha$ emitters discovered with MUSE do not have any counterpart in HST broadband images. 

\subsubsection{Selection criteria}
\label{sec:selection}

The parent sample of emission-line galaxies used for this study consists in 70 galaxies below a redshift cutoff $z\approx 1.5$, beyond which the \oii\ doublet leaves the spectral domain of MUSE. This sample includes 61 \oii\ emitters ($z=0.29-1.48$), seven nearby galaxies ($z=0.12-0.28$, whose \oii\ emission line is redshifted below the $4800$\,\am\ blue cutoff of MUSE), and two type-2 active galactic nuclei (AGN), identified by strong \neiii\ and \nev\ narrow emission lines,  both belonging to the same group of galaxies at $z\approx 1.284$.

We performed an emission-line fitting (with {\sc Camel}, see sect.~\ref{sec:linefitting}) on each of these galaxies to produce line flux and associated signal-to-noise ratio (S/N) maps. We used these measurements to select resolved galaxies by applying the following criterion: galaxies are included in our spatially resolved sample when they cover an area of at least 20 spaxels (corresponding to about twice the seeing disk of the MUSE-HDFS data cube) for the brightest emission line above a threshold in S/N\ per pixel of 15. This conservative selection ensures that the galaxy is spatially resolved and that the signal is good enough to trace the gas kinematics up to large galactocentric distances (i.e., ranging from $\sim 2.6$ to $13$ disk scale lengths with a median value of $5$ for our MUSE-HDFS sample). A total number of 28 spatially resolved galaxies in the MUSE FoV satisfy this criterion after rejecting galaxy ID\#545, which is at the margin of these criteria. Relaxing the constraint on resolution (e.g., decreasing to 1.5 times the seeing disk) or on S/N (threshold at 10) would increase the sample by $\approx 10$\%, and therefore not have a major impact on the final conclusions. Figure~\ref{fig:field} shows the location of these galaxies in the MUSE FoV, including one AGN  (ID\#10), overlaid on the HST/WFPC2 F814W image.

\subsubsection{Global properties: Redshift, stellar mass, and star formation rate}
\label{sec:globprop}

\begin{figure} \resizebox{\hsize}{!}{\includegraphics{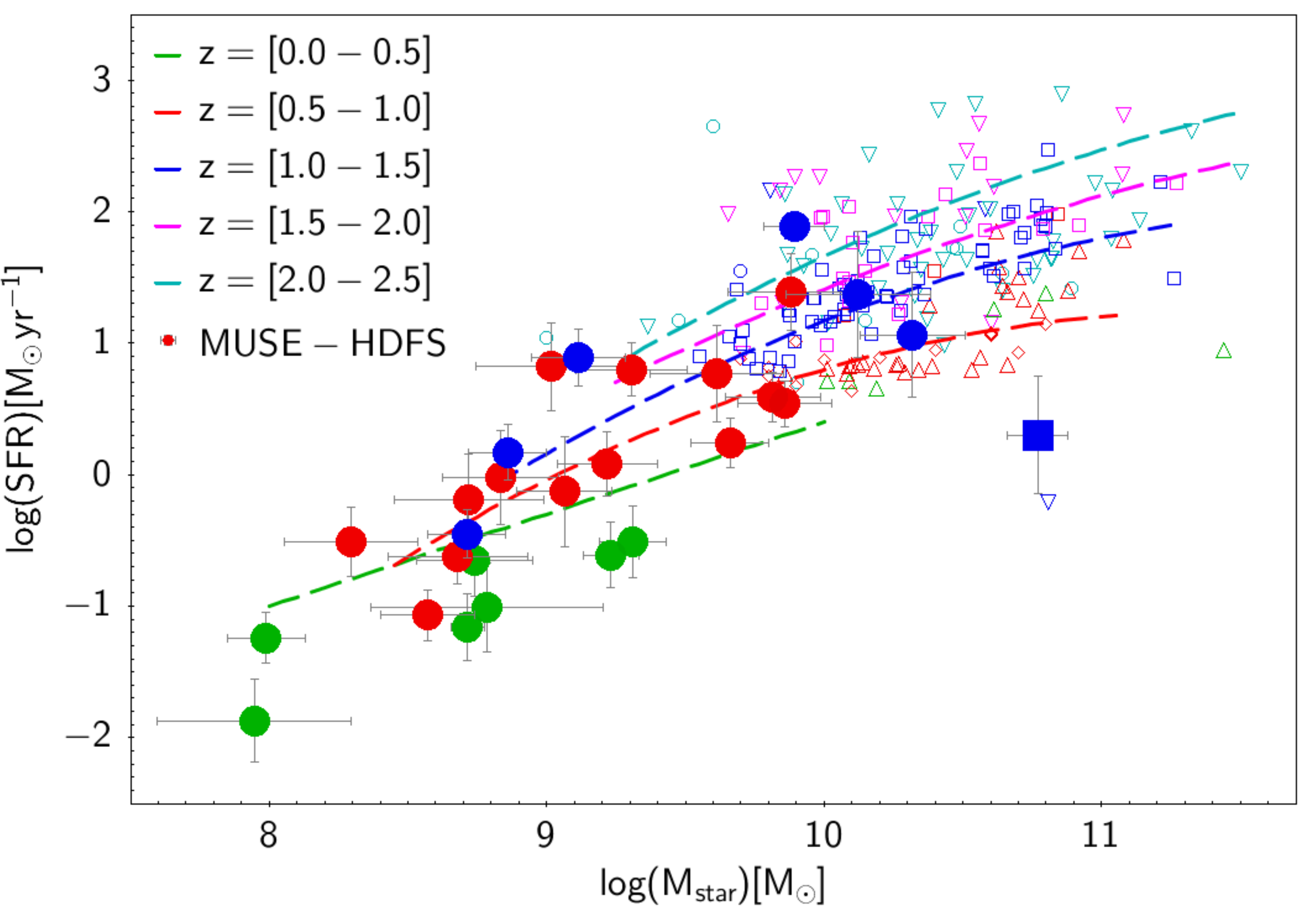}}
\caption{Sample of spatially resolved MUSE-HDFS galaxies with respect to the main sequence of star-forming galaxies. The MUSE-HDFS sample (large filled dots with error bars, 
except for the AGN ID \#10 indicated with a large filled square) is compared with previous IFS surveys of distant galaxies up to $z\sim 2.5$(open symbols, see text for references): IMAGES ($z \sim 0.4-0.75$, upward triangles), KMOS-HIZELS ($z \sim 0.8$, diamonds), MASSIV ($z \sim 0.9-1.8$, squares), SINS ($z \sim 1.4-2.6$, downward triangles), and OSIRIS ($z \sim 1.5-3.3$, circles). 
Dashed lines represent the empirical relations between SFR and stellar mass for different redshifts between $z=0$ (SDSS; \citealp{Brinchmann+2004}) and $z=[0.5-2.5]$ \citep{Whitaker+2014}. The MUSE-HDFS sample explores new territory in galaxy evolution studies, extending the analysis to stellar masses and SFR that are nearly two orders of magnitude lower than previous IFS samples.}
\label{fig:sfgms}
\end{figure}

The spatially resolved galaxies selected in MUSE-HDFS are spread over a redshift range $0.17 < z < 1.36$ with a median value of $z\approx 0.58$ (see Figure~\ref{fig:zmstarsfr}, left panel). Depending on their redshift, different sets of bright emission lines are accessible within the spectral range (4750-9300\,\AA) of MUSE. Half of our sample galaxies are in the redshift range $0.42 < z < 0.86$, for which we can observe emission lines from \oii\ to \oiiib\ with MUSE. Eleven galaxies are at higher redshift ($0.86 < z < 1.49$), four of which are part of a group at $z\approx 1.28$: ID \#10 (one of the two type-2 AGNs in MUSE-HDFS sample), ID \#13, 27, and 35 \citep[see][]{Bacon+2015}. For these higher redshift galaxies, the only bright emission line accessible with MUSE is the \oii\ doublet. Three galaxies are at relatively low redshift. At $z\approx 0.32$, galaxy ID \#28 shows the full set of bright emission lines from \oii\ to \halpha. However galaxies ID \#1 and 26 have redshifts that are too low ($z\approx 0.17$ and $0.22$ respectively) to include \oii. 

The stellar masses and star formation rates of the galaxies were estimated using the stellar population synthesis (SPS) code FAST \citep{Kriek+2009}. We $\chi^2$-fitted SPS templates 
 to the broadband visible and near-infrared (NIR) photometry. The visible photometry comes from the HST catalog \citep{Casertano+2000} in four filters (F300W, F450W, F606W, and F814W). Near-infrared photometry is derived from VLT/ISAAC deep exposures in the $J$, $H$, and $K$ bands \citep{Labbe+2003}. The templates were constructed from the stellar libraries of \cite{Bruzual+Charlot2003} assuming a \cite{Chabrier2003} IMF and an exponentially declining star formation history ($\text{SFR} \propto \exp(-t/\tau)$ with $8.5 < \log(\tau/{\rm yr}) < 10$). The redshifts were fixed to the accurate spectroscopic values determined from the MUSE spectra. Ages (in logarithmic units) were allowed to vary between $8$ and $10.2$ with a step of $0.2$. The dust attenuation curve was parameterized according to the \cite{Calzetti2001} dust law and dust extinction in the visual was tested between $0 < A_V < 3$ in steps of $0.1$ magnitudes. The best-fit stellar mass and SFR values are listed in Table~\ref{tbl:sample}. 
  
Galaxy stellar masses are spread over 3 orders of magnitude, ranging from $\log$(\mstarsmsun) $\sim 8$ to $\sim 11$ with a median value $\log$(\mstarsmsun) $\approx 9.1$ (see Figure~\ref{fig:zmstarsfr}, middle panel). Galaxy star formation rates are spread over 4 orders of magnitude, ranging from $\approx 0.01$ \msunpyr\ to $\approx 78$ \msunpyr, with a median value of SFR $\approx 1.2$ \msunpyr\ (see Figure~\ref{fig:zmstarsfr}, right panel). 
Figure~\ref{fig:sfgms} compares the SFR versus stellar mass relationship for the MUSE-HDFS sample to previous IFS samples at various redshifts up to $z\sim 2.5$ (IMAGES $z \sim 0.4-0.75$ \citealp{Puech+2010}; KMOS-HIZELS $z \sim 0.8$ \citealp{Sobral+2013}; MASSIV $z \sim 0.9-1.8$ \citealp{Contini+2012}; SINS $z \sim 1.4-2.6$ \citealp{Forster-Schreiber+2009}; and OSIRIS $z \sim 1.5-3.3$ \citealp{Law+2009}
) and to empirical relations of the so-called main sequence for normal star-forming galaxies at $z=0$ (SDSS; \citealp{Brinchmann+2004}) and higher redshifts ($z=[0.5-2.5]$, \citealp{Whitaker+2014}). This figure clearly shows that the MUSE-HDFS sample explores new territory in terms of galaxy stellar mass and SFR. Indeed, the previous IFS surveys almost exclusively probed massive galaxies (\mstar\ $\geq 10^{9.5}$ \msun) with relatively strong star formation activity (SFR $\geq 10$ \msunpyr). With this deep MUSE exposure in the HDFS, we extend the analysis to a galaxy sample that is two orders of magnitude lower in SFR and 1.5 orders of magnitude lower in stellar mass. Moreover, except for the AGN galaxy ID\#10 at $z\approx 1.3$ that is significantly below the main-sequence relation at this redshift, the MUSE-HDFS sample broadly follows the empirical relations for normal star-forming galaxies recently defined  from the 3D-HST observations in the CANDELS fields \citep{Whitaker+2014}.

\section{Morphology from HST images}
\label{sec:morpho}

The kinematic analysis of galaxies requires some knowledge of their morphologies (e.g., inclination, effective radius, and major axis position angle), in particular when determining  velocities in the galaxy plane. 
Therefore, it is highly recommended to use the highest spatial resolution and deepest observations available 
in order to constrain galaxy morphologies.  HST images acquired in several bands with the Wide Field Planetary Camera 2 (WFPC2) are available for the HDFS \citep{Williams+1996}. We  used the publicly available reduced F814W version 2 drizzled images with a sampling of 0.04$\arcsec$/pixel and a total exposure time of 31.2 hours. This band is the best compromise between depth, spatial resolution, and wavelength  necessary in our redshift range to avoid probing the rest-frame UV light from the youngest stellar populations.

Morphological parameters of some galaxies in HDFS were previously derived by \citet{Trujillo+2004}. In their study, they performed a bulge-disk decomposition on 2D surface brightness profiles. However, they did not study all of the galaxies we have in our sample because of both a redshift cut ($z<1$) based on photometric redshifts \citep{Labbe+2003} and a magnitude-limited selection ($I_{814}^{\rm AB}<24.5$ for $z<1$). In addition, they did not provide the PA of the major axis of the disk component that can be used to evaluate the level of kinematic disturbances (e.g., \citealp{Epinat+2012}).

We therefore decided to perform a similar analysis using {\sc Galfit} \citep{Peng+2002}. First we identified eight stars in the FoV based on the MUSE spectra, one of them was however saturated in the HST images. We extracted the seven non-saturated stellar PSFs from HST images and modeled them with a Moffat function using {\sc Galfit}. We noticed a trend in the variation of the Moffat parameters across the MUSE FoV (in particular the orientation of elongation seems to vary from north to south, whereas the FWHM seems to increase from east to west), but we did not have enough stars to characterize this variation or to adapt the modeled PSF to the location of each galaxy in the MUSE FoV. Since the variations were not dramatic (less than 5\%) and since the FWHM is small compared to the size of the galaxies in our sample (which are resolved with MUSE), we assumed the PSF to be circular.   We computed a model PSF using the average parameters deduced from the seven stars to reduce the noise in the
PSF, 
$FWHM=0.136\arcsec$ and $\beta=2.0$ (Moffat index).

Using this PSF, we modeled all the galaxies in our sample with a bulge and a disk (see Figure~\ref{fig:vfexample} for an example and Figures~\ref{fig:appen_figs_obj1} to~\ref{fig:appen_figs_obj88} for the full sample), as in \citet{Trujillo+2004}. 

The bulge is described by a \citet{Sersic+1968} profile
$$I(r)=I_b(0) e^{-b_n (r/r_{e})^{1/n}},$$
where $I_b(0)$ is the bulge central intensity, $r_{e}$ is the bulge effective radius, and $n$ is the Sersic index. The parameter $b_n$ depends on the index and is derived from $\Gamma$ and incomplete $\gamma$ functions \citep{Abramowitz+1964}, $\Gamma(2n)=2 \gamma(2n, b_n)$. To make the analysis simpler, we fixed this index to $n=4$, i.e., taking a classical de Vaucouleurs profile, and made the assumption that the bulge is spheroidal to first order by setting the axis ratio equal to unity.

The disk is described as an exponential disk \citep{Freeman+1970}
$$I(r)=I_d(0) e^{-r/R_d},$$
where $I_d(0)$ is the disk central intensity and $R_{\rm d}$ is the disk scale length. The axis ratio and position angle of the disk major axis are free parameters. In addition, the bulge and disk share the same center.
We investigated any dependence of our results with respect to i) various PSF models, such as a stacked PSF or a  Tiny Tim modeling \citep{Krist+2011} and ii) free Sersic parameters for  bulge modeling. 
The change of PSF only slightly affects the results. The larger the PSF, the smaller the disks, but the difference is negligible ($<2\%$).
The impact on the inclination is also negligible,$<5$\degr, which is in agreement with the expected uncertainties on the fit, and the same holds for the position angle, $<1$\degr, except for ID\#12, which is the smallest galaxy in our sample.
Using free parameters (index and axis ratio) for the bulge modeling does have a larger impact. Indeed, in some cases, the bulge model can fit what is expected to be a disk or it can efficiently fit  a bar and therefore provides a better fit of the underlying disk. In general, only strong bars are problematic and even with a bulge-free approach, it remains difficult for the disk not to be disturbed by the bar residuals. In spite of this, the fits (with free or fixed parameters) are not dramatically different. Inclinations agree within $10$\degr\ (except for galaxies ID \#11 and 13) with a typical dispersion of $\approx 4$\degr\ (excluding these two extreme galaxies). The disk sizes also agree within 20\%, except for five galaxies (IDs \#3, 4, 9, 20, and 37), with a typical dispersion of $\approx 8\%$ excluding these five galaxies. Values of the position angle also show very good agreement; these values are always better than $50$\degr\ and mainly better than $10$\degr\ (except for eight galaxies that are small, have a disk axis ratio $>0.8$, or have a bar).

\begin{figure} 
\resizebox{\hsize}{!}{\includegraphics{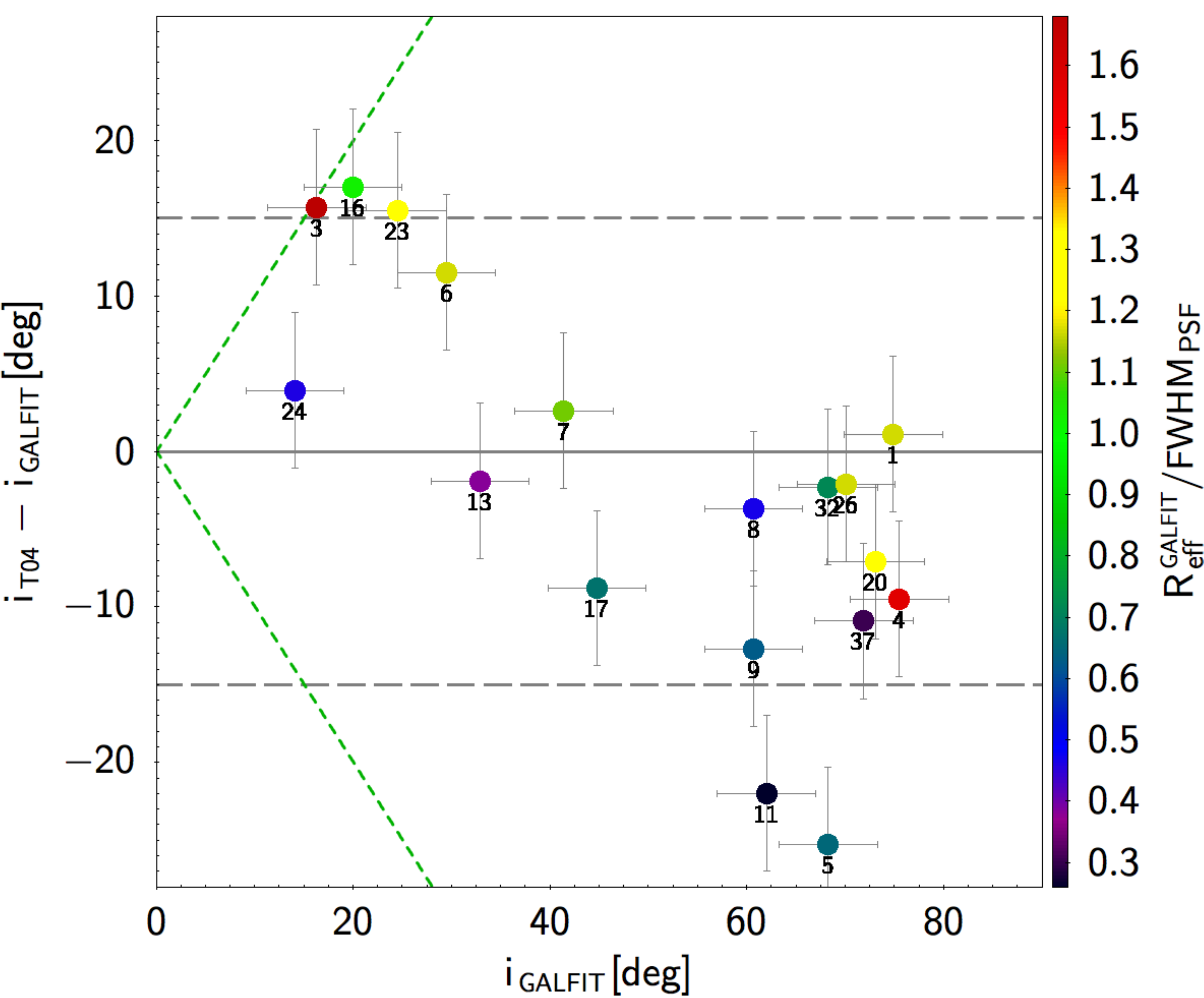}}
\caption{Comparison of disk inclinations derived with {\sc Galfit} ($i_{\rm GALFIT}$) for the MUSE-HDFS spatially resolved galaxies and previous measurements ($i_{\rm T04}$) performed by \citet{Trujillo+2004}. Green dotted lines are for $i_{\rm GALFIT}=0$ and $i_{\rm T04}=0$. Dots are color-coded as a function of the relative size of the disk (effective radius derived from {\sc Galfit}) with respect to the MUSE PSF FWHM. Labels indicate the galaxy ID.}
\label{fig:igalfitvsit04}
\end{figure}

In Figure~\ref{fig:igalfitvsit04}, we compare the disk inclination derived from our morphological decomposition to that provided in \citet{Trujillo+2004}. The inclination $i$ is derived from the axis ratio $b/a$ (from {\sc Galfit})  or from the ellipticity $\epsilon$ (from \citealp{Trujillo+2004}) assuming a thin disk
$$\cos{i} = \frac{b}{a} = 1 - \epsilon. $$
The inclinations agree within $\approx 20$\degr. We noticed, however, that low/high-inclination galaxies have a higher/lower inclination according to \citet{Trujillo+2004}. Position angles cannot be compared since they are not provided by \citet{Trujillo+2004}. We further checked that i) using \citet{Trujillo+2004} values for the disk inclination would not change the final results/conclusions of this paper and ii) that the distribution of inclinations in the MUSE-HDFS sample is compatible with the theoretical expectation for uniformly and randomly oriented thin disks \citep[see][sect.\,3.2.2 and Fig.\,2]{Epinat+2012}. Indeed, we found a median value of $62\degr$, which is very close to the theoretical value and a Kolmogorov-Smirnov test gives a probability higher than 75\% that the observed distribution is compatible with the theoretical distribution. 

All these comparisons indicate that the various analyses/assumptions are consistent within the uncertainties (depending on the model and PSF). Our model has the advantage that it is simple with a minimum of free parameters, making the analysis less sensitive to different interpretations. The derived morphological parameters of the disks (inclination, position angle, and  scale length) are listed in Table~\ref{tbl:sample}. 

\section{Kinematics from MUSE data cube}
\label{sec:kine}

\subsection{Emission-line flux and velocity maps}
\label{sec:linefitting}

\begin{figure} \resizebox{\hsize}{!}{\includegraphics{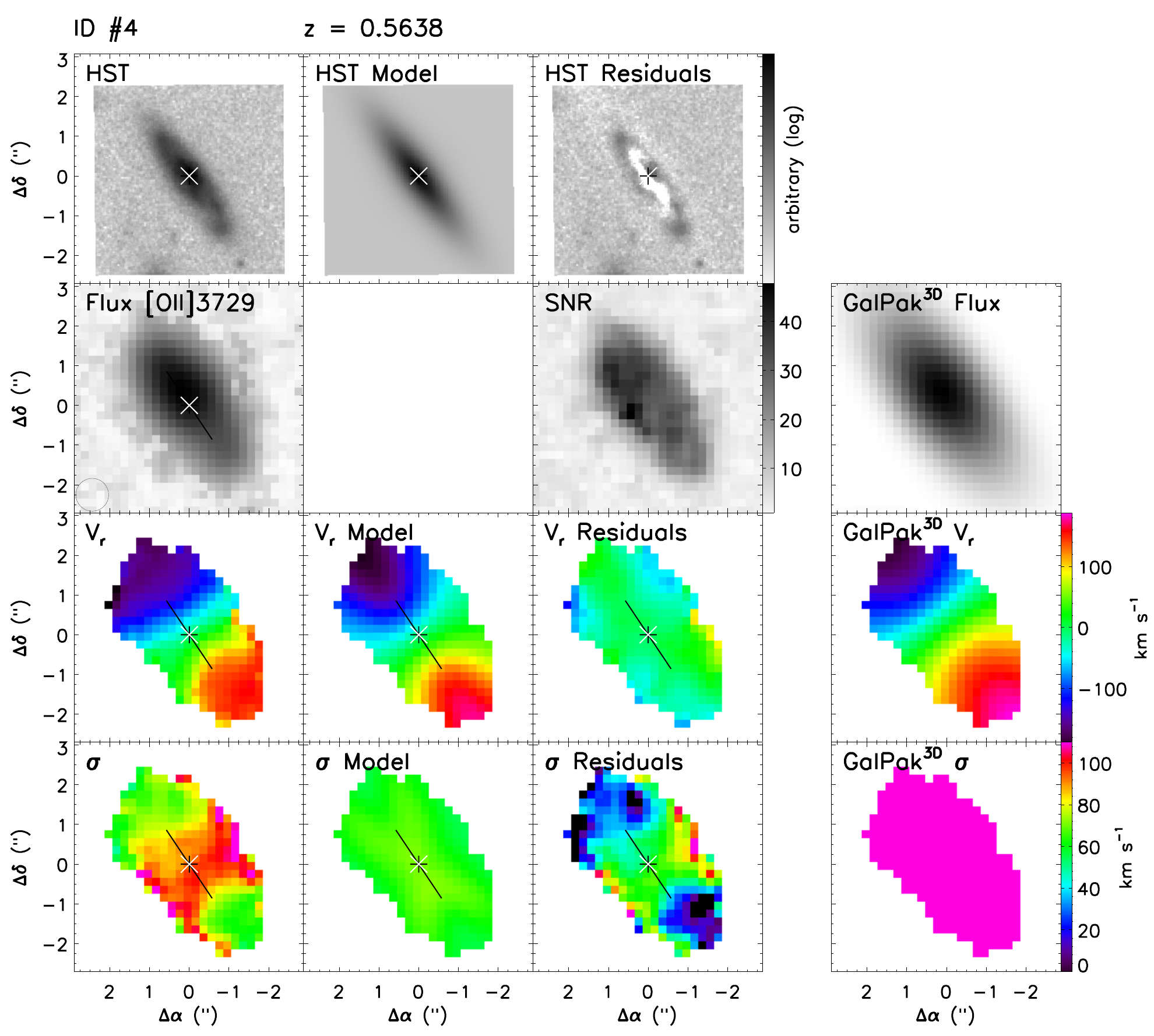}}
\caption{Example of morpho-kinematic analysis for galaxy ID \#4 at $z\approx 0.56$. Description is given from {\it left} to {\it right}. {\it Top} row: HST/WFPC2 F814W image, {\sc Galfit} model (disk$+$bulge), and residuals, all in the same arbitrary log-scale units. {\it Second} row: MUSE \oiib\ flux map (log scale) with the PSF FWHM size indicated with the black circle, corresponding S/N map (linear scale), and modeled flux map (log scale) from \gpk. {\it Third} row: MUSE observed velocity field from \oiib, velocity field of the 2D rotating disk model, associated residual velocity field, and rotating disk model from \gpk. {\it Bottom}: MUSE observed velocity dispersion map from \oiib, velocity dispersion map deduced from the 2D velocity field model (which takes into account beam-smearing effect and spectral PSF), convolved associated velocity dispersion map, and velocity dispersion map from \gpk. In each map, north is up and east is left. The center used for kinematic modeling is indicated with a white cross, the position angle is indicated with a black line that ends at the effective radius. 
}
\label{fig:vfexample}
\end{figure}

Emission-line fitting was performed in the MUSE-HDFS data cube for each of the 28 spatially resolved galaxies with the python code {\sc Camel}\footnote{\tt https://bitbucket.org/bepinat/camel.git}, as described in \cite{Epinat+2012}. To increase the S/N without significantly degrading the spatial resolution, a subresolution 2D spatial Gaussian smoothing of two pixels ($\approx 0.6 \times {\rm FWHM_{PSF}}$) was applied to the data cube extracted around each galaxy. The spectral range was divided into three distincts domains, depending on the redshift of the galaxy: \oii\ doublet only, \hbeta\ $+$ \oiii, and \halpha\ $+$ \nii. For each spaxel, the spectrum around the emission lines was fitted with a single Gaussian profile and a constant continuum. For each domain, all the emission lines are assumed to trace the same rotation velocity and velocity dispersion (i.e., the same FWHM). 
To minimize the effects of noise that sky line residuals introduce into the line parameter determination, the variance spectrum was used as an estimate of the noise to weight the contribution of each spectral element. This fitting technique produced the line flux maps, velocity field, and  velocity dispersion map for the sample 
galaxies, together with S/N maps for these quantities (see Figure~\ref{fig:vfexample} for an example and Figures~\ref{fig:appen_figs_obj1} to~\ref{fig:appen_figs_obj88} for the full sample); in the figures, only spaxels with a S/N $> 5$ are shown. We use different S/N thresholds for selection and kinematic measurements/analysis, as we want a robust selection as well as kinematic measurements that rely on both high quality data (S/N $>15$) and a statistical number of spaxels ($5<$ S/N $<15$).

\subsection{Disk kinematics modeling}
\label{sec:diskmodel}

We developed rotating disk models to reproduce the morpho-kinematics of the sample galaxies. 
Two methods are used in this process. The first "classical" method (Sect.~\ref{sec:2D}) compares the 2D velocity maps produced above to 2D kinematic models \citep{Epinat+2010, Epinat+2012}. The second method (Sect.~\ref{sec:galpak}), commonly used in radio astronomy but recently introduced to optical/NIR data by \citet{Bouche+2015}, directly compares the observed 3D data cube to a full 3D model data cube that takes the PSF and LSF convolutions  into account \citep[see also][]{Diteodoro+Fraternali2015}.

The classical method is robust and has been extensively used \citep[e.g.,][]{Forster-Schreiber+2009, Cresci+2009, Law+2009, Epinat+2012}, but it requires a priori knowledge of the disk inclination (from morphological analysis performed on high-resolution HST images as in Sect.~\ref{sec:morpho}) to retrieve the intrinsic deprojected rotation velocity. 
The second method (3D fitting) has the potential advantage that it can simultaneously recover both morphological (inclination, center, size)  and kinematic parameters. Hence, the 3D fitting algorithm can in principle provide additional information, especially in the regime of low signal-to-noise or if high-resolution HST images are not available.  However, a parametric 3D model may yield erroneous results in situations when the model is too simplified compared to complexities present in the data (such as mergers, bar morphology, kinematic and morphological offsets, and multiple Sersic components).
The following subsections describe the two methods.

\begin{figure*} 
\includegraphics[width=9.3cm]{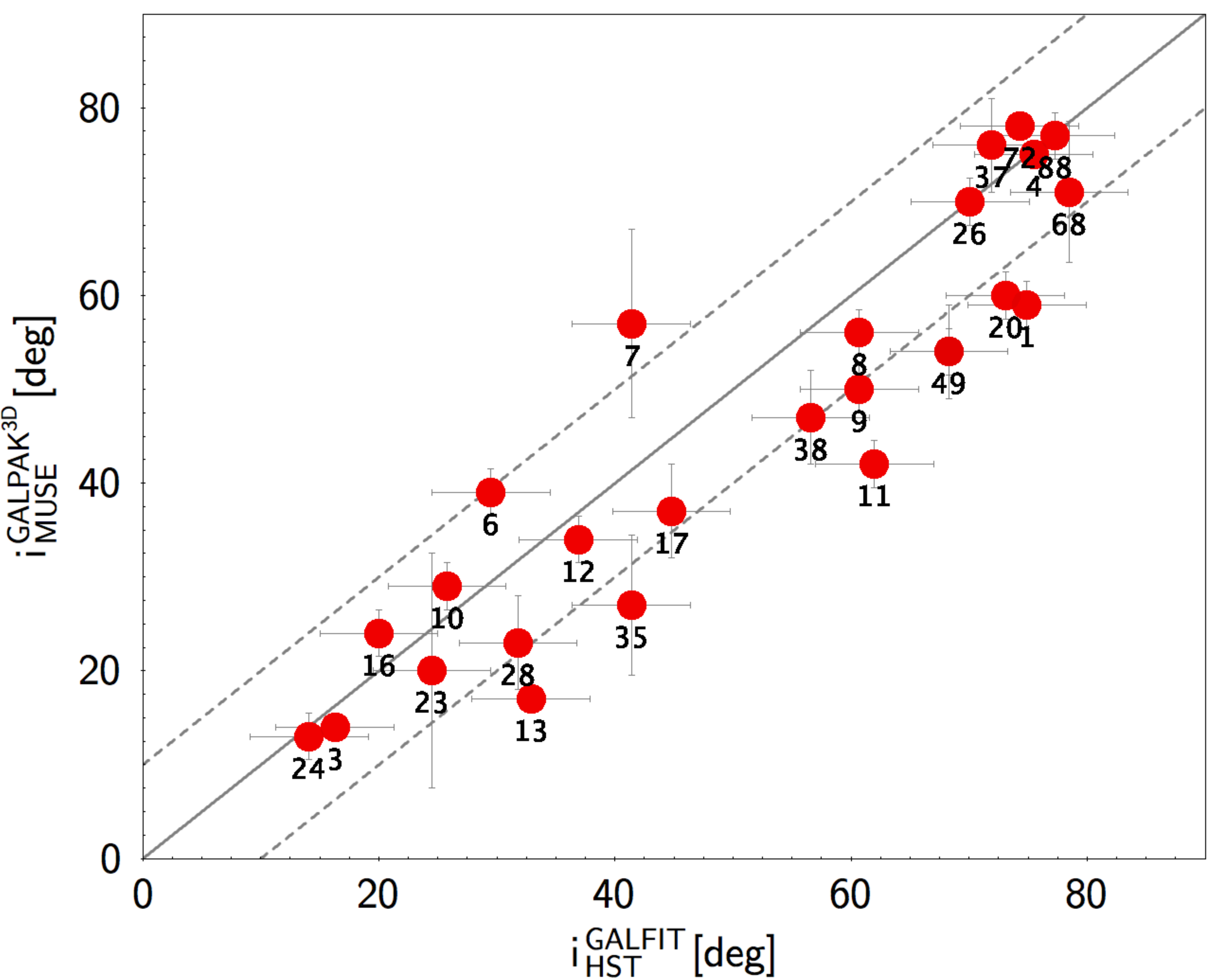}\includegraphics[width=9.3cm]{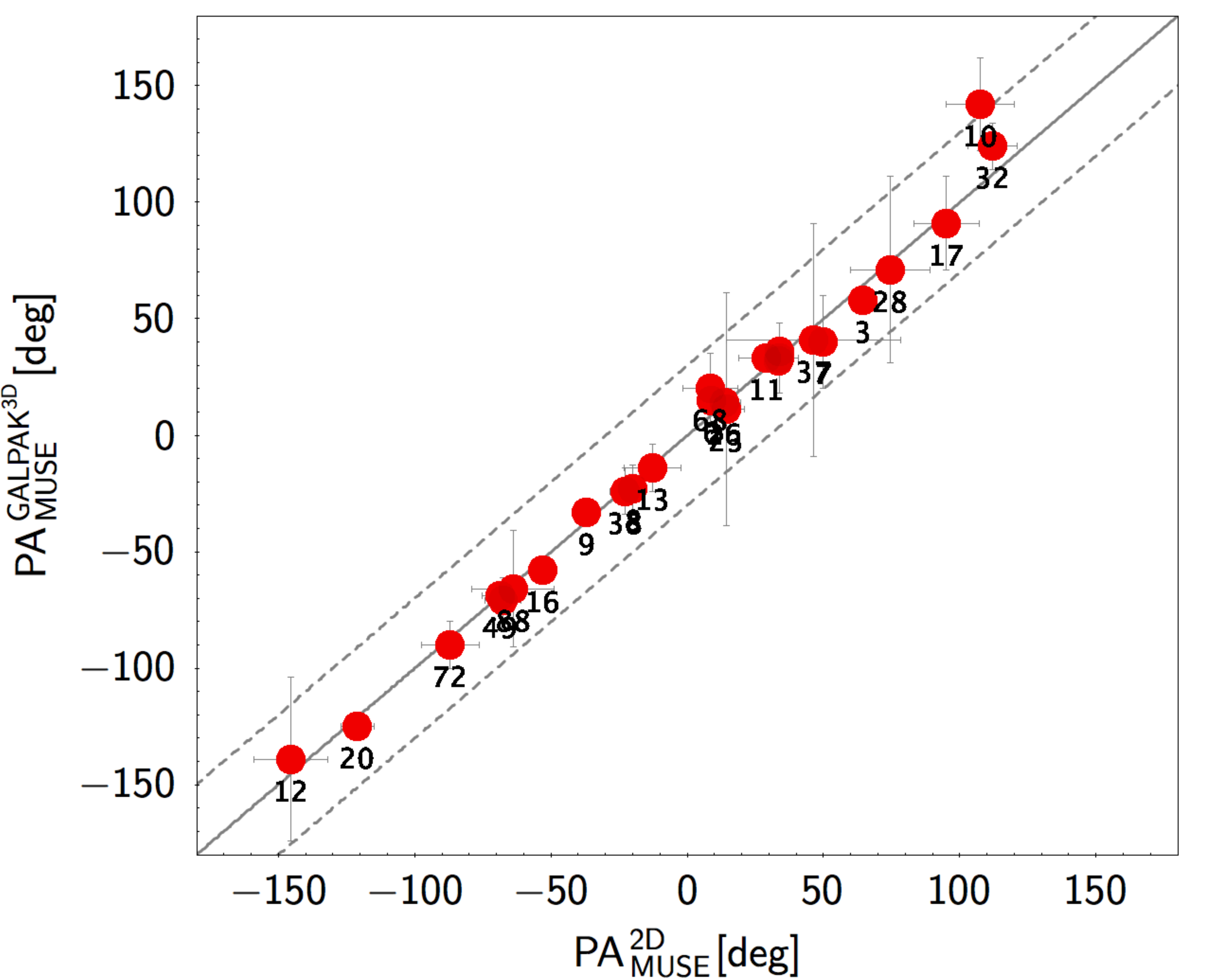}
\caption{{\it Left panel}: comparison of the inclination values derived with {\sc Galfit} on HST/F814W image to those obtained with \gpk\ on MUSE data for the spatially resolved MUSE-HDFS galaxies (red points). {\it Right panel}: comparison of the values obtained with 2D modeling to those obtained with \gpk\ for the disk major axis position angle for the spatially resolved MUSE-HDFS galaxies (red points). For nine galaxies (ID \#5, 6, 11, 16, 17, 23, 37, and 88), the kinematical PA was constrained in \gpk\ to be close to the morphological PA, i.e.,\, within $\pm 10\degr$. Labels indicate the galaxy IDs. The solid line represents the 1:1 relation and dashed lines indicate the typical scatter around this relation due to measurement uncertainties.}
\label{fig:compa2DGP_1}
\end{figure*}

\begin{figure*} 
\includegraphics[width=9.3cm]{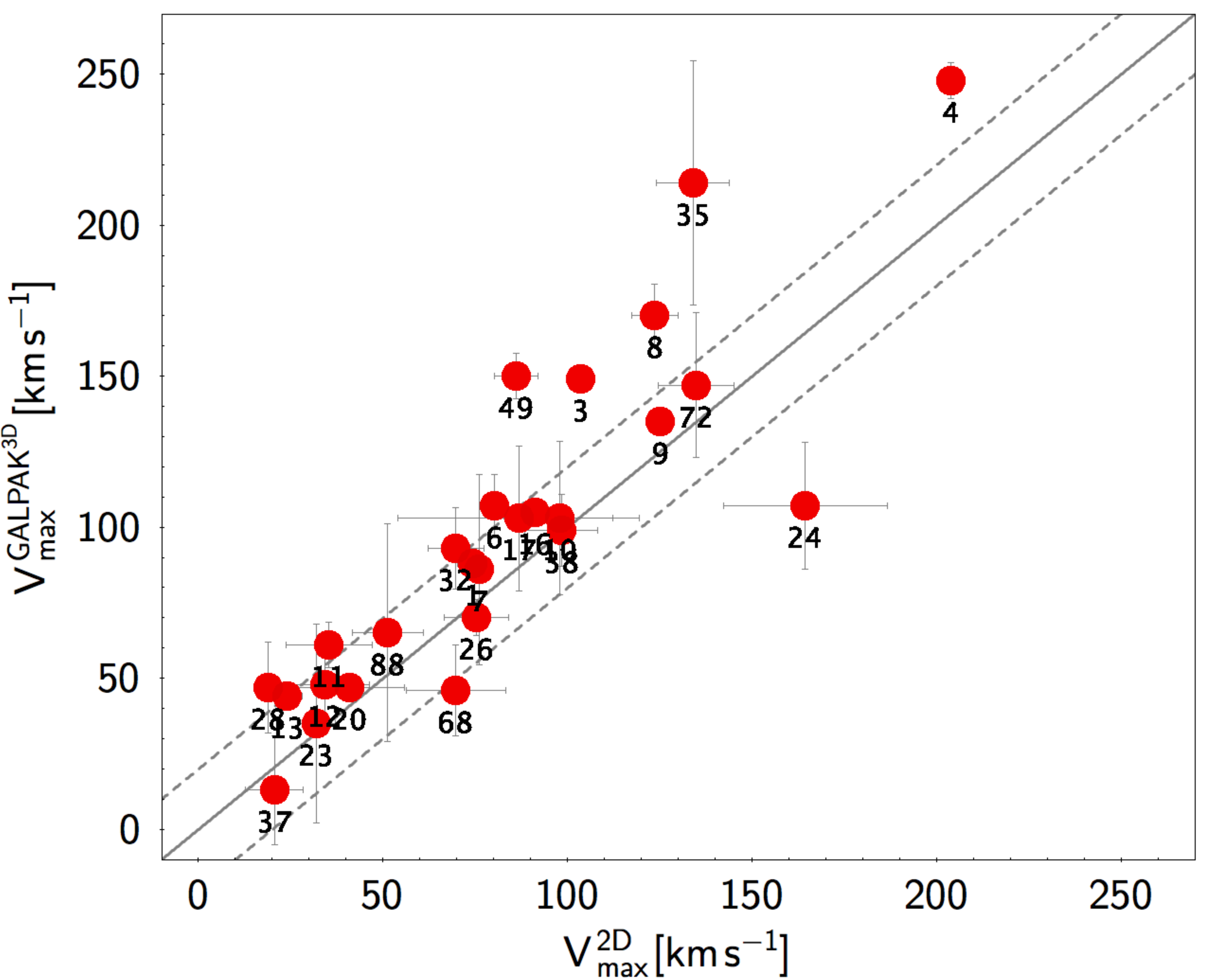}
\includegraphics[width=9.3cm]{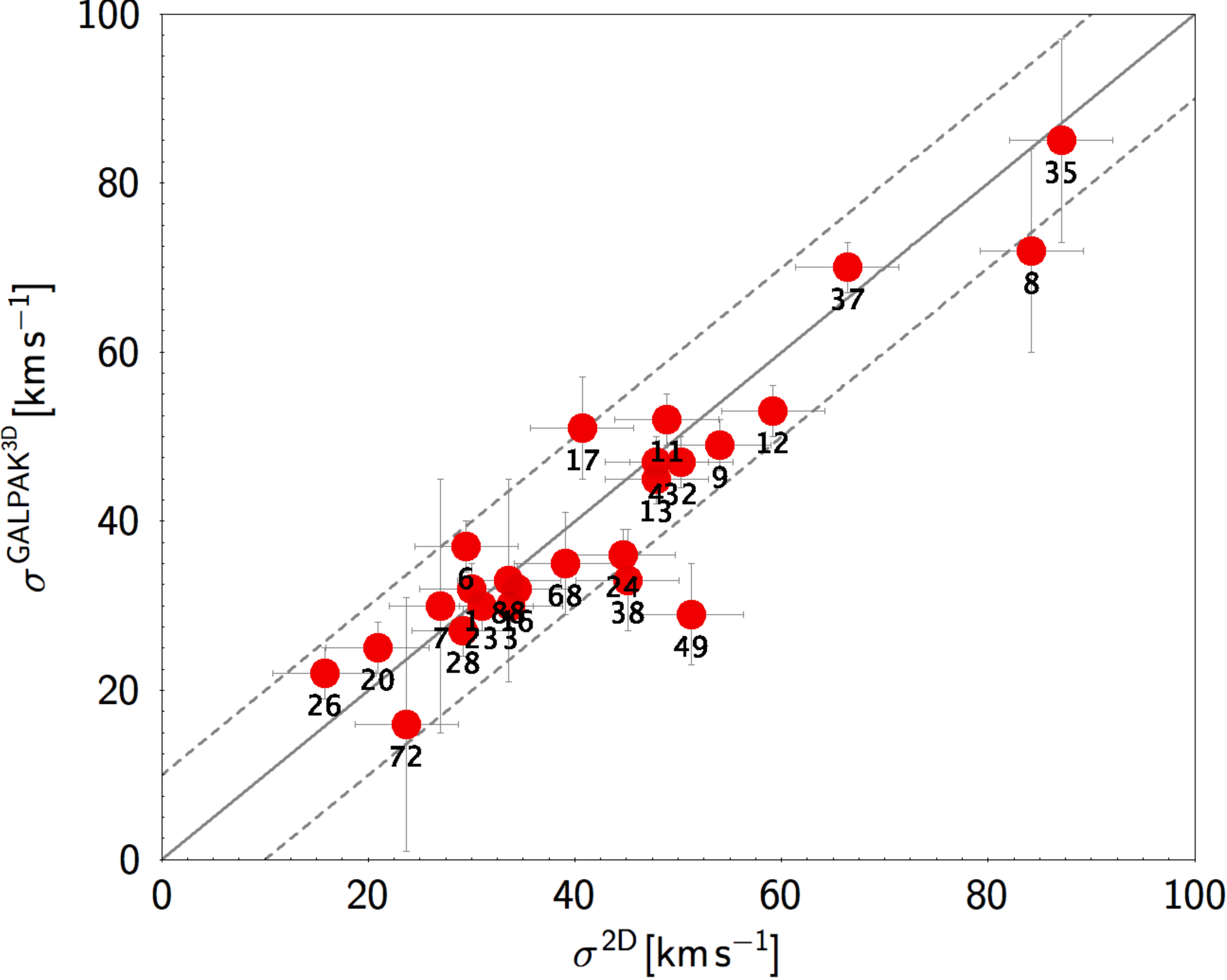}
\caption{Comparison of the values obtained with 2D modeling to those obtained with \gpk\ for disk maximum rotational velocity ({\it left panel}) and velocity dispersion ({\it right panel}) for the spatially resolved MUSE-HDFS galaxies (red points). Labels indicate the galaxy IDs. The solid line represents the 1:1 relation and the dashed lines indicate the typical scatter around this relation due to measurement uncertainties.}
\label{fig:compa2DGP_2}
\end{figure*}

\subsubsection{The 2D method}
\label{sec:2D}

Galaxy velocity fields are fitted with a model that assumes that the ionized gas is located in a thin rotating disk \citep{Epinat+2012}. The rotation curve is described by a linear slope in the inner parts and a plateau in the outer parts, which is reached at the turnover radius. The choice of this rotation curve shape is motivated by the determination of \vmax\ and relies on previous investigations \citep{Epinat+2010}. The model parameters are the position ($x,y$) of the galaxy center, disk inclination ($i$), position angle of the major axis ($PA$), plateau rotation velocity, and  turnover radius at which the plateau is reached. The method used to adjust the models is described in detail in \citet{Epinat+2010,Epinat+2012}. 
It is based on a $\chi^2$ minimization and uses the velocity error map to minimize the contribution of the regions with low S/N. The spatial PSF is taken into account in these models and is described with a 2D Gaussian. Its FWHM was computed at the corresponding wavelength (for \oii, \hbeta$+$\oiii, and \halpha $+$ \nii) using the bright star in the MUSE FoV \cite[see][]{Bacon+2015}.
To reduce the number of free parameters, we constrained the galaxy center and its inclination from the morphological analysis of the HST images (see section~\ref{sec:morpho}). Indeed, these parameters are the least well-constrained by the velocity field \citep{Epinat+2010}. The maximum rotation velocity used in the following analysis corresponds to the plateau rotation velocity, which makes sense for our sample of spatially resolved galaxies, as the turnover radius of the rotation curve is always smaller than the extent of the ionized gas. The intrinsic velocity dispersion map is obtained by subtracting in quadrature the map deduced from the model (which takes the instrumental spectral resolution into account and contains the beam-smearing effect on the velocity dispersion) and the observed velocity dispersion map. The spatially averaged values of the velocity dispersion are reported in Table~\ref{tbl:sample}. We emphasize that the median and mean values of the velocity dispersion measured in each spaxel of these residuals maps are very close. Therefore the spatially averaged value of the velocity dispersion is dominated neither by regions with large dispersions (often associated with clumps, see sect.~\ref{sec:voversig}) nor by low S/N regions. The model and residual maps are shown in Figure~\ref{fig:vfexample} and Figures~\ref{fig:appen_figs_obj1} to~\ref{fig:appen_figs_obj88}. 

\subsubsection{The \gpk\ method}
\label{sec:galpak}

Traditional disk modeling, as described previously, is based on 2D fitting of moment maps (line flux, position, and FWHM) extracted from IFS data. A recently developed tool\footnote{\tt http://galpak.irap.omp.eu/}, \gpk\ \citep{Bouche+2015}, makes it possible to directly compare the 3D data cubes to a 3D galaxy disk model that is convolved with the spatial PSF and instrumental line spread function (LSF). \gpk\ is particularly suited for extended objects when the size (half-light diameter) to seeing ratio is $0.5-1.0$ or greater \citep[discussed extensively in][]{Bouche+2015}, which is the case for the MUSE-HDFS sample; see \cite{Bouche+2013} and \cite{Schroetter+2015} for applications to SINFONI data. The model parameters are optimized using Monte Carlo Markov Chains (MCMC) from which we can compute the posterior distributions on each of the parameters. Similar to the 2D modeling described in the previous section, the shape of the rotation curve is chosen to be an arctangent function with two parameters: the turnover radius and the maximum rotation velocity reached in the plateau. This algorithm also assumes a single\footnote{Future versions of \gpk\ will include the possibility to model a galaxy with two components, such as a bulge and a disk.} and symmetric profile  for the light distribution of the galaxies, which
is assumed to be an exponential disk (Sersic profile with index $n=1$). Since \gpk\ takes the PSF and LSF into account, it returns the intrinsic deconvolved morpho-kinematics parameters, including the galaxy center and inclination, size (half-light radius), maximum rotation velocity, turnover radius, and velocity dispersion. The convolved flux and velocity (rotation and dispersion) maps computed with \gpk\ (version 1.4.5) are shown in Figure~\ref{fig:vfexample} and Figures~\ref{fig:appen_figs_obj1} to~\ref{fig:appen_figs_obj88}. 

\begin{figure*} 
\includegraphics[width=6cm]{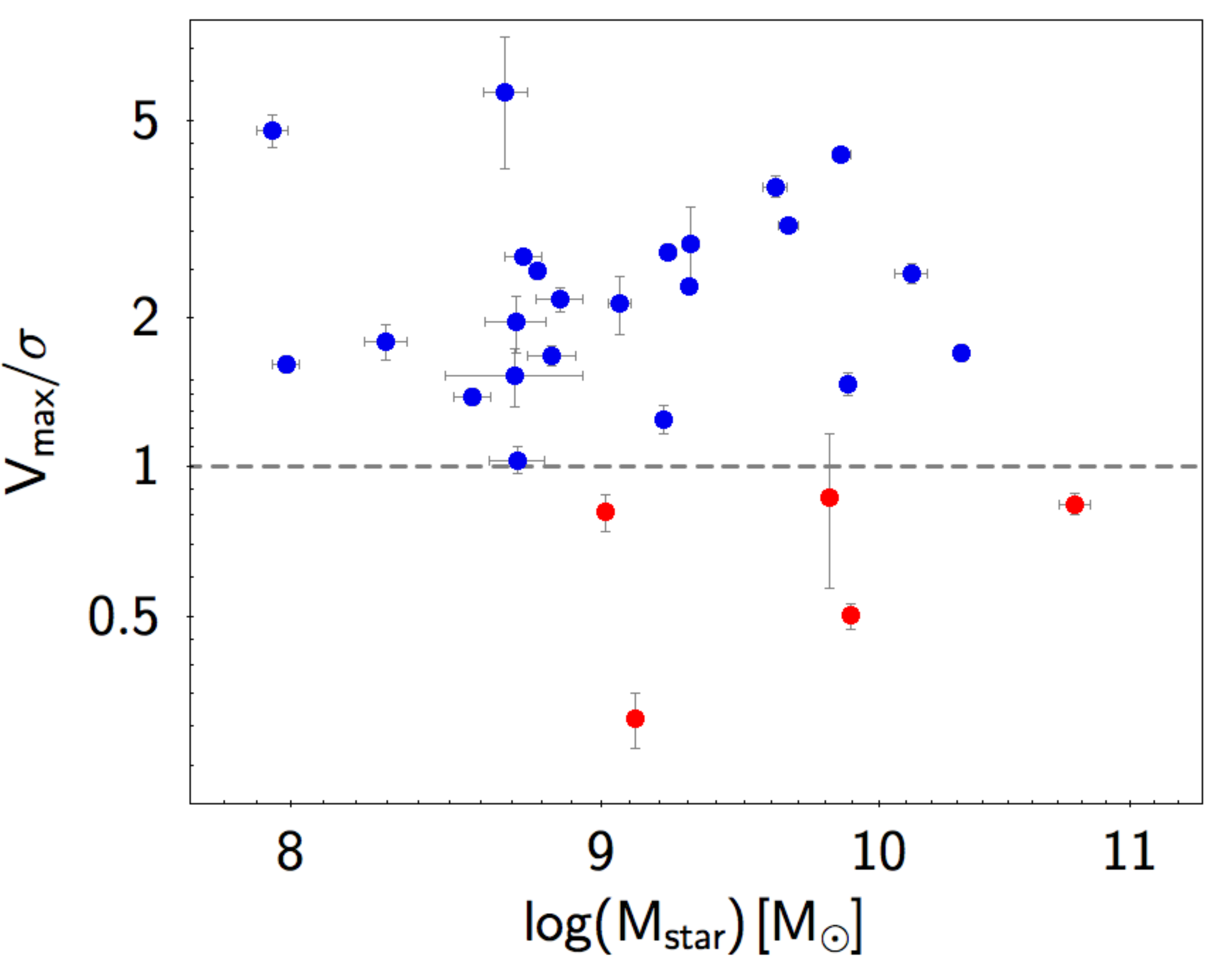}
\includegraphics[width=6cm]{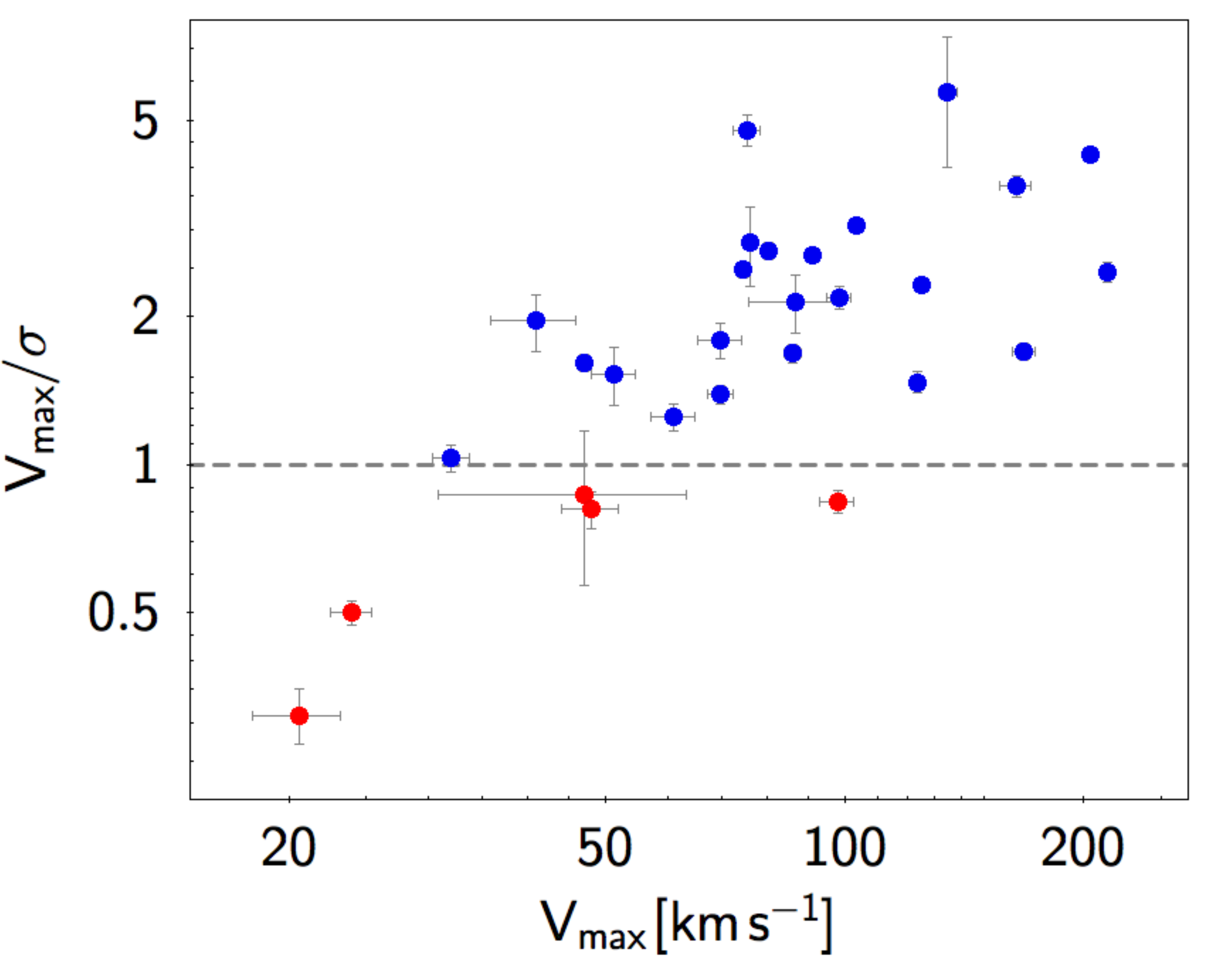}
\includegraphics[width=6cm]{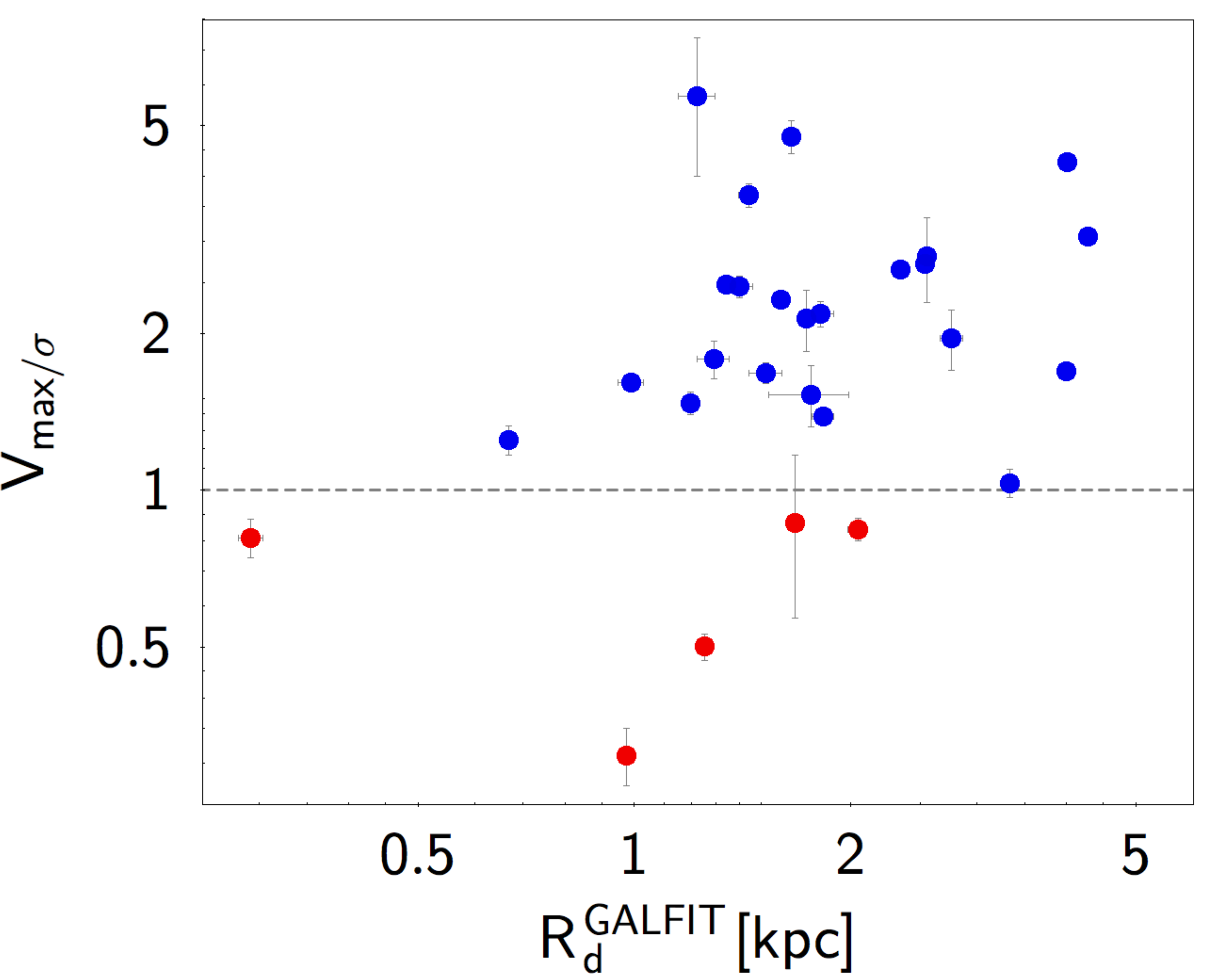}
\caption{Ratio \voversig\ as a function of galaxy stellar mass ({\it left panel}), maximum rotational velocity ({\it middle panel}), and size ({\it right panel}). The horizontal dashed lines set the limit between rotation-dominated (blue points above the line) and dispersion-dominated (red points below the line) galaxies. No clear correlation is seen but dispersion-dominated galaxies are mostly massive ($> 10^{9}$\msun), slowly rotating ($\leq 50$ \kms), and compact.}
\label{fig:voversig}
\end{figure*}

\subsection{Morpho-kinematic parameters}
\label{sec:morphokin}

Figures~\ref{fig:compa2DGP_1} and \ref{fig:compa2DGP_2} show the comparison of the morpho-kinematic parameters (inclination, position angle, maximum rotational velocity, and velocity dispersion) obtained with the two modeling techniques (2D and  \gpk), except for the inclination which is a comparison between the HST-based (using {\sc Galfit}) and the MUSE-based (using \gpk) values. Taking the intrinsic uncertainties of the different methods and the errors associated with the measurements into account, these parameters agree very well for most of the sample galaxies. 

Regarding the inclination, we notice that for some galaxies (e.g., galaxies ID \#11, 13, 35, and 49) with a size ($=$ effective radius) on the order of the FWHM of the MUSE PSF, \gpk\-derived values from the MUSE data cube are sometimes underestimated with respect to the values derived with {\sc Galfit} in the HST images. This could be explained by the fact that {\sc Galfit} parameters are constrained from HST images, which have a better spatial resolution (by a factor of $\approx 5$) than the MUSE observations. 

Values of the position angle of the disk major axis are in good agreement for most of the galaxies (right panel of Figure~\ref{fig:compa2DGP_1}). Taking measurement uncertainties into account, they all agree within $\pm 30\degr$. However, for nine galaxies (ID \#5, 6, 11, 16, 17, 23, 37, and 88), the kinematical PA was constrained in \gpk\ to be close to the morphological PA, i.e.,\, within $\pm 10\degr$. This was necessary for \gpk\ to converge toward realistic model parameters.

There is also a good agreement for the (deprojected) maximum rotational velocity derived from the two kinematic models; most of the values are distributed around the 1:1 line. 
However, there are six outliers with respect to this relation (galaxies ID \#3, 4, 8, 24, 35, and 49), 
which are all above the 1:1 relation with higher values of the rotation velocity obtained from \gpk\ with respect to the 2D modeling. For galaxies ID \#35 and 49, the discrepancy comes from a difference of about $\approx 15$\degr\ in the derived inclination. Indeed, the projected rotation velocities (with the $\sin i$ factor) are in good agreement for these two galaxies. Galaxies ID \#3 and 24 have the lowest inclinations of the sample ($< 20$\degr), which implies a higher intrinsic uncertainty in the deprojected maximum rotation velocity. Finally, galaxies ID \#4 and 8 show  strong morphological asymmetries both in the HST images and emission-line flux maps, which could explain why \gpk\ has difficulty converging and finishes at a high rotation velocity. Using an arctan function for the 2D modeling also does not significantly improve  the comparison.

The last comparison concerns the velocity dispersion that is shown in the right panel of Figure~\ref{fig:compa2DGP_2}. Again, the agreement is very good between the values obtained from the two different methods (2D and  \gpk) with no apparent systematic bias. The object with the largest discrepancy (galaxy ID\#49) still agrees to within $\pm 20$ \kms.

The final morpho-kinematic parameters are listed in Table~\ref{tbl:sample}. For the sake of homogeneity and completeness, and as the derived values from \gpk\ and 2D modeling agree within the measurement errors for the vast majority of galaxies/parameters, we choose to use the latter values for the subsequent analysis. 
The uncertainties listed in Table~\ref{tbl:sample} quantify the discrepancy between the different emission lines used for model fitting and/or between the 2D and \gpk\ modeling rather than the intrinsic error on the derivation of each parameter, which is usually smaller. 

\section{Results}
\label{sec:results}

\subsection{Dynamical state of the gas and close environment of galaxies}\label{sec:dynsup}

For the subsequent analysis (TFR, angular momentum, etc), it is essential to define a robust sample of (unperturbed) disk galaxies dominated by rotational motions. We thus review in the following subsections various criteria used to define such a sample, including the misalignment of kinematic and morphological position angles, the ratio between the ionized gas maximum circular velocity and its intrinsic velocity dispersion, and the local environment (minor/major pairs, interacting galaxies, etc).

\subsubsection{Misalignment between the kinematic and morphological major axes}
\label{sec:pa}

\begin{figure} \resizebox{\hsize}{!}{\includegraphics{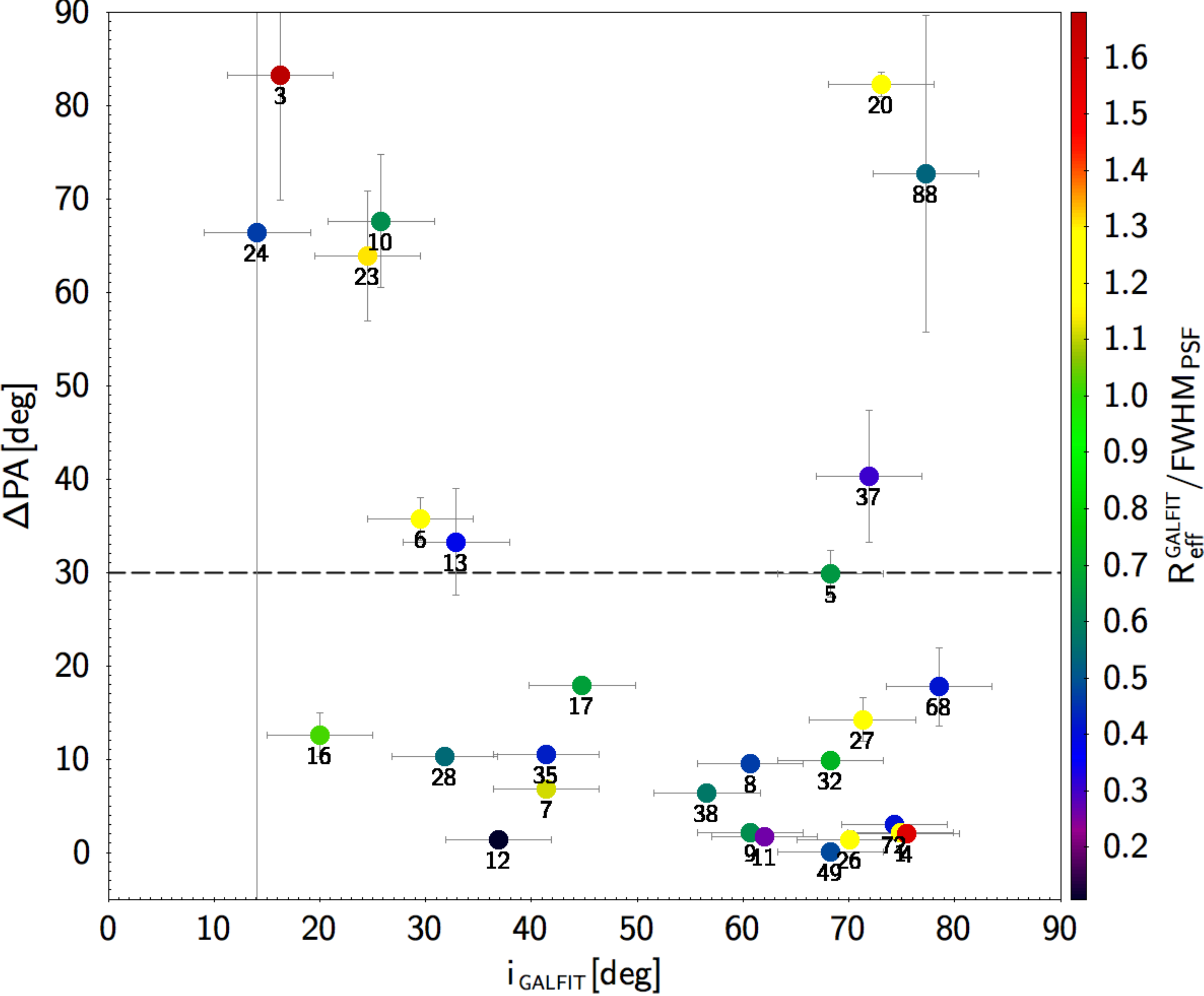}}
\caption{Difference between the morphological and kinematic position angles as a function of the disk inclination for the sample of spatially resolved galaxies in the MUSE-HDFS. Dots are color-coded as a function of the relative size of the disk (effective radius derived from {\sc Galfit}) with respect to the FWHM of the MUSE PSF. Labels indicate the galaxy ID. The dashed line indicate the mean value of the sample.}
\label{fig:diffPA}
\end{figure}

For rotating galaxies, we expect a rough agreement between the kinematic position angle derived from MUSE data (Sect.~\ref{sec:diskmodel}) and the morphological position angle determined from the HST/F814W optical images (Sect.~\ref{sec:morpho}). The difference between these two parameters, \DPA\ defined as an absolute value between $0$\degr\ and $90$\degr, is shown in Fig.~\ref{fig:diffPA} as a function of the galaxy inclination.  The mean and median values of \DPA\ for all the galaxies are $25$\degr\ and $13$\degr, respectively, within the expected errors of both measurements. For half of the galaxies the agreement is better than $15$\degr\ and for $\approx 70$\% of the sample galaxies the agreement is better than 30\degr. Among the nine galaxies showing a higher difference (\DPA\ $> 30$\degr), six are at relatively low inclination ($i < 40$\degr) and four are dispersion dominated (see below). Since these low-inclination galaxies appear rounder, it is more difficult to define a major axis and therefore to determine the morphological position angle. Among the  \DPA\ $> 30$\degr\ cases, the misalignment can be attributed to morphological substructures, such as spiral arms (ID \#3, 10), central bar (ID \#6, 20), or even clumps in a low-surface brightness disk (ID \#13, 23), which may introduce significant errors in the disk inclination estimate. Though galaxies ID \#37 and 88 look like highly inclined disks from their broadband morphology, they are among the farthest ($z > 1$) in our sample with a ratio between the projected size (disk effective radius derived from {\sc Galfit}) and the FWHM\ of the MUSE PSF, which is also among the lowest in our sample. This could explain the disagreement between the kinematic and morphological position angles for these two galaxies. 

\subsubsection{Dynamical state of the ionized gas}
\label{sec:voversig}

A commonly used criterion for describing the dynamical state of ionized gas in high-redshift, star-forming galaxies \citep[e.g.,][]{Forster-Schreiber+2009,Epinat+2012} is the ratio \voversig\ between its maximum circular velocity and its intrinsic velocity dispersion. It has been shown that this ratio is much lower at high redshift  than in the intermediate redshift ($z\sim 0.5$) and local universe \citep[e.g.,][]{Forster-Schreiber+2009,Law+2009,Gnerucci+2011,Epinat+2012,Wisnioski+2015}. 

To quantify the dynamical state of the ionized gas in our sample of MUSE-HDFS galaxies, we assume that in all of these systems the rotating-disk model adequately describes the ionized gas velocity (i.e., the first velocity moment; see sect.~\ref{sec:diskmodel}). As for using the measured velocity dispersion (i.e., the second centered velocity moment) of the gas clouds as a direct tracer of the gravitational potential (mostly driven by the distribution of stars and dark matter), it is strictly valid only if the non-gravitational contributions to the total velocity dispersion are small. Indeed, the intrinsic velocity dispersion can be separated into a thermal component ($\sigma_{\rm th}$), a turbulent component ($\sigma_{\rm tr}$), and the gravitational component ($\sigma_{\rm gr}$). For ionized gas at a typical temperature of about $10^4$ K, $\sigma_{\rm th}$ should not be larger than 10 \kms. The contribution from turbulent motion is more complex to assess, as it can originate from internal motions within the gas clouds and from shocks induced by non-axisymmetric perturbations such as bars, inflows/outflows, or mass compression due to gravitational interactions. As in other studies of high-redshift galaxies \citep[e.g.,][]{Forster-Schreiber+2006,Cresci+2009,Genzel+2011,Epinat+2012,Wisnioski+2015}, our velocity dispersion maps are often clearly clumpy and not as uniform as some models, such as \gpk\ , predict. However, the turbulence should not significantly influence the observed circular velocity, which primarily reflects the gravitational potential. Assuming that the clumps in the velocity dispersion maps that are not reproduced by our \gpk\ models come from turbulence, while the bulk of the measured velocity dispersion is associated with the gravitational potential, we additionally assume that the local velocity dispersion of the gas estimated by our models is representative of the gravitationally induced random motions in the galaxy. 

The median/average value of the \voversig\ ratio for the full sample of MUSE-HDFS galaxies is $\approx 2$ with values ranging from $\approx 0.3$ to $5.7$. Applying the commonly used criterion to distinguish rotation-dominated (\voversig\ $> 1$) from dispersion-dominated (\voversig\ $\leq 1$) gaseous systems, we find that five galaxies (18\%) fall into the last category. Figure~\ref{fig:voversig} shows the \voversig\ ratio as a function of galaxy stellar mass, rotation velocity, and size.  Although we see no clear correlation with galaxy stellar mass, we note that the dispersion-dominated galaxies are all above $10^{9}$ \msun\ and thus constitute about one-third of the MUSE-HDFS galaxy population in this mass range. In the very low-mass regime ($< 10^{9}$ \msun), all of the sample galaxies have gas kinematics that are dynamically dominated by rotation. Similar trends are found as a function of rotation velocity, taking $\sim 50$ \kms\ as a threshold. The apparent correlation between \voversig\ and \vmax\ is mainly driven by the large range of values for the rotation velocity seen in the MUSE-HDFS sample compared to the smaller range of dispersion velocities. Four of the five dispersion-dominated objects are relatively slow rotating galaxies (\vmax\ $\leq 50$ \kms) with the extreme case of ID \#37 (lower left point in the figure) being a very compact galaxy that also shows a misalignment between the kinematic and morphological position angles (see above). The rightmost diagram of Fig.~\ref{fig:voversig} shows also that all the dispersion-dominated galaxies are rather small with maximum disk scale lengths on the order of $\sim 2$ kpc. 

To summarize, the MUSE-HDFS dispersion-dominated galaxies are mostly massive, slowly rotating, and compact.

\subsubsection{Close pairs and interacting galaxies}
\label{sec:pairs}

As a result of its large FoV and $0.2$\arcsec/pixel spatial sampling, MUSE is well suited to explore the small-scale environment of galaxies at intermediate redshifts. We performed such an analysis in the deep MUSE-HDFS field (Ventou et al., in prep) and identified six spatially resolved galaxies in close pairs, i.e., within a projected separation of 25$h^{-1}$ kpc and a velocity difference of 500 \kms
. Among these close pairs, only one at $z \approx 1.36$ is classified as a major pair with a mass ratio 1:4. This pair is composed of ID \#88 (resolved) and ID \#589\footnote{This ID does not exist in the catalog of \cite{Bacon+2015} as this object was discovered after publication.} (unresolved) separated by $\approx$ 5 kpc and 20 \kms. The other five pairs are all classified as minor pairs with mass ratios ranging from $\approx$ 1:10 to 1:100. These pairs involve the following spatially resolved galaxies that are always the most massive members: ID \#6, 7, and 8 at $z\approx 0.4-0.6$, ID \#49 at $z\approx 1$ and ID \#27 at $z\approx 1.28$. This last galaxy, ID \#27, is a special case. Indeed, in addition to a minor interaction with ID \#124, this highly perturbed galaxy is also interacting\footnote{This system does not count as a close pair, since the two galaxies have a projected separation higher than the maximum value of  25$h^{-1}$ kpc used to define close pairs.} with the AGN ID \#10 (with a mass ratio $\sim$ 1:1), as revealed by low-surface brightness tidal tails connecting the two objects (see Appendix~\ref{sec:appen_notes} for a detailed description). Another galaxy that shows signs of merging/interactions is ID \#12, a very compact galaxy with an extended low-surface brightness jet-like structure, which could be a relic of past gravitational interactions. Considering all these cases (minor/major close pairs, interacting galaxies, and merger remnants), we find a $\sim$ 30\% fraction of interacting/merging galaxies (14\% of major and 18\% of minor interactions) in the MUSE-HDFS sample of spatially resolved galaxies. Following \cite{Moster+2011}, an uncertainty of about 20\% affects this fraction owing to cosmic variance. It is beyond the scope of this paper to investigate any larger scale environment (e.g., galaxy groups) dependence on galaxy kinematics properties. Such studies will be performed on larger galaxy samples during the course of the MUSE-GTO. 

To conclude regarding the dynamical state of the ionized gas in our sample of spatially resolved MUSE-HDFS galaxies, the misalignment between the kinematic and morphological major axes, defined as \DPA\ $> 30$\degr, is mostly due to i) the difficulty of accurately measuring the PA in nearly face-on galaxies and ii) morphological substructures, such as spiral arms, bars, and clumps. About 18\% of the galaxies have gaseous disks that are dynamically dominated by random motions, as revealed by low \voversig\ ratios. These galaxies are mostly massive and slowly rotating. This fraction of dispersion-dominated galaxies is roughly consistent with the 17\% found at $z\approx 1-2$ in the first-year KMOS$^{\rm 3D}$ sample \citep{Wisnioski+2015}. Finally, about 30\% of the sample galaxies are part of a close pair and/or show clear signs of recent gravitational interactions. Restricting this fraction to major interactions only, we find a value of 14\%, which is slightly below  what was found in previous IFS surveys at $z\approx 0.9-3$ \cite[e.g.,][]{Forster-Schreiber+2009, Epinat+2012, Lopez-Sanjuan+2013}, but still compatible taking  the $\approx 20$\% uncertainty due to cosmic variance into account \citep{Moster+2011}.

\subsection{The Tully-Fisher relation}
\label{sec:TFR}

\begin{figure} \resizebox{\hsize}{!}{\includegraphics{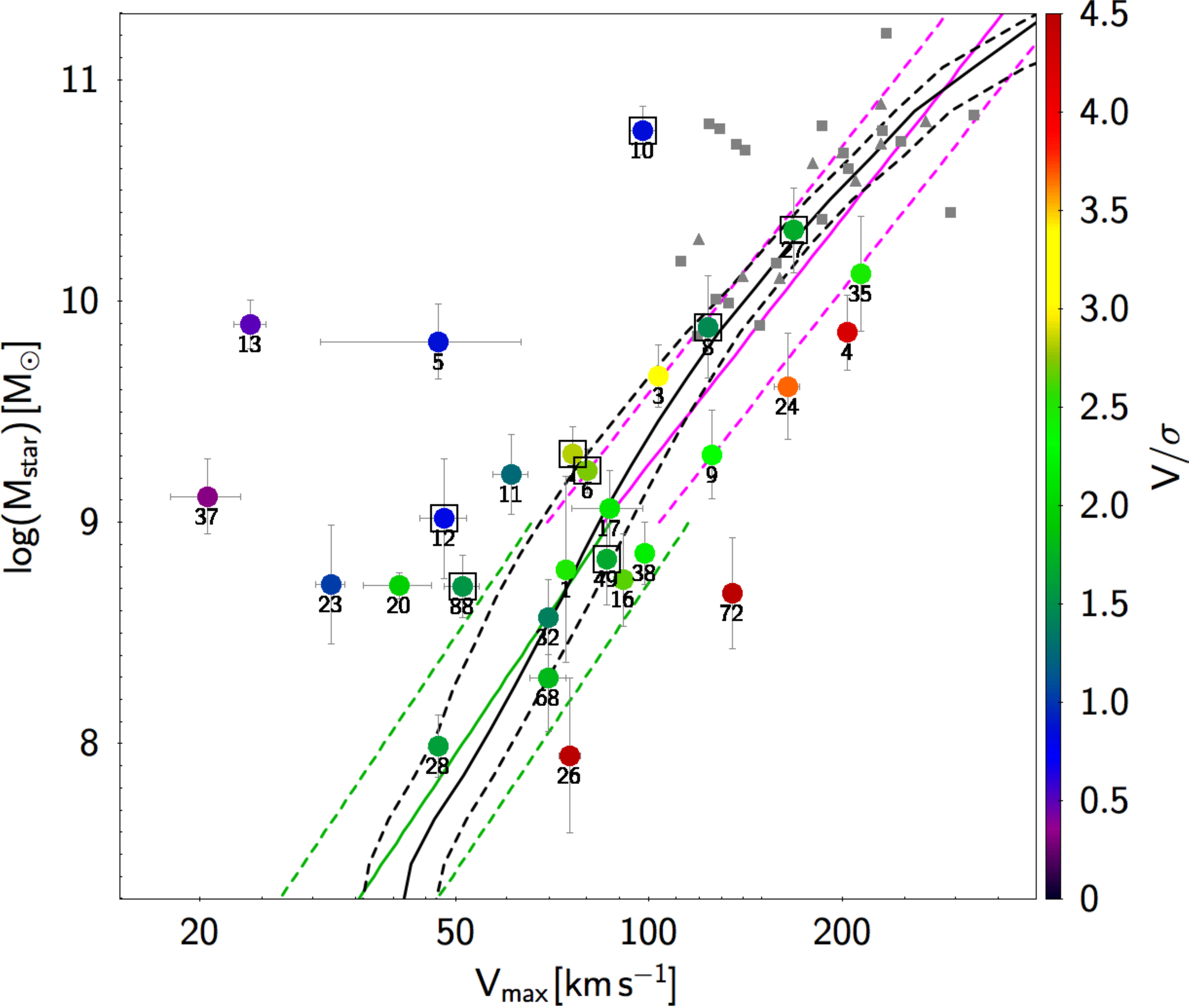}}
\caption{Tully-Fisher relation for the sample of spatially resolved galaxies in the MUSE-HDFS. Labels indicate the galaxy ID. The points are color-coded as a function of the \voversig\ ratio. Galaxies in close pairs and/or showing signatures of recent gravitational interactions are identified with black squares. Previous IFS samples of (massive) star-forming galaxies in similar redshift ranges: IMAGES \citep[gray triangles;][]{Puech+2010} and MASSIV \citep[gray squares;][]{Epinat+2012} are also shown for comparison. The TFR defined in the high-mass regime (\mstar\ $> 10^{9}$\msun) from previous surveys in a similar redshift range \cite[$z=0.2-1.4$;][]{Miller+2011, Vergani+2012} is shown with a magenta line. The TFR defined in the low-mass regime \cite[\mstar\ $< 10^{9}$ \msun;][$z=0.2-1.0$]{Kassin+2007,Kassin+2012,Miller+2014} is shown with a green line. The dotted lines show the dispersion around these relations. A comparison with the predictions from the EAGLE cosmological hydrodynamical simulations for $z\approx 0.1$ galaxies \citep{Schaye+2015} is also shown (black line with the dashed lines representing the 16\% and 84\% percentiles).}
\label{fig:tfr}
\end{figure}

We use the intrinsic, inclination-corrected, maximum rotation velocities, and stellar masses of the MUSE-HDFS galaxies to investigate the Tully-Fisher relation (TFR). In Figure~\ref{fig:tfr} the MUSE-HDFS sample galaxies are color-coded as a function of the gas dynamical state and parametrized by the \voversig\ ratio (see sect.~\ref{sec:voversig});  galaxies in close pairs and/or showing signs of gravitational interactions (see sect.~\ref{sec:pairs}) are also identified with large squares. 

In the high-mass regime (\mstar\ $> 10^{9.5}$\msun), all the galaxies with internal gas dynamics dominated by random motions (i.e., the dark blue-purple dots in the upper left corner of the diagram, which include the AGN ID \#10) are clearly offset from the TFR for rotation-dominated disks.  Our sample is small in this mass range, but the five rotation-dominated galaxies perfectly follow the TFR defined from previous surveys in a similar redshift range \cite[$z=0.2-1.4$][]{Miller+2011, Vergani+2012}. 

The situation is somewhat different in the low-mass regime (\mstar\ $< 10^{9.5}$ \msun). Indeed, rotation-dominated galaxies seem to follow the TFR defined so far with multislit spectroscopy in this lower masses/velocities regime \citep{Kassin+2007,Kassin+2012,Miller+2014} but with a higher scatter compared to more massive objects. Among the three galaxies classified as dispersion dominated with \voversig\ $\leqslant 1$, two galaxies (IDs \#37 and \#23) depart significantly from the TFR. However, the dispersion-dominated galaxy ID \#12 does not really stand out from rotation-dominated galaxies that are scattered around the TFR.

Except for the AGN dispersion-dominated and massive galaxy ID \#10, all the objects that are part of a pair and/or show signatures of recent gravitational interactions follow the TFR defined with isolated galaxies; no distinction is seen between the two samples. As a sanity check, we investigated any dependence of the TFR on our resolving power (parametrized by the ratio between the size of the galaxy and the effective seeing; see section~\ref{sec:selection}) and on the galaxy inclinations. No significant effect was found; slowly rotating galaxies are not particularly likely to be barely resolved or  have low inclination angles. However, most of the slowly rotating galaxies that are offset of the TFR show also a significant misalignment (\DPA\ $\geq 30$\degr) between their morphological and kinematic major axes.

We also investigated any redshift evolution of the TFR. Dividing the galaxies into three or even two redshift bins does not reveal any significant evolution of the TFR. This is not so surprising, as we know that the zero-point evolution of this scaling relation is very mild from $z\sim 1$ \citep[e.g.,][]{Miller+2011, Vergani+2012}. Our sample is definitely too small to derive robust conclusions over such a large redshift interval. No doubt that the growing sample built from the various (deep) MUSE datasets will allow us to address this issue in the coming years, especially in the low-mass regime. 

Figure~\ref{fig:tfr} also compares the TFR defined for the MUSE-HDFS sample to data and models from the literature. Data come from previous IFS surveys in similar and complementary redshift ranges: IMAGES at $z\sim 0.4-0.8$ \citep{Puech+2010} and MASSIV at $z\sim 0.9-1.4$ \citep{Epinat+2012,Vergani+2012}. If needed, stellar masses were corrected to take into account any different IMF used in the SED modeling. 
As previously stated, the rotation-dominated galaxies selected in the MUSE-HDFS field follow the TFR defined with previous IFS surveys in the high-mass regime.  It is clear, however, that deep observations with MUSE make it possible to extend the TFR studies to much lower stellar masses, a regime that was almost inaccessible with the previous instrumentation. A comparison with recent predictions from the EAGLE cosmological hydrodynamical simulations \citep{Schaye+2015, Crain+2015} is also shown in Figure~\ref{fig:tfr}. The EAGLE predictions\footnote{Predicted values below $1.8 \times 10^8$ \msun\ should be treated with caution since they are below the resolution limit.} for the TFR are based on $z\approx 0.1$ simulated galaxies selected to be late type (Sersic index $< 2.5$) and use maximum circular velocities. The agreement between the MUSE-HDFS dataset and the EAGLE predictions is quite impressive, even if the scatter around the TFR seen in the data is higher than that predicted by the simulations (delineated with dashed lines in Fig.~\ref{fig:tfr}). 

Very recently, based on the analysis of DEEP2 emission-line galaxies 
at $0.1<z<0.4$, \cite{Simons+2015} argue that there is a transition in the TFR at a galaxy mass of $10^{9.5}$\msun. Below this mass, the TFR  has a very significant scatter for slowly rotating (\vmax\ $\leq 30$ \kms) objects for their
sample. In this mass regime, galaxies can be either rotation-dominated (\voversig\ $> 1$) on the TFR or much more compact (or asymmetric) and dispersion dominated, which scatter around the TFR. \cite{Simons+2015} conclude that below this transition mass, a galaxy may or may not settle into a disk. We do not find such a large population of dispersion-dominated slow rotating galaxies in the low-mass regime of our sample. Indeed, 90\% of the MUSE-HDFS galaxies with stellar masses below $10^{9.5}$\msun\ are rotation dominated and have thus already settled into a disk. We checked that this difference is not due to our selection criterion (see sect.~\ref{sec:selection}), which could bias our sample against compact low-mass galaxies. Indeed, among the low-mass galaxy sample of \cite{Simons+2015}, the vast majority (73\% and 90\% of the slow rotating galaxies and rotation-dominated galaxies, respectively) would have been selected according to our criterion.

\subsection{Angular momentum}
\label{sec:angmom}

\begin{figure} \resizebox{\hsize}{!}{\includegraphics{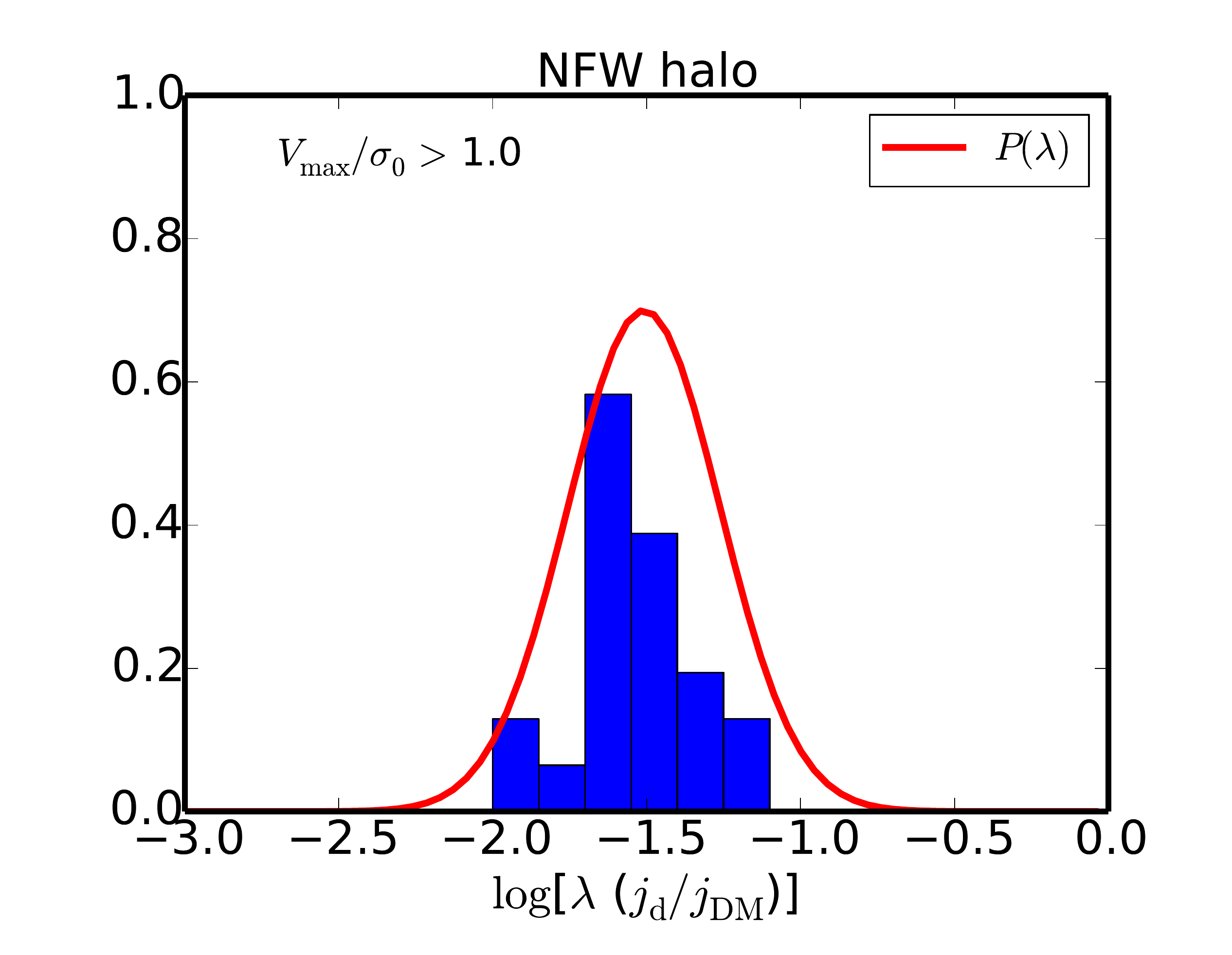}}
\caption{Distribution of the inferred spin parameter proxy  [$\lambda \times (j_{\rm d}/j_{\rm DM})$] computed from Equation~\ref{eq:MMW}  for the subset of 23 rotation-dominated galaxies with a ratio \voversig\ $>1$ (see Sect.~\ref{sec:voversig}). The red curve shows the theoretical lognormal $P(\lambda)$ distribution for dark matter haloes with mean $\overline \lambda=0.031$ and dispersion $\sigma=0.57$ \citep{Bullock+2001,Bett+2007,Munoz-Cuartas+2011}.}
\label{fig:spin}
\end{figure}

From our measurements of disk sizes $R_{\rm d}$ (see sect.~\ref{sec:morpho}) and maximum rotation velocities \vmax\ (see sect.~\ref{sec:morphokin}), we can constrain the specific angular momentum $j$ of the ionized gas in our sample of MUSE-HDFS galaxies, since $j\propto R_{\rm d} \times$ \vmax. Another constraint on the dynamical state of the baryons comes from the ratio $R_{\rm d}/V_{\rm max}$, which is proportional to the spin parameter $\lambda$, as described in \citet{Bouche+2007} and \citet{Burkert+2015}, following the theoretical work of \citet{Mo+1998}.

Indeed, under the scenario of spherical collapse of the baryons within their dark matter (DM) halo, where the baryons retain a fraction of the halo angular momentum, the galaxy disk scale length $R_{\rm d}$ is proportional to the spin parameter $\lambda$ times the halo circular velocity $V_c$ \citep{Fall+Efstathiou1980,Fall1983,Mo+1998}
\begin{equation}
R_{\rm d}\propto \lambda \frac{f_{\rm J}}{m_{\rm d}} V_{\rm c} H(z)^{-1} \label{eq:MMW}
,\end{equation}
where $H(z)$ is the Hubble constant at redshift $z$, $V_{\rm c}$ is the halo circular velocity, $f_{\rm J}$ is the fraction of the total angular momentum contained in the disk ($J_{\rm d} = f_{\rm J}\times J_{\rm DM}$), and $m_{\rm d}$ is the disk-to-dark halo mass fraction. The dimensionless spin parameter is $\lambda \equiv J_{\rm DM}\sqrt E/GM_{\rm DM}^{5/2}$ \citep[e.g.,][]{Peebles1969,Bullock+2001}, with $E$ representing the total energy, $M_{\rm DM}$ the mass of the DM halo, and $J_{\rm DM}$ the DM angular momentum. In N-body simulations, the spin parameter $\lambda$ is robustly measured and follows a lognormal distribution with $\lambda\approx0.031$, which is almost independent of mass and redshift \citep[e.g.,][]{Bett+2007,Bett+2010}.

Generally speaking, the ratio ${f_{\rm J}}/{m_{\rm d}}$ in Equation~\ref{eq:MMW} is equivalent to $j_{\rm d}/j_{\rm DM}$  \citep{Mo+1998,Burkert+2015} where $j_{\rm d}$ ($j_{\rm DM}$) is the disk (halo) specific angular momentum, respectively. Naturally, if the baryons retain their specific angular momentum,  then this ratio $j_{\rm d}/j_{\rm DM}$ is unity.
Equation~\ref{eq:MMW} has several limitations, as discussed in \citet{Burkert+2015}. It neglects the self-gravity of the disk, adiabatic contraction, and the contribution of the velocity dispersion \citep{Burkert+2010}. 
Nonetheless, one can investigate the distribution of the spin parameter $\lambda$ for our MUSE-HDFS galaxy sample, using Equation~\ref{eq:MMW} and the \cite{Mo+1998} formalism. 
Differences between observed and predicted $\lambda$ distributions with these assumptions might bring to light differences from these assumptions.   

With our measurements of galaxy sizes and maximum rotation velocities ($R_{\rm d}$ and \vmax; see Table~\ref{tbl:sample}), we can estimate the quantity $\lambda\times(j_{\rm d}/j_{\rm  DM})$ with Eq.~\ref{eq:MMW}  for the MUSE-HDFS rotation-dominated galaxies. Disk sizes measured from ionized gas are good proxies for high-redshift star-forming galaxies, 
as results from the 3D-HST survey \citep[e.g.,][]{Nelson+2015} have shown that the ratio between the gas and stellar continuum sizes is $\approx 1.1$ with a very weak dependence on stellar mass. Regarding the halo circular velocity $V_{\rm c}$, we estimate this parameter from the galaxy rotation velocity at $R_{\rm 1/2}$ (which is equal to $1.68\times R_{\rm d}$), using the correction for the turbulent pressure gradient of \citet{Burkert+2010}; see Eq. B4 of \citet{Burkert+2015}.

The proportionality constant in Equation~\ref{eq:MMW} is equal to $0.1/\sqrt{2}$ for an isothermal halo \citep{Mo+1998,Burkert+2015}, and  for a NFW halo \citep{NFW1996} it is given by Equ.\,28 of \citet{Mo+1998}. In the case of a NFW profile, this requires knowledge of the disk-to-halo mass fraction $m_{\rm d}$.  As discussed in \citet{Burkert+2015}, one can use either a constant value for $m_{\rm d}$~\footnote{The majority of our MUSE-HDFS resolved galaxies have a disk-to-halo mass ratio $m_{\rm d}$  between 0.1 and 0.2.}, or independent scaling relations from abundance matching \citep{Moster+2013,Behroozi+2013}, and equivalently the baryonic TFR. In this work (Figure~\ref{fig:spin}), we assume a NFW profile with $m_{\rm d}$ estimated from the baryonic TFR of \citet{Trachternach+2009} and its evolution from \citet{Dutton+2011}, but found little difference between these approaches. 

Figure~\ref{fig:spin} shows the resulting  distribution of $\lambda\times(j_{\rm d}/j_{\rm  DM})$. Here, we restrict the MUSE-HDFS sample to the 23 galaxies that show rotation-dominated gas kinematics with a ratio \voversig\ $>1$ (see Sect.~\ref{sec:voversig}). The red solid curve represents the DM expectation \citep{Bullock+2001,Bett+2007,Munoz-Cuartas+2011}. This figure shows that the spin distribution of the MUSE-HDFS (rotation-dominated) galaxies is broadly consistent with that of the dark matter, which is similar to \citet{Burkert+2015} who also found $\lambda\times(j_{\rm d}/j_{\rm  DM})$ distributions broadly consistent with the theoretically predicted distributions, using a larger sample of $\approx300$ star-forming galaxies from the KMOS$^{\rm 3D}$ and SINS surveys and also using the ionized gas as a dynamical tracer. However, our MUSE-HDFS sample of galaxies covers a stellar mass range of $10^8$ to $10^{10}$ \msun\ (with $\approx 70$\% of the sample below $10^{9.5}$ \msun), whereas the KMOS$^{\rm 3D}$ sample covers a range of higher stellar masses with only a small fraction ($\approx 20$\%) of galaxies in the sample below $10^{10}$ \msun, i.e., where our sample ends. In subsequent papers, with larger samples drawn from MUSE-GTO datasets, we will be able to refine this sort of analysis for a larger number of low-mass galaxies, following \citet{Burkert+2015}.

Our results in Figure~\ref{fig:spin}, like those in \citet{Burkert+2015}, imply that the disk specific angular momentum is similar to that of the DM halos, i.e., $j_{\rm d}\sim j_{\rm DM}$, in agreement with the most recent state-of-the-art hydrodynamical simulations \citep{Genel+2015,Zavala+2015,Teklu+2015}. This agreement seems natural in the context of spherical collapse,  where accretion is isotropic and the baryon angular momentum is closely related to that of the DM halo. However, this agreement looks surprising given that galaxies retain only a small fraction of their cosmic share of baryons and that galactic winds may boost the disk net angular momentum \citep[e.g.,][]{VanDenBosch+2001,Maller+2002,Dutton+2012,Genel+2015}. Moreover,  gas accretion can be highly anisotropic (from the filaments in the cosmic web) in massive galaxies \citep{Kimm+2011,Pichon+2011,Stewart+2011,Stewart+2013,Danovich+2012,Danovich+2015}, and thus one might expect the disk angular momentum to not follow the halo primordial angular momentum distribution. \citet{Danovich+2015}, however, showed that in $M_{\rm h}\approx 10^{12}$ \msun\ halos the spin parameter of the cold baryons starts at 2-3 times that of the DM at large radii ($r \sim R_{\rm vir}$) and ends up being 0.03 at $r<0.1 R_{\rm vir}$,  i.e., close to the expectation for spherical collapse.
 
We should note that our dynamical tracer, i.e., the ionized gas, certainly does not probe the outermost regions of galactic disks, where high angular momentum gas might still be present in a neutral phase. Indeed, many galaxies in the local universe have large and extended \hi\ disks, such as M81, M83, and M33 \citep{Huchtmeier+1981,Yun+1994,Putman+2009,Bigiel+2010} that appear to be a common feature, as seen in the recent HI survey of \cite{Reeves+2016}.  At high redshifts, recent studies of circum-galactic gas around galaxies, as traced by low-ionization lines in background quasar spectra, reveal mounting evidence that the gas distribution is highly anisotropic \citep{Bordoloi+2011,Bouche+2012,Kacprzak+2012,Bordoloi+2014,Lan+2014}, coplanar, and, in some cases, rotating with the galaxy host  \citep{Bouche+2013,Bouche+2016}.

 
As mentioned earlier, our disk size and kinematic measurements can also characterize a more directly measurable quantity of the MUSE-HDFS galaxies, namely their specific angular momentum $j_{\rm d}\propto R_{\rm d}\times V_{\rm max}$. The ratio between the specific angular momentum of the baryons to that of the dark matter halo yields insight into the quantity $j_{\rm d} / j_{\rm DM}$, the fraction of specific angular momentum retained by the disk. The specific DM angular momentum scales with the DM halo mass as \citep{Fall+Efstathiou1980}
\begin{equation}
j_{\rm DM}\propto  M_{\rm DM}^{2/3}
,\end{equation}
which follows from the virial DM relations \citep{Burkert+2015}.
\citet{Romanowsky+Fall2012} showed that the stellar component of local massive ($> 10^{10}$ \msun) disks follows a similar relation between $j_{\star}$, the stellar specific angular momentum, and the stellar mass \mstar. 

Following \citet{Romanowsky+Fall2012}, for thin exponential disks with flat rotation curves, the disk specific angular momentum is defined as $j_{\rm d} \equiv 2\times V_{\rm max} \times R_{\rm d}$, where $V_{\rm max}$ is the disk maximum rotation velocity and $R_{\rm d}$ the disk scale length. In the local universe, essentially all star-forming galaxies with stellar disks also have a rotationally supported gaseous structure. Some galaxies with gaseous disks have little stellar angular momentum, but these are early types, i.e., non-star forming. Hence, our measurement of $j_{\rm d}$ should not be viewed as an estimate of the stellar angular momentum, but of the baryonic angular momentum.

\begin{figure} \resizebox{\hsize}{!}{\includegraphics{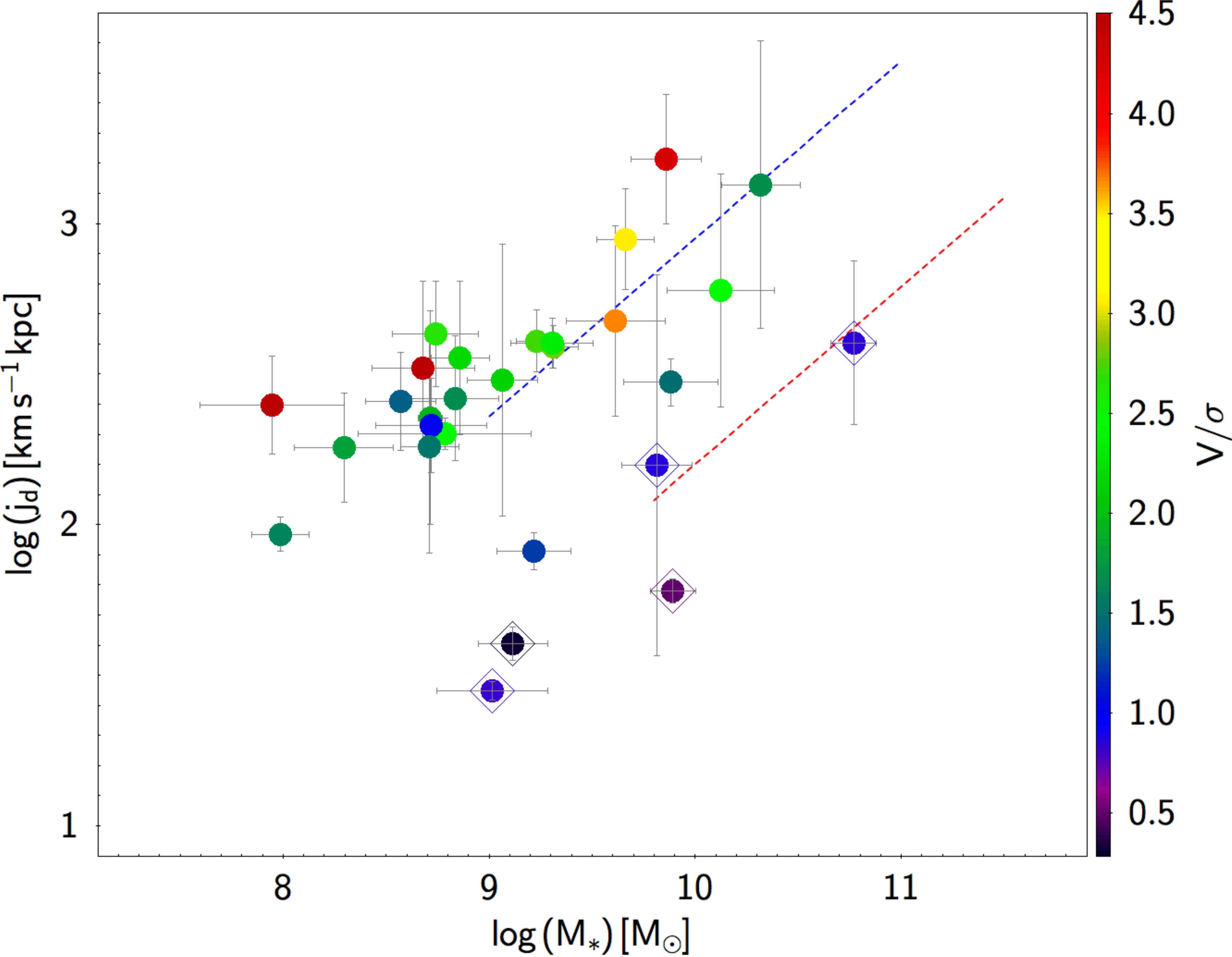}}
\caption{Specific angular momentum of disks $j_{\rm d}$ as a function of galaxy stellar mass for the MUSE-HDFS sample of spatially resolved galaxies. Data points are color-coded according to the \voversig\ ratio. Diamond outlines indicate dispersion-dominated galaxies with \voversig\ $< 1$. The dashed lines show the relations defined for massive galaxies at $z=0$ \citep{Fall+Romanowsky2013}, distinguishing between spheroids (red line, \mstar\ $>10^{10}$~\msun) and disks (blue line, \mstar\ $>10^{9}$~\msun).}
\label{fig:jM}
\end{figure}

Figure~\ref{fig:jM} shows the resulting baryonic specific angular momentum $j_{\rm d}$ as a function of galaxy stellar mass \mstar\ for the MUSE-HDFS sample, with the data points color-coded according to the ratio $V/\sigma$. The deep MUSE observation of the HDFS allows us to probe the relation between $ j_{\rm d}$ and \mstar\ down to low stellar masses below $10^{10}$ \msun. The blue/red lines show the scaling relations for $j_\star$ of local $z=0$ massive spiral (\mstar\ $>10^{9}$~\msun) / elliptical (\mstar\ $>10^{10}$~\msun) galaxies from \citet{Fall+Romanowsky2013}.  Our sample of intermediate-redshift ($z$ median $\approx 0.6$) galaxies appear to follow the local $z=0$ scaling relations, indicating little evolution with cosmic time. The redshift evolution, expected to follow $\propto(1+z)^{-0.5}$, might be masked by the ratio between the stellar and gas disk sizes.  However, the residuals of our $j_{\rm d}$ estimates around the
$M^{2/3}$ fiducial line appear to show a redshift dependence that is consistent with the scaling $(1+z)^{-0.5}$. Further analysis of this potential redshift dependence is warranted with the inclusion of the mass completeness on a larger sample.

In the $j_{\rm d}$-\mstar\ plane (Fig.~\ref{fig:jM}), we see that intermediate-redshift, star-forming galaxies fill a continuum transition from the local spiral (rotation-dominated galaxies) to elliptical (dispersion-dominated galaxies) scaling relations, thus shifting from rotation- to dispersion-dominated (parametrized with \voversig) gas kinematics. 
This transition from star-forming, rotation-dominated galaxies to dispersion-dominated is in qualitative agreement with \citet{Genel+2015}, who showed that some galaxies may experience processes that change their specific angular momentum values in a roughly mass-independent way. We thus speculate here, assuming gas kinematics does reflect the underlying  dynamical state of the galaxy, that the lowest \voversig\ ratios measured in the above-mentioned dispersion dominated galaxies (galaxies with \voversig\ $<1$, indicated with diamonds in Fig.~\ref{fig:jM}) correspond to a global loss of angular momentum. Major mergers could be responsible for some of this evolution, as could compaction through violent disk instabilities 
\citep{Dekel+2014,Bournaud2016,Tacchella+2016}. Hence, a picture emerges in which most galaxies initially have high angular momentum and, as some lose their specific angular momentum, a sequence parallel to that for late-type galaxies forms over time, with galaxies getting closer to the early-type scaling relation in the $j_{\rm d}$-\mstar\ diagram of \citet{Fall+Romanowsky2013}.

\section{Conclusions}
\label{sec:conclu}

This paper presents the morpho-kinematic analysis of the first sample of (mainly) low-mass (\mstar\ $\leq 10^{9.5}$ \msun) star-forming galaxies at intermediate redshift ($0.2 < z < 1.4$). 
Such a study is now possible thanks to the unique capabilities of MUSE, the new IFS available on the ESO-VLT. The unprecedented sensitivity and large field of view of MUSE open unexplored 
territory for galaxy evolution, allowing for deep and blind exposures, which, complemented with high-resolution and deep-enough HST images, offers the possibility to unveil and characterize the faint end of the 
galaxy mass function at significant lookback times. 

MUSE observations of the HDFS, the deepest to date, allowed us to identify a sample of 28 spatially resolved, star-forming galaxies over a broad redshift range. The vast majority of these galaxies follow 
the so-called $z$-dependent main sequences previously defined for massive objects, but here the galaxies extend to much lower stellar masses and star formation rates.  The detailed analysis of their morphology (from HST images), gas 
kinematics (from the MUSE data cube), and close environments (from both) shed light on their structural properties and dynamical state. 

Most of the MUSE-HDFS galaxies have gas kinematics consistent with rotating disks. However, about 20\% of these galaxies are dynamically dominated by random motions, as revealed by low (gaseous) \voversig\ ratios. 
This fraction of dispersion-dominated galaxies is similar to the one found at $z\sim 1-2$ in the first-year KMOS$^{\rm 3D}$ sample \citep{Wisnioski+2015}. 
About 30\% of the sample galaxies are also part of a close pair and/or show clear signs of recent gravitational interactions, similar to what was found in previous IFS surveys of more 
massive galaxies \cite[e.g.,][]{Forster-Schreiber+2009, Epinat+2012, Lopez-Sanjuan+2013}. This indicates that the dynamical state and the level of gravitational interactions of 
star-forming galaxies is not a strong function of their stellar mass. 

We use the (inclination-corrected) maximum rotation velocities and stellar masses of the MUSE-HDFS galaxies to investigate their location relative to the TFR. 
The five rotation-dominated galaxies in the high-mass regime follow the TFR defined from previous surveys in a similar redshift range \cite[$z=0.2-1.4$][]{Miller+2011, Vergani+2012}. 
The situation is somewhat different in the low-mass regime (\mstar\ $< 10^{9.5}$ \msun). Indeed, rotation-dominated galaxies broadly follow the TFR so far defined in this 
lower mass/velocity regime using slit spectroscopy \citep{Kassin+2007,Kassin+2012,Miller+2014} but with a higher scatter compared to the more massive objects.
Contrary to \cite{Simons+2015}, we do not find a large population of dispersion-dominated, slow rotating galaxies in the low-mass regime; 90\% of the MUSE-HDFS galaxies with stellar masses below $10^{9.5}$ \msun\ are rotation dominated and thus have already settled into a disk.  

Finally, we take advantage of this unique dataset to characterize the angular momentum and spin parameter of intermediate-$z$ low-mass galaxies and explore the link with their dark matter haloes. 
We find that the spin distribution for the haloes of the MUSE-HDFS galaxies is broadly consistent with predictions for the dark matter in the stellar mass range $10^8$ to $10^{10}$ \msun. 
The MUSE-HDFS galaxies follow the scaling relations defined in the local universe 
\citep{Romanowsky+Fall2012, Fall+Romanowsky2013} between the specific angular momentum and stellar mass. However, we found that intermediate-redshift, star-forming galaxies 
fill a continuum transition from the spiral to elliptical local scaling relations, according to their dynamical (rotation- or dispersion-dominated) state. This suggests that some galaxies may lose 
their angular momentum and become dispersion dominated prior to becoming passive. These new observational results appear in line with recent theoretical arguments \citep[e.g.,][]{Dekel+2014,Genel+2015, Bournaud2016,Tacchella+2016}. 

Even though the MUSE-HDFS sample is already impressive in terms of data quality and uniqueness, it is still a small sample. More robust conclusions will be drawn from 
subsequent analyses based on the much larger datasets that will be acquired over the course of the MUSE-GTO.

\begin{acknowledgements}
We thank the referee for providing useful and constructive comments on the submitted version of this paper.
We warmly thank Joel Vernet, Fernando Selman, and all Paranal staff for their enthusiastic support of MUSE during the commissioning runs. 
We thank M. Fall, S. Genel, and R. Genzel for useful comments on this paper.
This work has been carried out thanks to the support of the ANR FOGHAR (ANR-13-BS05-0010-02), the OCEVU Labex (ANR-11-LABX-0060) 
and the A*MIDEX project (ANR-11-IDEX-0001-02) funded by the ``Investissements d'avenir'' French government program managed by the ANR. 
RB acknowledges support from the ERC advanced grant 339659-MUSICOS. 
JR acknowledges support from the ERC starting grant CALENDS. 
JS acknowledges support from the ERC grant agreement 278594-GasAroundGalaxies.
LMD acknowledges support from the Lyon Institute of Origins under grant ANR-10-LABX-0066.
TPKM acknowledges support from the Spanish Ministry of Economy
and Competitiveness (MINECO) under grant number AYA2013-41243-P.
BE acknowledges financial support from ``Programme National de Cosmologie and Galaxie'' (PNCG) of CNRS/INSU, France.
\end{acknowledgements}

\bibliographystyle{aa}
\bibliography{biblio.bib}


\begin{table*}[]
\caption{The sample of spatially resolved galaxies in the MUSE-HDFS.}
\begin{center}
\begin{tabular}{rrrrrrrrrrr}
\hline
\hline
\multicolumn{1}{c}{ID} &
\multicolumn{1}{c}{$z$} &
\multicolumn{1}{c}{$i$} &
\multicolumn{1}{c}{$PA_{\rm mor}$} &
\multicolumn{1}{c}{$R_{\rm d}$} &
\multicolumn{1}{c}{$PA_{\rm kin}$} &
\multicolumn{1}{c}{$V_{\rm max}$} &
\multicolumn{1}{c}{$\sigma$} &
\multicolumn{1}{c}{Int.} &
\multicolumn{1}{c}{$\log(M_{\rm star}$)} &
\multicolumn{1}{c}{$\log(SFR)$} \\
\multicolumn{1}{c}{} &
\multicolumn{1}{c}{} &
\multicolumn{1}{c}{($\degr$)} &
\multicolumn{1}{c}{($\degr$)} &
\multicolumn{1}{c}{($\arcsec$/kpc)} & 
\multicolumn{1}{c}{($\degr$)} &
\multicolumn{1}{c}{(\kms)} &
\multicolumn{1}{c}{(\kms)} &
\multicolumn{1}{c}{} &
\multicolumn{1}{c}{(\msun)} &
\multicolumn{1}{c}{(\msunpyr)} \\
\multicolumn{1}{c}{[1]} &
\multicolumn{1}{c}{[2]} &
\multicolumn{1}{c}{[3]} &
\multicolumn{1}{c}{[4]} &
\multicolumn{1}{c}{[5]} & 
\multicolumn{1}{c}{[6]} &
\multicolumn{1}{c}{[7]} &
\multicolumn{1}{c}{[8]} &
\multicolumn{1}{c}{[9]} &
\multicolumn{1}{c}{[10]} &
\multicolumn{1}{c}{[11]} \\
\hline
    1 & 0.1723 & 75 & $31.3\pm0.1$ & {$0.46/1.34\pm0.01$} & $33\pm2$ & $74\pm4$ & $30\pm3$ & 0 & $8.78\pm0.42$ & $-1.01\pm0.67$\\
    3 & 0.5637 & 16 & $-18.8\pm3.0$ & {$0.66/4.28\pm0.04$} & $64\pm2$ & $103\pm20$ & $34\pm1$ & 0 & $9.66\pm0.14$ & $0.24\pm0.37$\\
    4 & 0.5638 & 75 & $35.8\pm0.1$ & {$0.62/4.01\pm0.03$} & $34\pm1$ & $204\pm20$ & $48\pm3$ & 0 & $9.86\pm0.17$ & $0.54\pm0.35$\\
    5 & 0.5798 & 68 & $6.9\pm0.2$ & {$0.25/1.67\pm0.02$} & $37\pm6$ & $47\pm16$ & $54\pm3$ & 0 & $9.82\pm0.17$ & $0.59\pm0.38$\\
    6 & 0.4230 & 29 & $45.0\pm2.0$ & {$0.46/2.54\pm0.03$} & $9\pm7$ & $80\pm4$ & $29\pm3$ & 1 & $9.23\pm0.10$ & $-0.61\pm0.50$\\
    7 & 0.4637 & 41 & $43.0\pm0.7$ & {$0.44/2.55\pm0.03$} & $50\pm3$ & $76\pm2$ & $27\pm15$ & 1 & $9.31\pm0.12$ & $-0.51\pm0.54$\\
    8 & 0.5772 & 61 & $-29.4\pm0.4$ & {$0.18/1.20\pm0.01$} & $-20\pm2$ & $124\pm20$ & $84\pm12$ & 1 & $9.88\pm0.23$ & $1.39\pm0.58$\\
    9 & 0.5638 & 61 & $-34.8\pm0.4$ & {$0.25/1.60\pm0.02$} & $-37\pm2$ & $125\pm3$ & $54\pm3$ & 0 & $9.31\pm0.20$ & $0.80\pm0.41$\\
  10 & 1.2840 & 26 & $39.9\pm4.1$ & {$0.24/2.05\pm0.06$} & $107\pm15$ & $98\pm14$ & $116\pm3$ & 2 & $10.77\pm0.11$ & $0.30\pm0.89$\\
  11 & 0.5779 & 62 & $27.3\pm0.5$ & {$0.10/0.67\pm0.01$} & $29\pm3$ & $61\pm12$ & $49\pm1$ & 0 & $9.22\pm0.18$ & $0.08\pm0.49$\\
  12 & 0.6701 & 37 & $-144.0\pm4.9$ & {$0.04/0.29\pm0.01$} & $-145\pm30$ & $48\pm12$ & $59\pm1$ & 2 & $9.02\pm0.27$ & $0.82\pm0.67$\\
  13 & 1.2900 & 33 & $-45.9\pm1.9$ & {$0.15/1.25\pm0.02$} & $-13\pm2$ & $24\pm4$ & $48\pm1$ & 0 & $9.89\pm0.11$ & $1.89\pm0.18$\\
  16 & 0.4647 & 20 & $-40.0\pm7.6$ & {$0.40/2.35\pm0.06$} & $-53\pm10$ & $91\pm3$ & $34\pm3$ & 0 & $8.74\pm0.21$ & $-0.65\pm0.55$\\
  17 & 0.5810 & 45 & $77.3\pm1.2$ & {$0.26/1.73\pm0.04$} & $95\pm15$ & $87\pm30$ & $41\pm6$ & 0 & $9.06\pm0.17$ & $-0.13\pm0.84$\\
  20 & 0.4275 & 73 & $-38.8\pm0.5$ & {$0.49/2.77\pm0.10$} & $-121\pm4$ & $41\pm15$ & $21\pm3$ & 0 & $8.71\pm0.06$ & $-1.16\pm0.50$\\
  23 & 0.5641 & 24 & $78.3\pm4.9$ & {$0.51/3.38\pm0.09$} & $14\pm8$ & $32\pm6$ & $31\pm3$ & 0 & $8.72\pm0.27$ & $-0.19\pm0.70$\\
  24 & 0.9723 & 14 & $-1.2\pm13.8$ & {$0.18/1.44\pm0.05$} & $-68\pm3$ & $164\pm21$ & $45\pm3$ & 0 & $9.61\pm0.24$ & $0.77\pm0.73$\\
  26 & 0.2249 & 70 & $15.0\pm0.4$ & {$0.46/1.66\pm0.05$} & $14\pm4$ & $75\pm9$ & $16\pm3$ & 0 & $7.95\pm0.35$ & $-1.87\pm0.63$\\
  27 & 1.2853 & 71 & $-9.4\pm0.5$ & {$0.48/4.00\pm0.05$} & $5\pm2$ & $168\pm15$ & $99\pm6$ & {1,2} & $10.32\pm0.19$ & $1.06\pm0.95$\\
  28 & 0.3179 & 32 & $84.9\pm3.0$ & {$0.21/0.99\pm0.04$} & $75\pm22$ & $47\pm10$ & $29\pm3$ & 0 & $7.99\pm0.14$ & $-1.24\pm0.38$\\
  32 & 0.5644 & 68 & $122.0\pm0.5$ & {$0.28/1.83\pm0.06$} & $112\pm8$ & $70\pm7$ & $50\pm3$ & 0 & $8.57\pm0.17$ & $-1.07\pm0.39$\\
  35 & 1.2806 & 41 & $44.3\pm2.5$ & {$0.17/1.40\pm0.06$} & $34\pm7$ & $214\pm30$ & $87\pm12$ & 0 & $10.12\pm0.26$ & $1.37\pm0.96$\\
  37 & 1.0978 & 72 & $6.1\pm0.7$ & {$0.12/0.98\pm0.02$} & $46\pm20$ & $21\pm7$ & $66\pm3$ & 0 & $9.12\pm0.17$ & $0.89\pm0.43$\\
  38 & 1.0217 & 57 & $-29.0\pm1.2$ & {$0.23/1.82\pm0.08$} & $-23\pm6$ & $98\pm10$ & $45\pm6$ & 0 & $8.86\pm0.14$ & $0.17\pm0.42$\\
  49 & 0.9991 & 68 & $-68.8\pm1.0$ & {$0.19/1.52\pm0.08$} & $-69\pm6$ & $86\pm20$ & $51\pm6$ & 1 & $8.83\pm0.21$ & $-0.02\pm0.72$\\
  68 & 0.9715 & 78 & $26.3\pm0.9$ & {$0.16/1.29\pm0.06$} & $9\pm7$ & $70\pm13$ & $39\pm6$ & 0 & $8.29\pm0.24$ & $-0.51\pm0.53$\\
  72 & 0.8391 & 74 & $-90.0\pm1.0$ & {$0.16/1.22\pm0.07$} & $-87\pm8$ & $135\pm9$ & $24\pm21$ & 0 & $8.68\pm0.25$ & $-0.62\pm0.42$\\
  88 & 1.3600 & 77 & $8.8\pm2.0$ & {$0.21/1.76\pm0.22$} & $-64\pm12$ & $51\pm9$ & $34\pm12$ & 2 & $8.71\pm0.14$ & $-0.45\pm0.37$\\
\hline\end{tabular}
\tablefoot{
Column [1]: Identification number according to \cite{Bacon+2015}; 
Column [2]: Spectroscopic redshift from MUSE according to \cite{Bacon+2015}; 
Columns [3-4-5]: Disk inclination (with typical uncertainties of $\pm 5$\degr), position angle of the disk major axis, and disk scale length (both in arcsec and kpc, but uncertainties are given in kpc only) derived from {\sc Galfit} modeling on HST F814W images (see sect.~\ref{sec:morpho}); 
Columns [6-7-8]: Position angle of the disk major axis, maximum rotation velocity, and velocity dispersion derived from kinematic modeling of MUSE data cubes (see sect.~\ref{sec:kine}); 
Column [9]: Flag indicating the level of gravitational interactions: $0=$ isolated, $1=$ minor pair/interaction, $2=$ major pair/interaction or clear merger remnant (see sect.~\ref{sec:pairs}); 
Columns [10-11]: Stellar mass and star formation rate derived from SED fitting on visible$+$NIR photometry (see sect.~\ref{sec:globprop}).
}
\end{center}
\label{tbl:sample}
\end{table*}

\clearpage

\appendix
\section{Notes on individual galaxies and associated morpho-kinematics maps}
\label{sec:appen_notes}

\paragraph{\bf ID \#1}
Nearly edge-on ($i=75\degr$) and extended ($\approx 4$\arcsec\ in diameter) galaxy disk at $z\approx 0.17$, the lowest redshift of the sample, enabling the observation of both \oiiib\ and \halpha\ strong emission lines with MUSE. The morphology is hardly reproduced by an exponential disk because the intensity is too low in the center. This might be because of some absorption not taken into account in the model. Both sets of emission lines have an asymmetric distribution probably due to an off-centered star-forming region in the southwest side. This asymmetry is also seen in the morphology model residuals. Velocity fields derived from these emission lines are consistent and show an observed regular pattern within $\pm 60$ \kms. Disk modeling (both in 2D and with \gpk) reproduces well the observed velocity fields. Kinematic parameters reported in Table~\ref{tbl:sample} are those derived from \halpha. The gas dynamics of this galaxy is clearly dominated by the rotation with a ratio $V/\sigma \approx 2.5$. The velocity dispersion map is rather 
flat except far from the center along the minor axis. We note also that the \oiiib\ emission is more extended than \halpha\ along the galaxy minor axis, which could be the signature for outflows.

\begin{figure} \resizebox{\hsize}{!}{\includegraphics{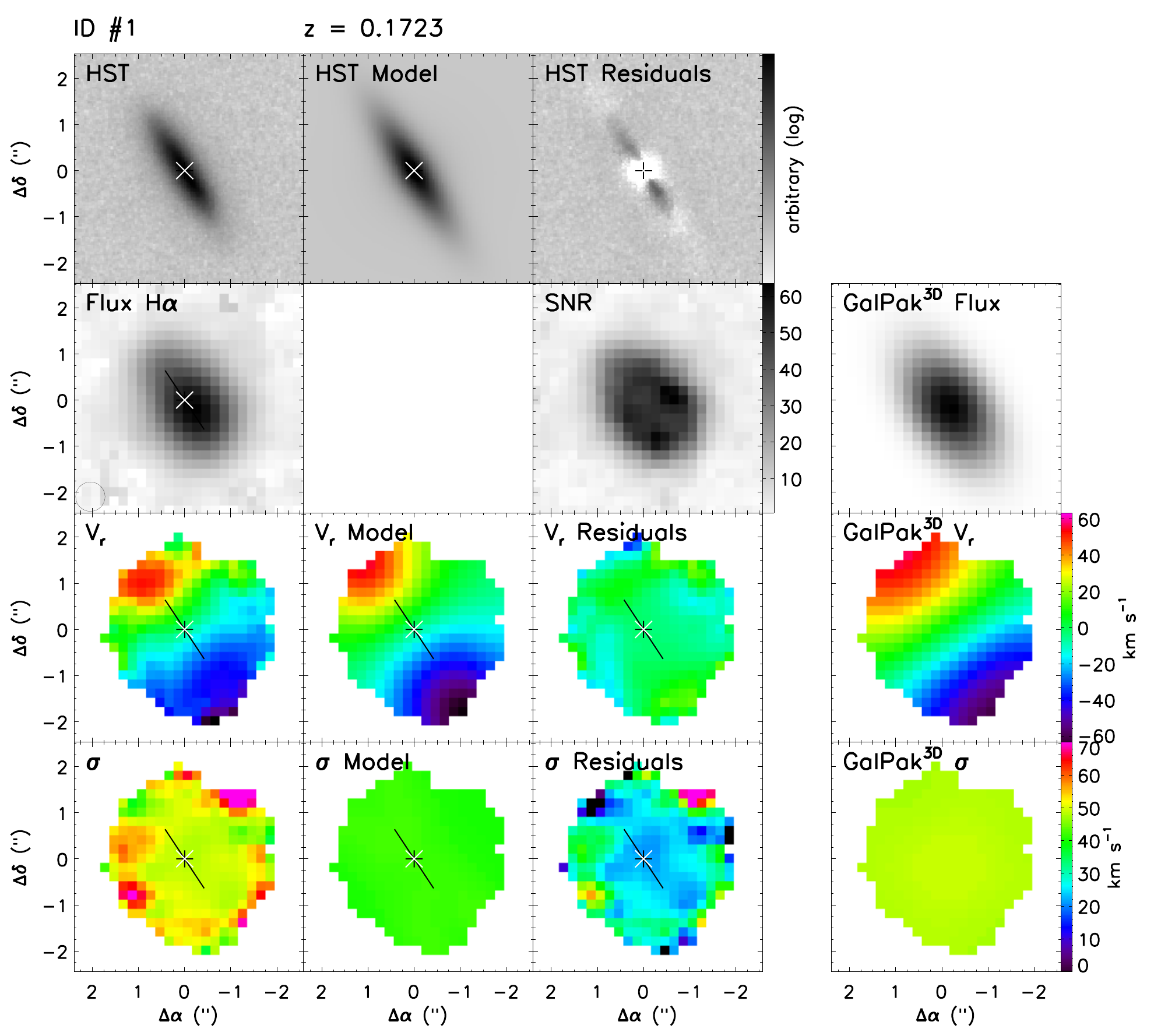}}
\caption{Morpho-kinematics maps for galaxy ID \#1. Same caption as Fig.~\ref{fig:vfexample} but emission-line fitting in the MUSE data cube is based on \halpha.}
\label{fig:appen_figs_obj1}
\end{figure}

\paragraph{\bf ID \#3}
This well-resolved ($\approx 6$\arcsec\ in diameter) galaxy is a nearly face-on ($i=16\degr$) disk at $z\approx 0.56$, allowing us to observe both \oii\ (the brightest) and \oiiib\ strong emission lines with MUSE. The disk of this galaxy is rather clumpy with bright \hii\ regions along spiral arms. The flux distribution seems more clumpy and asymmetric for the \oiiib\ line than for the \oii\ doublet. Velocity fields derived from \oii\ and \oiiib\ emission lines are consistent and show an observed low gradient regular pattern within $\pm 40$ \kms. The velocity dispersion map is rather flat. Disk modeling (both in 2D and with \gpk) reproduces well the observed velocity fields, but the \gpk\ analysis returns an overestimated disk inclination. This value has thus been further constrained in \gpk\ to be within $16\pm5\degr$. The kinematic position angle is almost orthogonal to the morphological position angle, probably owing to an ill-constrained morphology since it has several asymmetric arms. Kinematic parameters reported in Table~\ref{tbl:sample} are 
those derived from \oii. The gas dynamics of this galaxy is clearly dominated by the rotation with a ratio $V/\sigma \approx 3$. 

\begin{figure} \resizebox{\hsize}{!}{\includegraphics{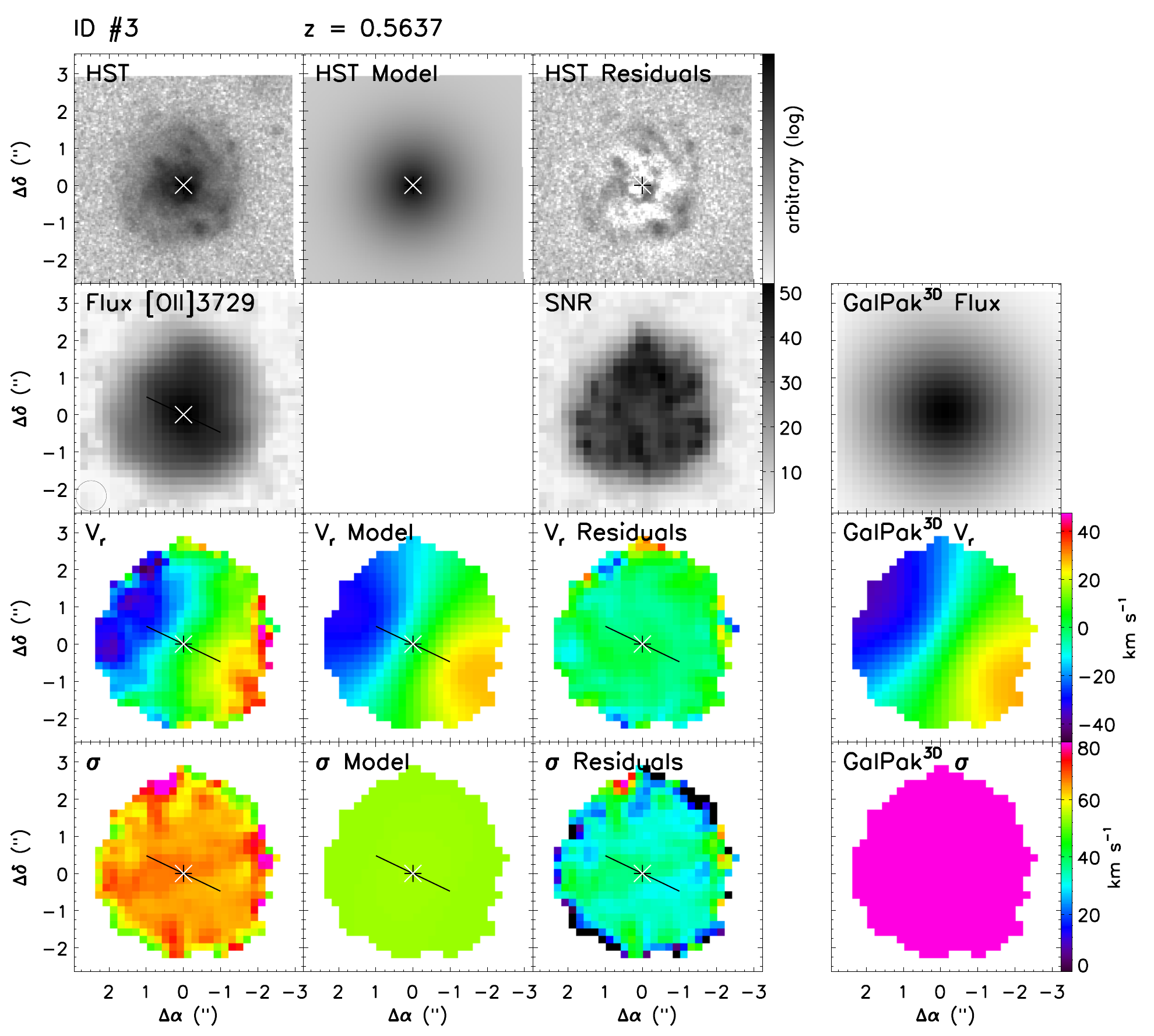}}
\caption{Morpho-kinematics maps for galaxy ID \#3. Same caption as Fig.~\ref{fig:vfexample}}
\label{fig:appen_figs_obj3}
\end{figure}

\paragraph{\bf ID \#4}
Highly inclined ($i\approx 75\degr$) and large ($\approx 4$\arcsec\ in diameter) disk at $z\approx 0.56$, allowing us to observe both \oii\ (the brightest) and \oiiib\ strong emission lines with MUSE. The morphological center is offset with respect to the center of external isophotes by less than $0.1$\arcsec, which might be due to some absorption in this inclined disk. The morphological model residuals display a spiral pattern.
The emission of this galaxy is dominated by a bright \hii\ region located in the northeast direction with respect to the galaxy morphological/kinematic center. Velocity fields derived from these emission lines are consistent and show an observed regular pattern within $\pm 150$ \kms. Disk modeling (both in 2D and with \gpk) reproduces well the observed velocity fields but the \gpk\ analysis returns an underestimated disk inclination. This value has thus been further constrained in \gpk\ to be within $76\pm 5\degr$. The velocity dispersion map shows an increase along the minor axis, as expected from beam smearing effects for such a large rotation velocity. 
Kinematic parameters reported in Table~\ref{tbl:sample} are those derived from \oii. The gas dynamics of this galaxy is clearly dominated by the rotation with a ratio $V/\sigma \approx 4.2$. 

\paragraph{\bf ID \#5}
The inclination of this galaxy at $z\approx 0.58$ is quite difficult to constrain with a value of $70\degr$ returned with {\sc Galfit,} whereas both \cite{Trujillo+2004} and \gpk\ give a value around $45\degr$. This discrepancy probably comes from the fact that the emission of this galaxy is rather compact with a bright nucleus and two open arms extending north and south. Velocity fields derived from both \oii\ and \oiiib\ (the brightest) emission lines are consistent and show an observed regular pattern within $\pm 40$ \kms. Disk modeling in 2D reproduces quite well the observed velocity fields, but \gpk\ has some difficulty to converge, probably because of asymmetries seen in the HST images of this galaxy. Constraining the kinematical position angle to be close to the morphological one, i.e.,\, within $35\pm 10\degr$, improves significantly the 3D modeling. The kinematic parameters reported in Table~\ref{tbl:sample} are derived from \oiiib\ emission line. The velocity dispersion is relatively large, which could indicate that the velocity gradient is not correctly resolved. With a ratio $V/\sigma \approx 0.9$, this object seems to be dominated by the dispersion, especially in its central region.

\begin{figure} \resizebox{\hsize}{!}{\includegraphics{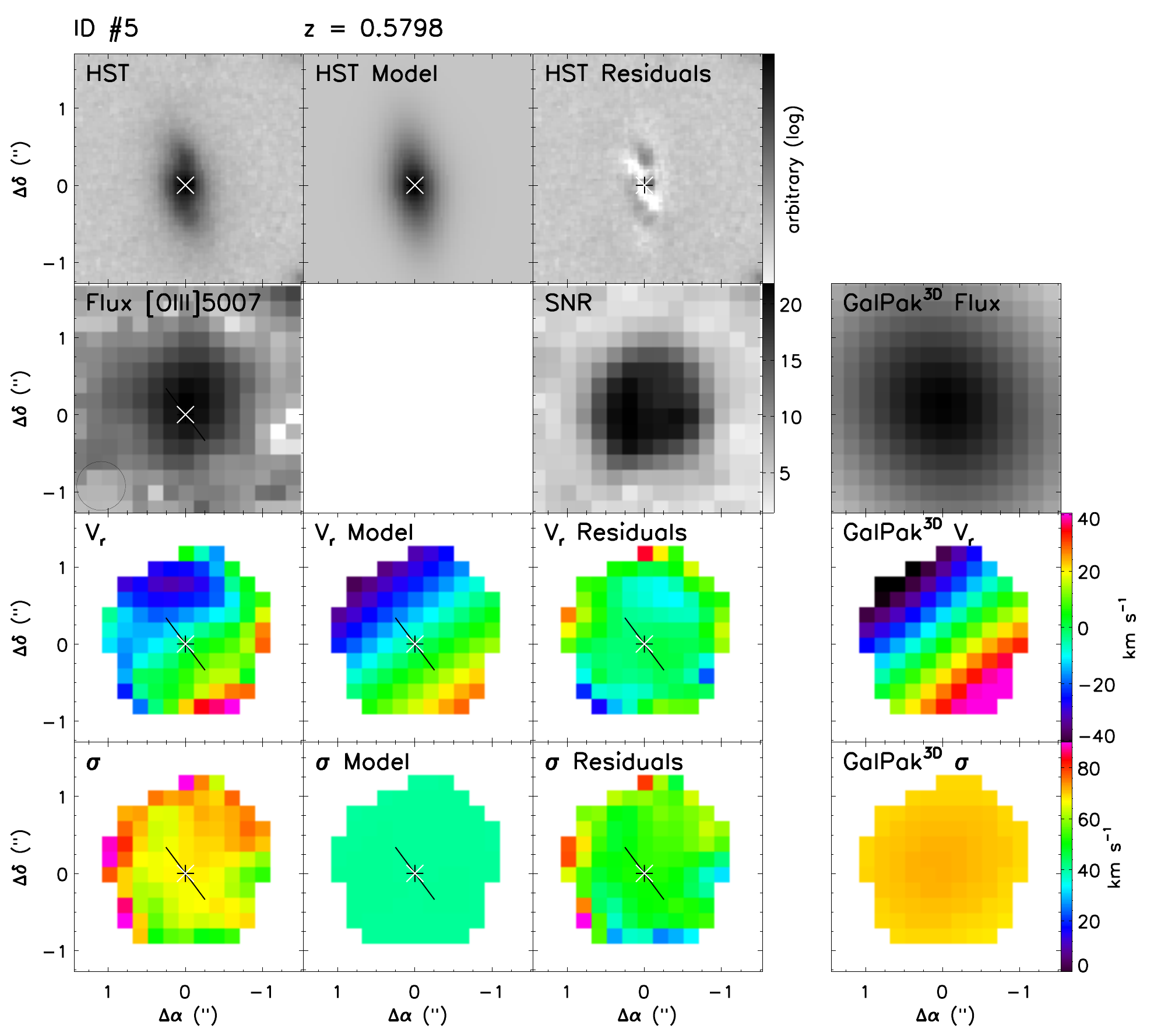}}
\caption{Morpho-kinematics maps for galaxy ID \#5. Same caption as Fig.~\ref{fig:vfexample} but emission-line fitting in the MUSE data cube is based on \oiiib.}
\label{fig:appen_figs_obj5}
\end{figure}

\begin{figure} \resizebox{\hsize}{!}{\includegraphics{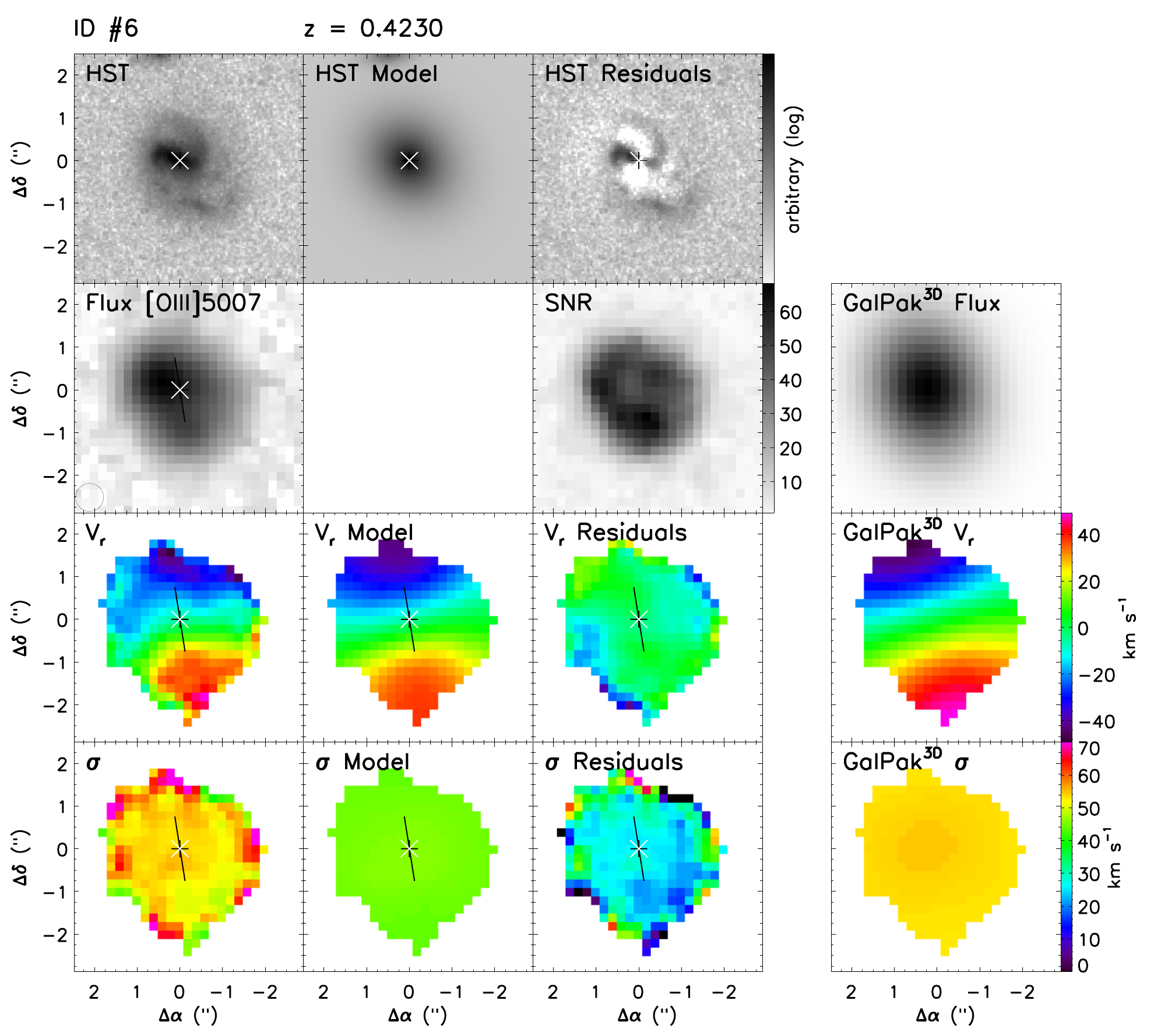}}
\caption{Morpho-kinematics maps for galaxy ID \#6. Same caption as Fig.~\ref{fig:vfexample} but emission-line fitting in the MUSE data cube is based on \oiiib.}
\label{fig:appen_figs_obj6}
\end{figure}

\paragraph{\bf ID \#6}
The properties of this asymmetric barred galaxy at $z\approx 0.42$ have already been discussed in \cite{Bacon+2015}. It is a rather large ($>2$\arcsec\ in diameter) galaxy that clearly displays an elongated and asymmetric spiral morphology (extending southward) with a strong bar, which makes its morphological model difficult to adjust. In particular, the bar affects the determination of the axis ratio as well as of the true position angle. The barred structure is clearly detected in the MUSE velocity field with its typical S-shape on top of a rotating disk velocity field. The maximum rotation velocity derived with the disk modeling (both in 2D and with \gpk) is compatible with the $\approx 70$ \kms\ estimated previously with GIRAFFE \citep{Puech+2008}. The velocity dispersion map is rather flat. For the 3D modeling, we constrained the kinematical PA to be close to the morphological PA, i.e.,\, within $10\pm 10\degr$. Kinematic parameters reported in Table~\ref{tbl:sample} are those derived from \oiiib. The gas dynamics of this galaxy is  dominated by the rotation with a ratio $V/\sigma \approx 2.7$. This galaxy has a low-mass ($\approx 10^8$\msun) satellite, ID \#101, located at a projected distance of 30 kpc (barely seen northward in Fig.~\ref{fig:appen_figs_obj6} owing to the limited size of the thumbnail) and with a systemic velocity difference of $11$ \kms. With a stellar mass ratio of  $\sim 1/30$, it can be considered a possible future minor merger (Ventou et al., in prep.).  

\paragraph{\bf ID \#7}
Moderately inclined ($i\approx 40\degr$) and large ($\approx 3$\arcsec\ in diameter) disk at $z\approx 0.46$, allowing us to observe both \oii\ and \oiiib\ (the brightest) strong emission lines with MUSE. This galaxy displays an asymmetric spiral structure that makes its morphology difficult to model accurately. The emission of this galaxy is dominated by two bright \hii\ regions located toward the northeast and southwest directions with respect to the galaxy morphological/kinematic center. This galaxy has a very low-mass ($\approx 10^7$\msun) satellite, not identified in the redshift catalog of \cite{Bacon+2015}, located southward at a projected distance of 9 kpc and with a systemic velocity difference of $310$ \kms. With a stellar mass ratio of  $\sim 1/100$, it can be considered a possible future minor merger (Ventou et al., in prep.). Velocity fields derived from these emission lines are consistent and show an observed regular pattern within $\pm 60$ \kms. Disk modeling (both in 2D and with \gpk) reproduces well 
the observed velocity fields. The velocity dispersion map is rather flat. Kinematic parameters reported in Table~\ref{tbl:sample} are those derived from \oiiib. The gas dynamics of this galaxy is clearly dominated by the rotation with a ratio $V/\sigma \approx 3$. 

\begin{figure} \resizebox{\hsize}{!}{\includegraphics{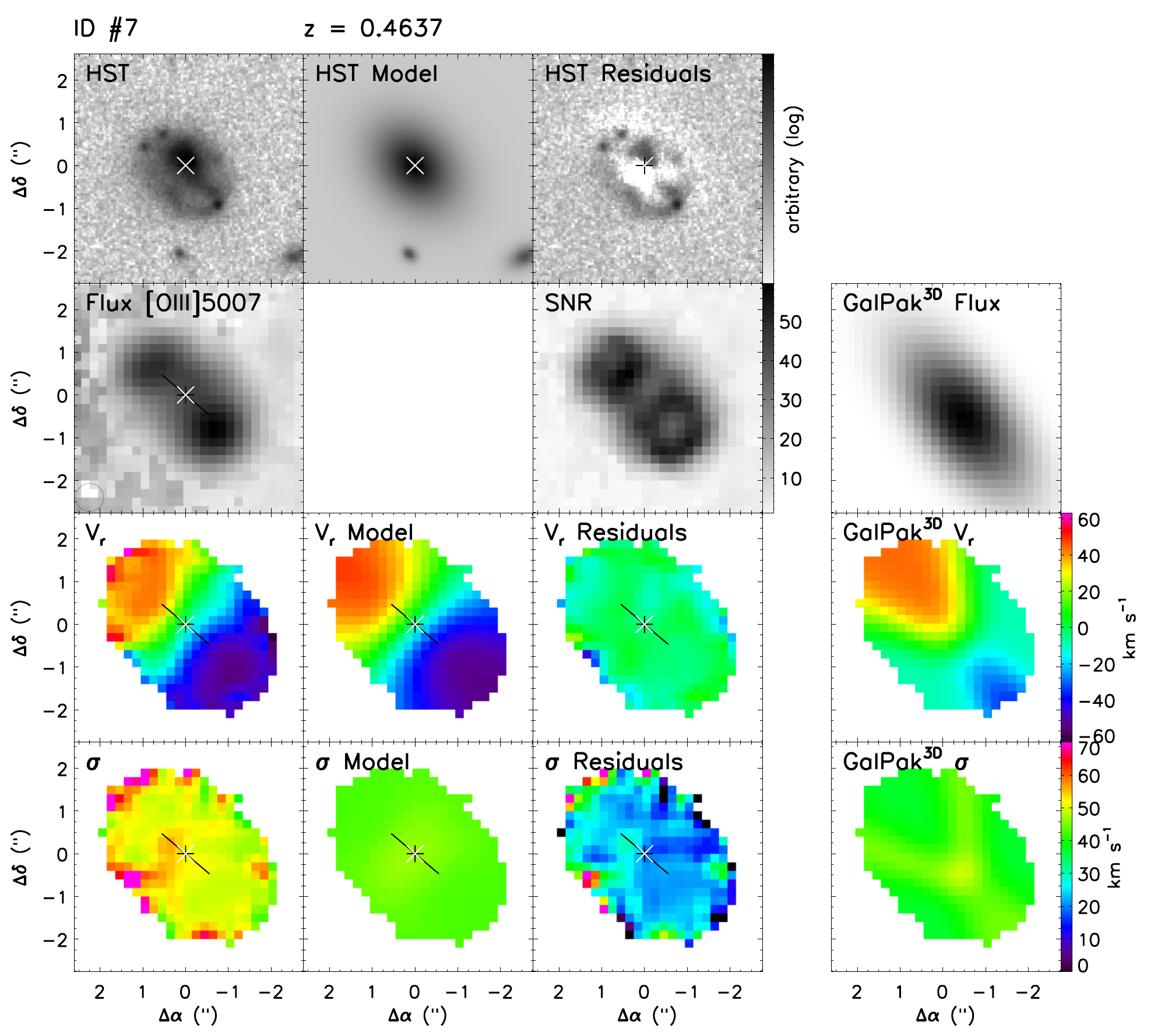}}
\caption{Morpho-kinematics maps for galaxy ID \#7. Same caption as Fig.~\ref{fig:vfexample} but emission-line fitting in the MUSE data cube is based on \oiiib.}
\label{fig:appen_figs_obj7}
\end{figure}

\paragraph{\bf ID \#8}
This moderately inclined ($i\approx 60\degr$) disk at $z\approx 0.58$ shows both \oii\ (the brightest) and \oiiib\  emission lines with MUSE. This galaxy displays an asymmetric broadband morphology as revealed by the model residuals. Velocity fields derived from these emission lines are consistent and show a main rotating component within $\pm 130$ \kms and an extension toward the east that may be linked to a (passive) small object seen in the HST image. This galaxy has a low-mass ($\approx 10^{8.5}$\msun) satellite, not identified in the redshift catalog of \cite{Bacon+2015}, located westward at a projected distance of 6 kpc and with a systemic velocity difference of $125$ \kms. With a stellar mass ratio of  $\sim 1/30$, it can be considered a possible future minor merger (Ventou et al., in prep.). Disk modeling (both in 2D and with \gpk) reproduces well the observed velocity fields, but the \gpk\ analysis returns an underestimated disk inclination. This value has thus been further constrained in \gpk\ to be within $61\pm 5\degr$. 
The velocity dispersion map shows an increase at the edges. This could also be due to beam smearing effect, in particular in the east extent, but this could also be related to in- and outflowing material. Kinematic parameters reported in Table~\ref{tbl:sample} are those derived from \oii. Even if the velocity dispersion is high ($\sigma \approx 84$ \kms) the gas dynamics of this galaxy is still dominated by the rotation with a ratio $V/\sigma \approx 1.5$. 

\begin{figure} \resizebox{\hsize}{!}{\includegraphics{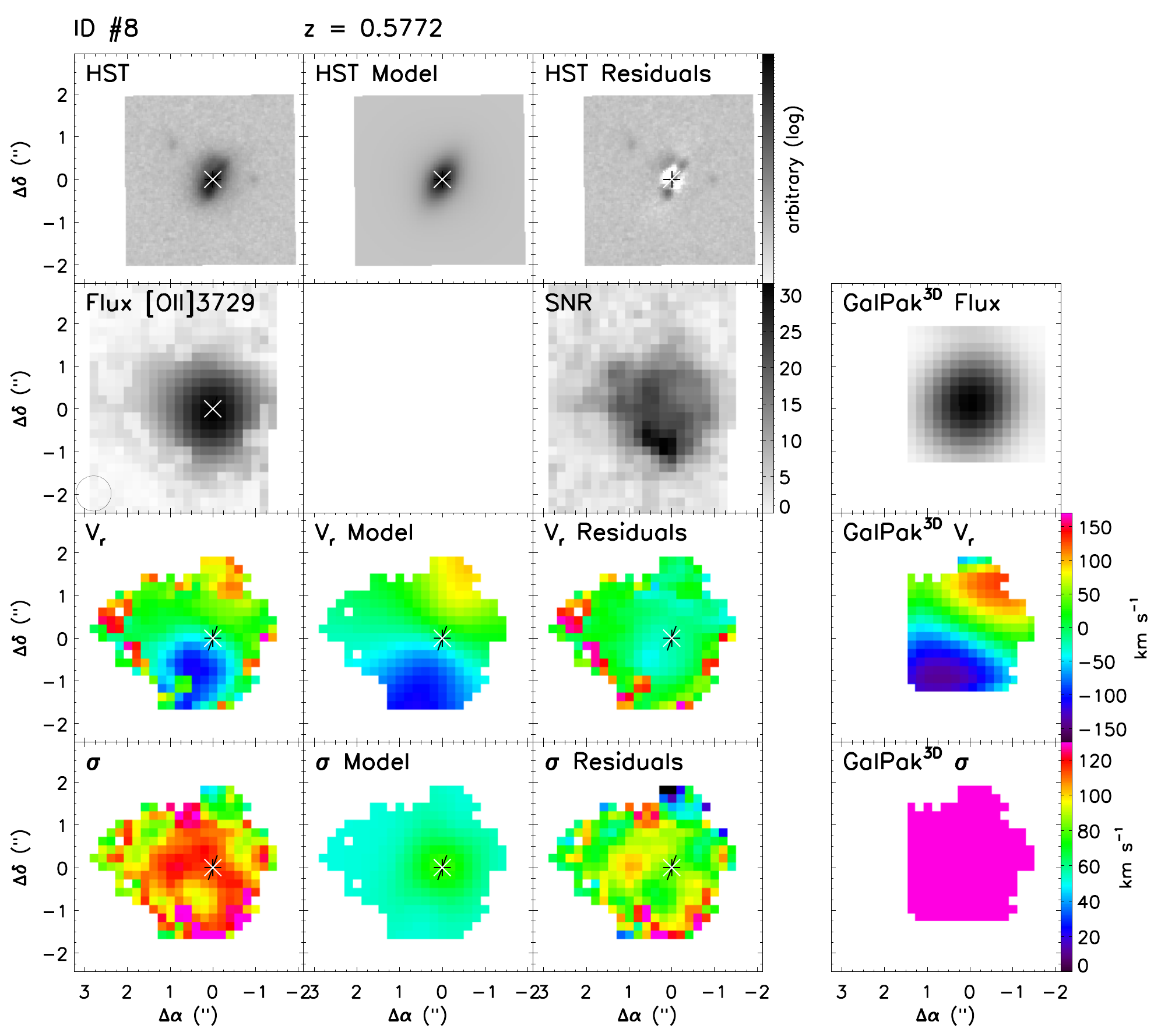}}
\caption{Morpho-kinematics maps for galaxy ID \#8. Same caption as Fig.~\ref{fig:vfexample}}
\label{fig:appen_figs_obj8}
\end{figure}

\paragraph{\bf ID  \#9}
The properties of this large ($\approx 2$\arcsec\ in diameter) galaxy at $z\approx 0.56$ have already been discussed in \cite{Bacon+2015}. The maximum rotation velocity of $\approx 125$ \kms\ derived with the disk modeling (both in 2D and with \gpk) is significantly lower than the value ($\approx 220$ \kms) obtained with GIRAFFE data \citep{Puech+2008}. The velocity field of this galaxy is rather regular and typical of early-type disks with a prominent bulge. The velocity dispersion ($\approx 55$ \kms) is clearly peaked at the center of this galaxy and is higher than the value of $32$ \kms\ reported in the IMAGES survey \citep{Puech+2006}. We note, however, some structures in the velocity dispersion residual map of this galaxy with velocity dispersion on the order of $\sim 80-100$ \kms. Such broad component, aligned along the southwest side of the minor axis could well be produced by superwind-driven shocks \citep{Bacon+2015}. 
Kinematic parameters reported in Table~\ref{tbl:sample} are those derived from \oii. The gas dynamics of this galaxy is dominated by the rotation with a ratio $V/\sigma \approx 2.3$.

\begin{figure} \resizebox{\hsize}{!}{\includegraphics{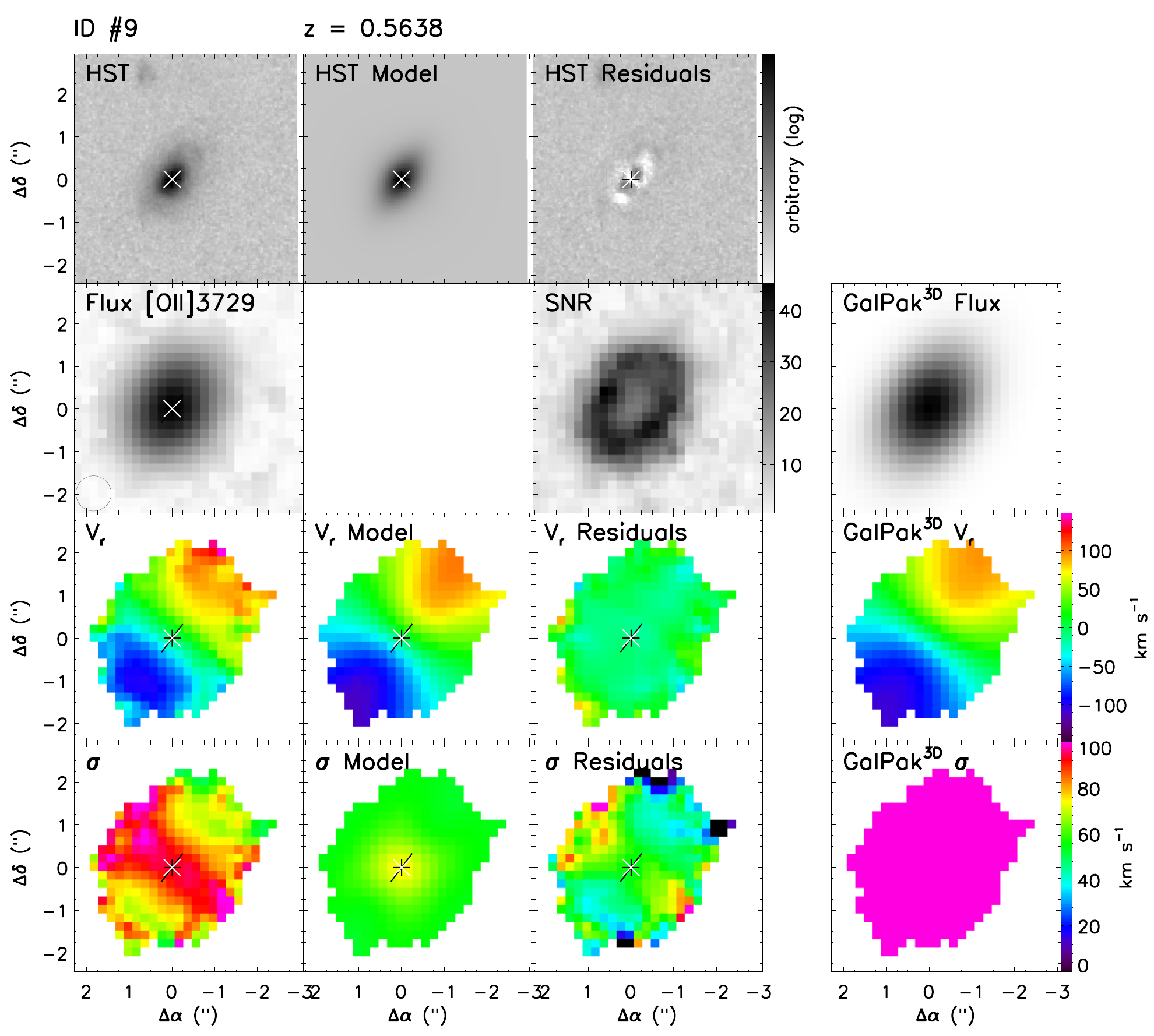}}
\caption{Morpho-kinematics maps for galaxy ID \#9. Same caption as Fig.~\ref{fig:vfexample}}
\label{fig:appen_figs_obj9}
\end{figure}

\paragraph{\bf ID \#10}
This galaxy is the most massive among our sample ($M\approx 10^{10.8}$\msun), is a member of a group at $z\approx 1.28$, and is probably interacting with the galaxy ID\#27 as revealed by low-surface brightness tidal tails in between the two objects. The galaxy is rather compact ($\approx 1$\arcsec\ in diameter) in the HST images and its spectrum is typical of an AGN with prominent \neiii, \nev, and \mgii\ emission lines. Its broadband morphology clearly shows a central component and two low-surface brightness arms. This galaxy is seen nearly face-on with an inclination of $\approx 25 \degr$ derived both with {\sc Galfit} and \gpk. The velocity field derived from the \oii\ doublet seems quite perturbed with a strange feature on the northwest side (which is probably due to a night sky line residual), but also shows a low gradient within $\pm 40$ \kms. The velocity dispersion map peaks at high value ($\approx 120$ \kms), which is quite common for AGNs. With a ratio $V/\sigma \approx 0.8$, this object is clearly dominated by non-circular motions. 

\begin{figure} \resizebox{\hsize}{!}{\includegraphics{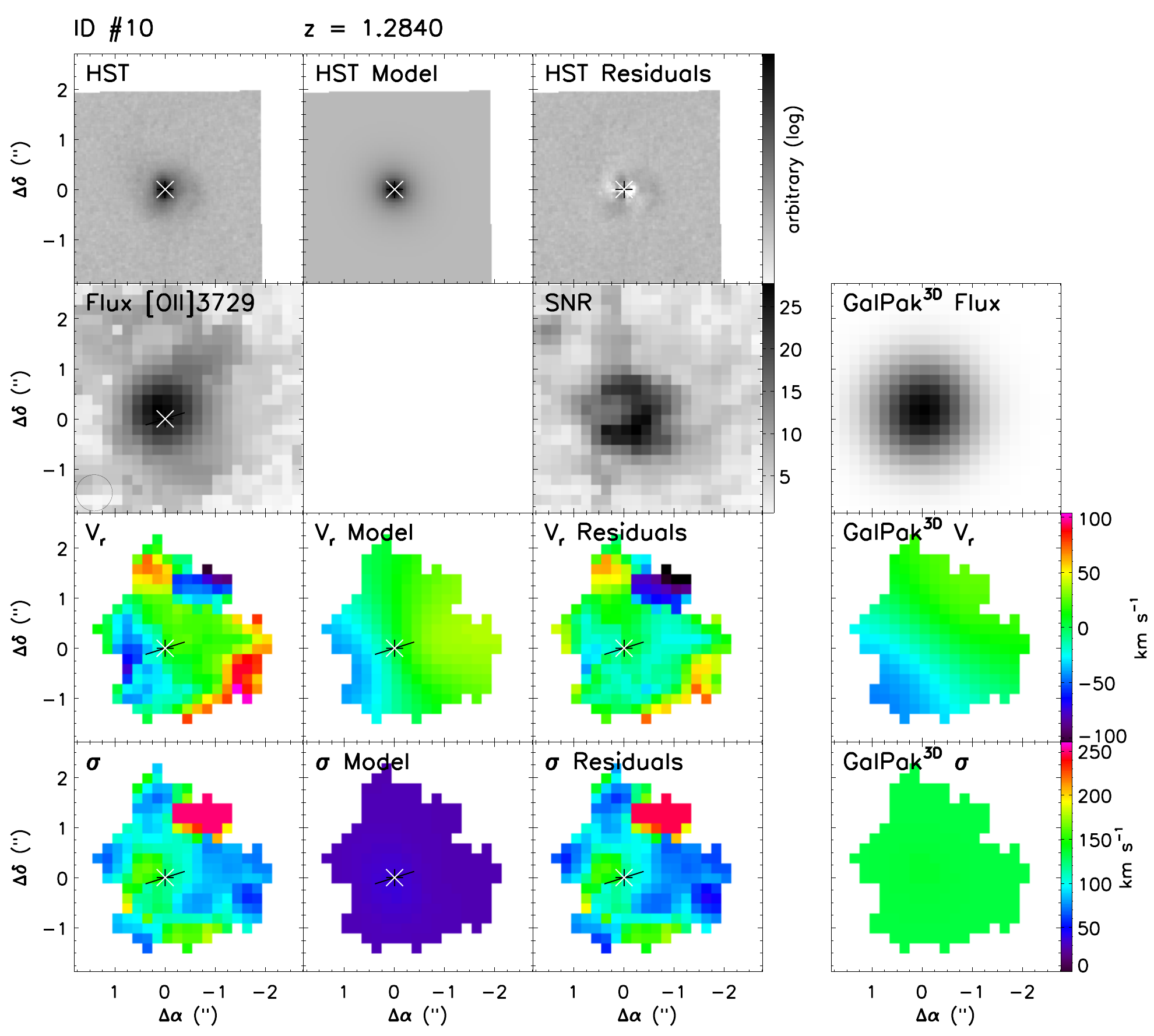}}
\caption{Morpho-kinematics maps for galaxy ID \#10. Same caption as Fig.~\ref{fig:vfexample}}
\label{fig:appen_figs_obj10}
\end{figure}

\paragraph{\bf ID \#11}
This moderately inclined ($i\approx 60\degr$) and rather small ($\approx 1$\arcsec\ in diameter) disk at $z\approx 0.58$ shows both \oii\ and \oiiib\ emission lines with MUSE. The line flux maps seem rather circular for such an inclined galaxy as seen at high resolution in the HST images. This might be because of the size of the beam with respect to the size of the galaxy. However, velocity fields derived from these emission lines are consistent and show an observed regular pattern within $\pm 30$ \kms. Disk modeling in 2D reproduces well the observed velocity fields. However the kinematical position angle is not very well constrained with \gpk. Constraining this parameter to be close to the morphological one, i.e.,\, within $30\pm 10\degr$, significantly improves the 3D modeling. Kinematic parameters reported in Table~\ref{tbl:sample} are those derived from \oiiib. Even if the rotation velocity is rather low (\vmax\ $\approx 61$ \kms) and the velocity dispersion is relatively high (which may indicate that the velocity gradient is not well resolved) the gas dynamics of this galaxy is still dominated by its rotation with a ratio $V/\sigma \approx 1.2$. 

\begin{figure} \resizebox{\hsize}{!}{\includegraphics{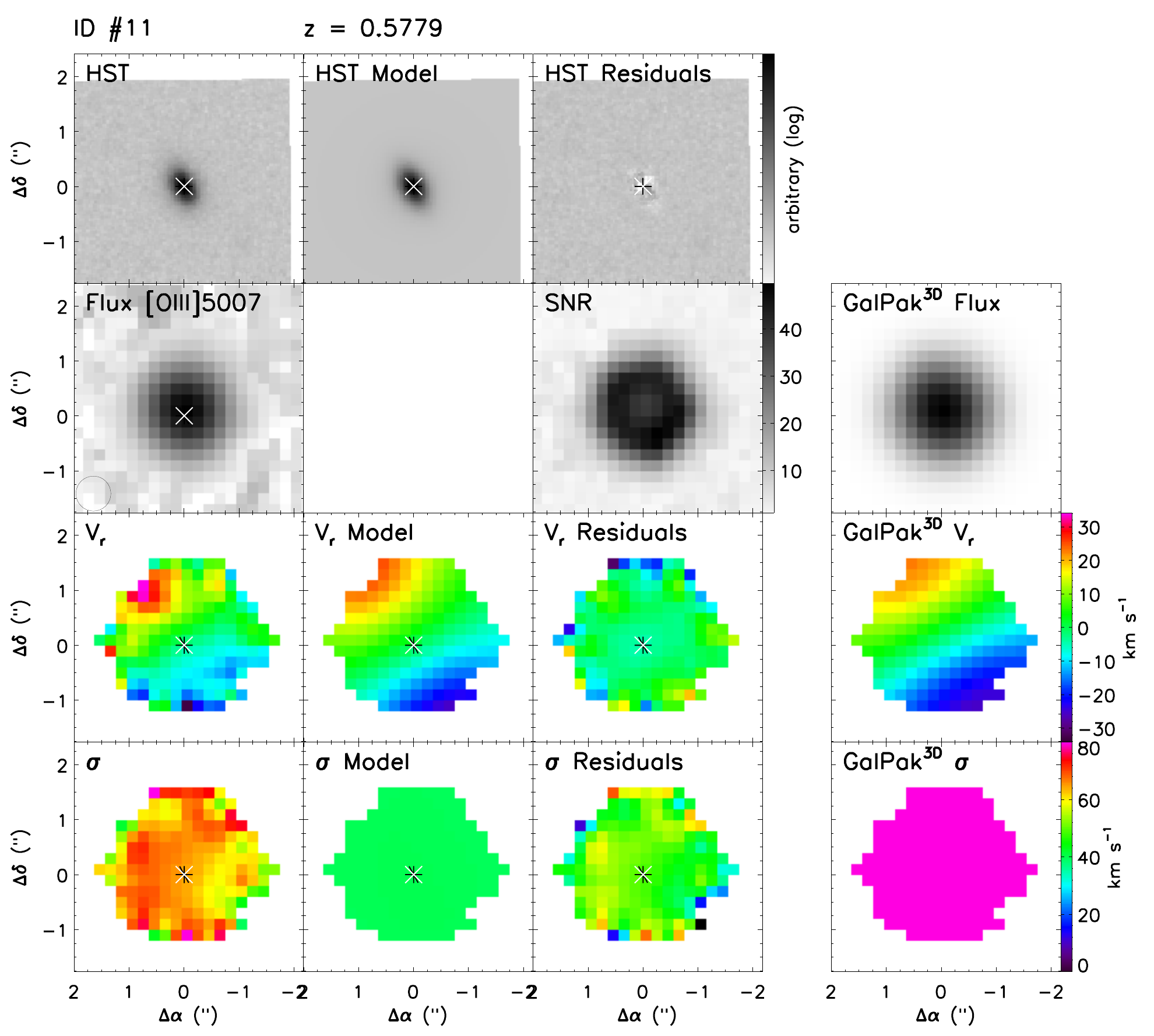}}
\caption{Morpho-kinematics maps for galaxy ID \#11. Same caption as Fig.~\ref{fig:vfexample} but emission-line fitting in the MUSE data cube is based on \oiiib.}
\label{fig:appen_figs_obj11}
\end{figure}

\paragraph{\bf ID \#12}
This galaxy at $z\approx 0.67$ is very compact ($\approx 0.5$\arcsec\ in diameter) in the broadband HST images but shows (also in MUSE white-light image) an extended low-surface brightness jet-like structure eastward, which could be a relic of past gravitational interactions. The \oiiib\ and \oii\ emission-line flux maps are rather circular and their extent ($>2$\arcsec\ in diameter) is probably only due to beam smearing and the very high S/N. 
With a rather low inclination ($i\approx 37\degr$), derived both with {\sc Galfit} and \gpk, the velocity gradient within $\pm 20$ \kms\ is very low. Kinematic parameters reported in Table~\ref{tbl:sample} are those derived from \oii. With a ratio $V/\sigma \approx 0.8$, this object seems to be dominated by non-circular motions. 

\begin{figure} \resizebox{\hsize}{!}{\includegraphics{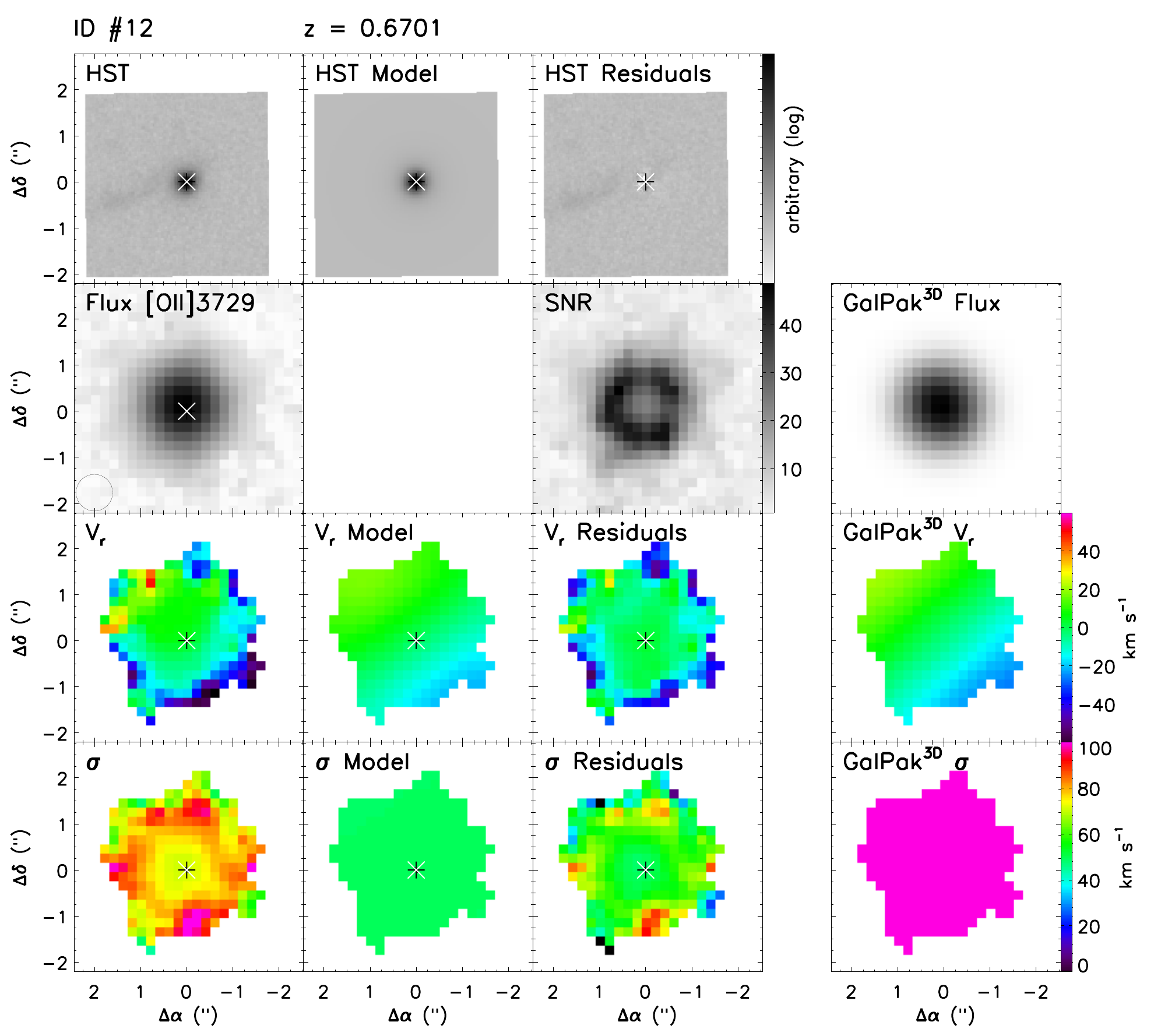}}
\caption{Morpho-kinematics maps for galaxy ID \#12. Same caption as Fig.~\ref{fig:vfexample}}
\label{fig:appen_figs_obj12}
\end{figure}

\paragraph{\bf ID \#13}
This galaxy is also a member of the group at $z\approx 1.28$. It is rather compact ($\approx 1.5$\arcsec\ in diameter) in the HST images and its spectrum shows strong \oii\ emission together with prominent \mgii\ and \feii\ absorption lines. This galaxy is seen nearly face-on with an inclination of $\approx 33 \degr$ derived with {\sc Galfit}. Its broadband morphology shows an asymmetric central morphology. At low surface brightness, it has an arm-like structure or some overdensities at the eastern side. The velocity field derived from the \oii\ doublet shows a very low gradient within $\pm 10$ \kms, aligned with its morphological major axis, which confirms that this galaxy is  seen almost face-on. The velocity dispersion map is rather flat with values around $\sim 50 - 40$ \kms. 
With a ratio $V/\sigma \approx 0.5$, this object is clearly dominated by non-circular motions.

\begin{figure} \resizebox{\hsize}{!}{\includegraphics{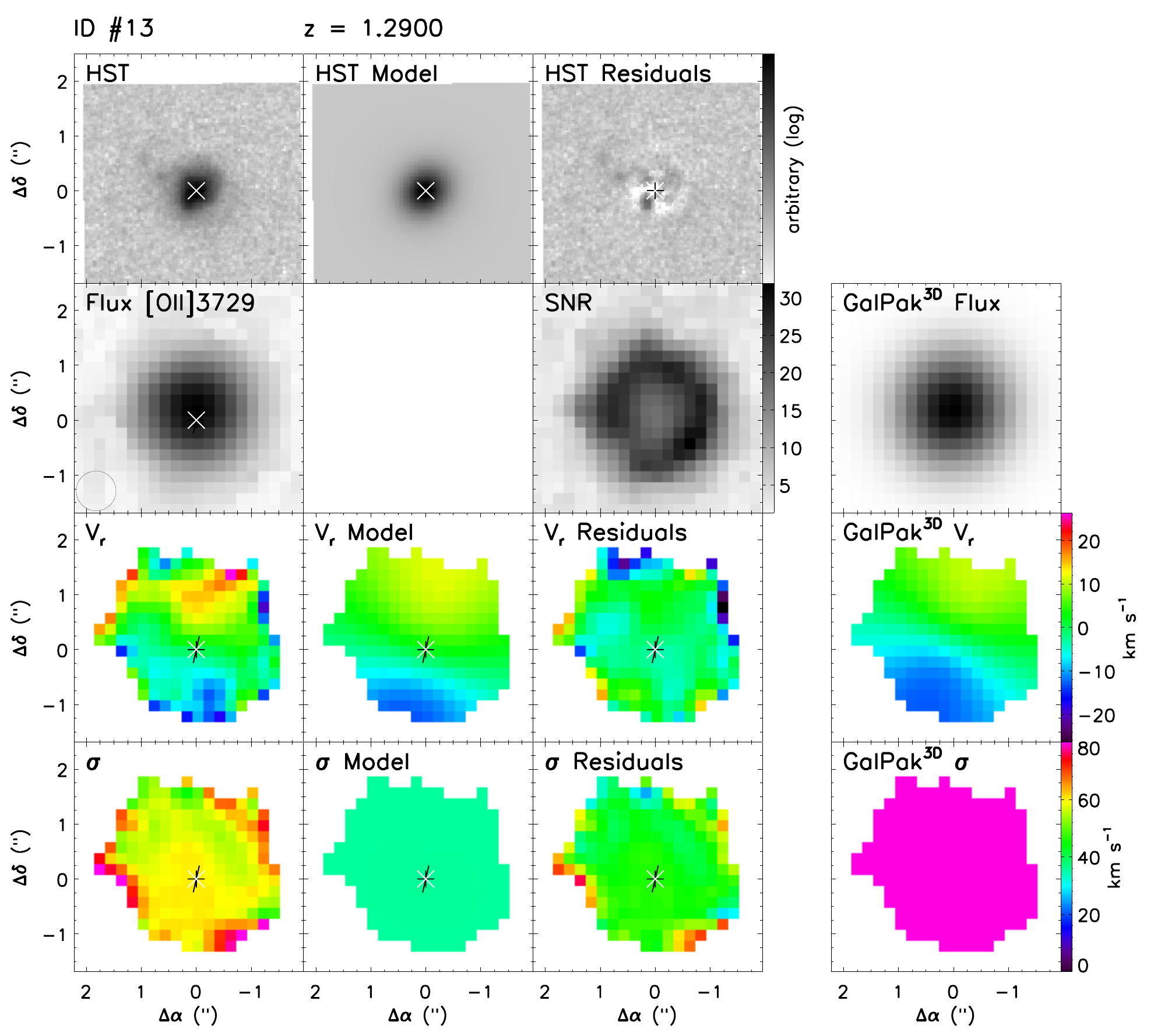}}
\caption{Morpho-kinematics maps for galaxy ID \#13. Same caption as Fig.~\ref{fig:vfexample}}
\label{fig:appen_figs_obj13}
\end{figure}

\paragraph{\bf ID \#16}
This galaxy at $z\approx 0.46$ has a bright and compact central component embedded in a large, diffuse and moderately inclined ($i\sim 20-30\degr$) low-surface brightness disk, allowing us to observe both \oii\ and \oiiib\ (the brightest) strong emission lines with MUSE. One bright compact object is also seen at about $1$\arcsec\ eastward in the HST image but it is measured at a different redshift in the MUSE data. Velocity fields derived from these emission lines are consistent and show an observed regular pattern within $\pm 30$ \kms. Disk modeling in 2D reproduces well the observed velocity fields. However \gpk\ has some difficulty determining the kinematical position angle and thus the velocity fields. Constraining this parameter to be close to the morphological one, i.e.,\, within $-55\pm 10\degr$, improves the 3D modeling significantly. The velocity dispersion map is rather flat. Kinematic parameters reported in Table~\ref{tbl:sample} are those derived from \oiiib. The gas dynamics of this galaxy is clearly dominated by the rotation with a ratio $V/\sigma \approx 2.6$. 

\begin{figure} \resizebox{\hsize}{!}{\includegraphics{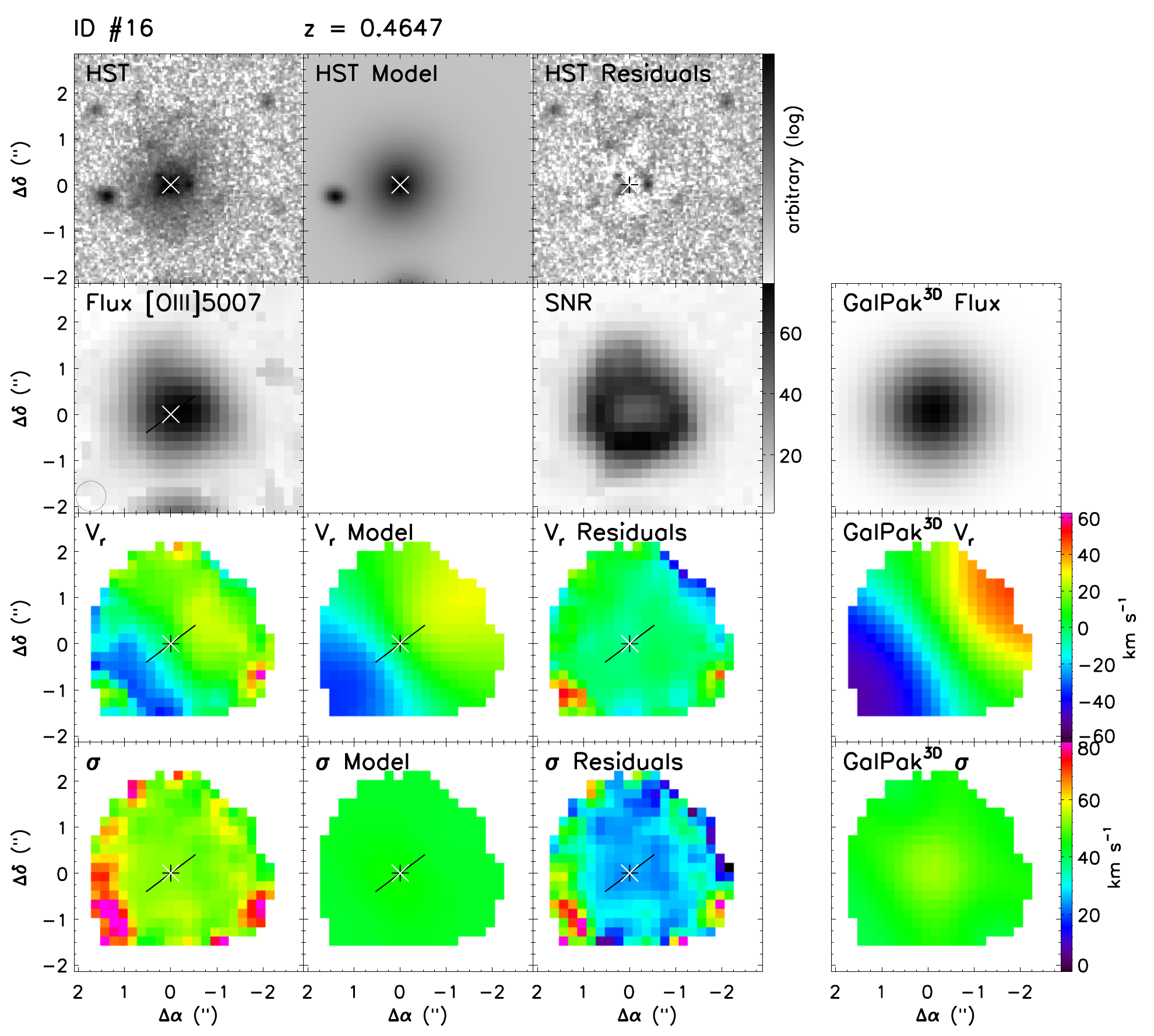}}
\caption{Morpho-kinematics maps for galaxy ID \#16. Same caption as Fig.~\ref{fig:vfexample} but emission-line fitting in the MUSE data cube is based on \oiiib.}
\label{fig:appen_figs_obj16}
\end{figure}

\paragraph{\bf ID \#17}
This galaxy at $z\approx 0.58$ exhibits an asymmetric broadband morphology. It is located at the border of the MUSE FoV and thus 
have \oii, \hbeta, and \oiiib\ emission lines more noisy in the MUSE data. Velocity fields derived from these emission lines are however consistent and show an observed pattern within $\pm 60$ \kms. 
Disk modeling in 2D reproduces quite well the observed velocity fields, but \gpk\ has some difficulty to converge, probably because of asymmetries seen in the HST images of this galaxy. Constraining the kinematical position angle to be close to the morphological one, i.e.,\, within $95\pm 10\degr$, improves  the 3D modeling significantly. The velocity dispersion map is also quite noisy, showing a peak on the northeast side both in \oiiib\ and \oii\ emission lines.
Kinematic parameters reported in Table~\ref{tbl:sample} are those derived from the 2D modeling with the \hbeta\ emission line. The gas dynamics of this galaxy is dominated by the rotation with a ratio $V/\sigma \approx 2.1$. 

\begin{figure} \resizebox{\hsize}{!}{\includegraphics{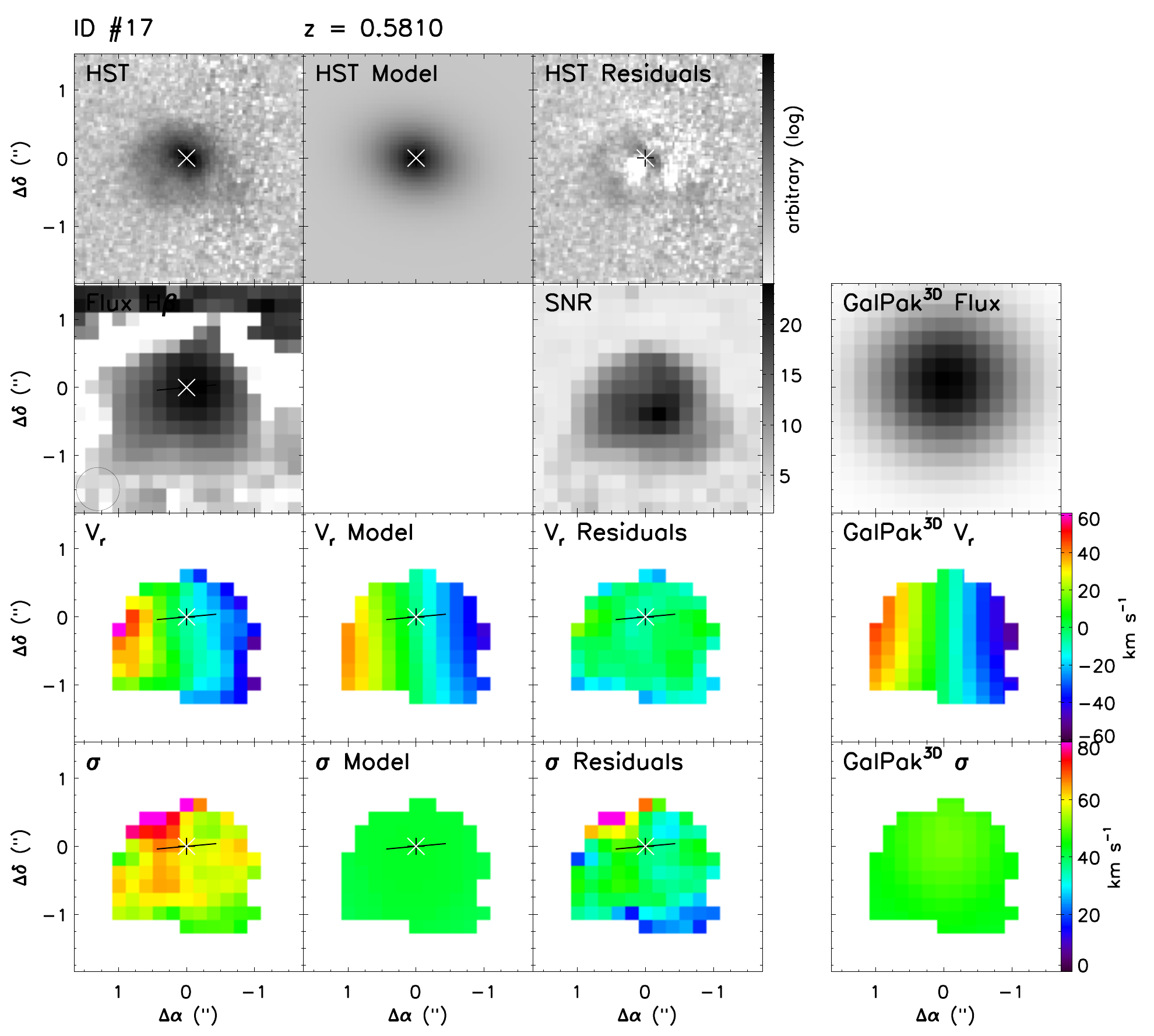}}
\caption{Morpho-kinematics maps for galaxy ID \#17. Same caption as Fig.~\ref{fig:appen_figs_obj5} but emission-line fitting in the MUSE data cube is based on \hbeta.}
\label{fig:appen_figs_obj17}
\end{figure}

\paragraph{\bf ID \#20}
This galaxy at $z\approx 0.43$ is rather large ($>2$\arcsec\ in diameter). Its morphology seems to indicate that it is dominated by a strong bar with arms starting at the end of the bar. An extended low-surface brightness component (elongated in the east-west direction) seems to be observed slightly above the noise in the HST/F814W image. This faint component is difficult to model. Therefore, the inclination provided by morphological modeling is not representative of the true inclination. Its kinematics is also completely dominated by the bar, where the bulk of the ionized gas emission in \hii\ regions is concentrated. This interpretation of morphology is further supported by the dynamics. Indeed, the velocity gradient (derived both from \oii\ and \oiiib\ emission lines) along the bar is almost null, whereas we would have expected a strong gradient along this elongation in the case of an edge-on galaxy. In addition, it seems that a velocity gradient almost orthogonal to the main elongation is observed, which supports the idea of a low-surface brightness disk dominated by a strong bar. Disk modeling in 2D reproduces well the observed velocity fields, but \gpk\ fails as a result of the irrealistic light profile assumption. The velocity dispersion map is quite flat. Kinematic parameters reported in Table~\ref{tbl:sample} are those derived from \oiiib. The gas dynamics of this galaxy is clearly dominated by the rotation with a ratio $V/\sigma \approx 2$.

\begin{figure} \resizebox{0.74\hsize}{!}{\includegraphics{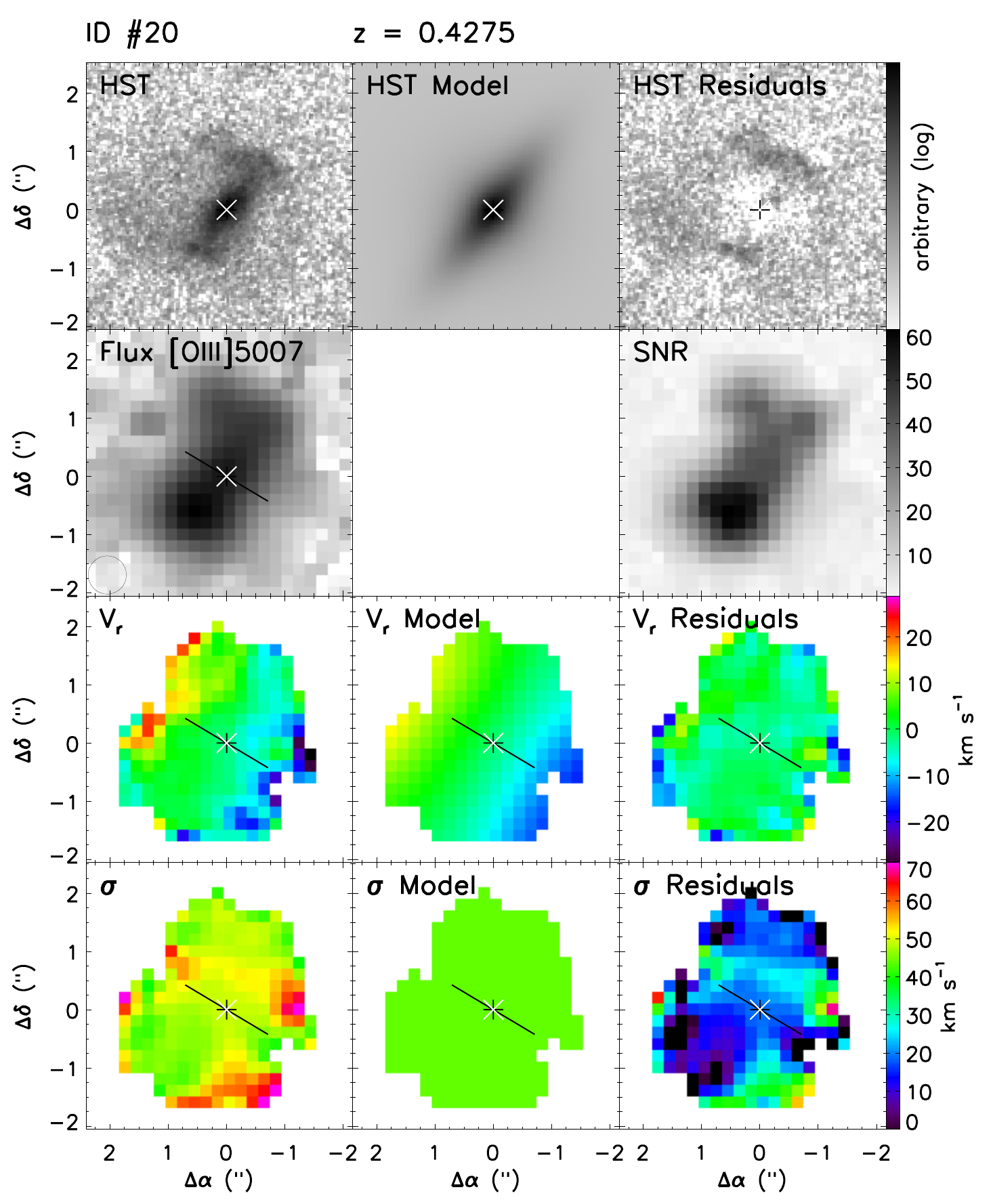}}
\caption{Morpho-kinematics maps for galaxy ID \#20. Description is given from {\it left} to {\it right}. {\it Top} row: HST/WFPC2 F814W image, {\sc Galfit} model (disk$+$bulge), and residuals, both in arbitrary log-scale units. {\it Second} row: MUSE \oiiib\ flux map (log scale) with the PSF FWHM size indicated with the black circle, and corresponding S/N map (linear scale). {\it Third} row: MUSE observed velocity field from \oiiib, velocity field of the 2D rotating disk model, and associated residual velocity field. {\it Bottom}: MUSE observed velocity dispersion map from \oiiib, velocity dispersion map deduced from the 2D velocity field model (which takes into account the beam smearing effect and spectral resolution), and deconvolved associated velocity dispersion map. In each map, north is up and east is left. The center used for kinematic modeling is indicated with a white cross, the position angle is indicated by the black line that ends at the effective radius.}
\label{fig:appen_figs_obj20}
\end{figure}

\paragraph{\bf ID \#23}
HST images show a low surface brightness, rather large ($\approx2$\arcsec\ in diameter) and nearly face-on disk with numerous \hii\ regions for this low-mass galaxy at $z\approx 0.56$. Velocity fields derived from \oii\ and \oiiib\ emission lines slightly differ in the orientation of the major axis ($\approx 20$\degr) as well as between the 2D and \gpk\ models. This might be due to the combination of an asymmetric line flux distribution with a better spatial resolution for the \oiiib\ line. However both velocity fields are consistent and show a very low gradient pattern within $\pm 20$ \kms. The morphological major axis position angle differs by $\approx 60\degr$ from the kinematic position angle, probably because of its almost face-on orientation combined with its low surface brightness, which makes the morphological orientation determination not much accurate. Disk modeling in 2D reproduces well the observed velocity fields. However this is not the case for \gpk, which does not converge to the correct major axis position angle. Constraining this parameter to be within $25\pm 10\degr$ improves  the 3D modeling significantly. The velocity dispersion map is rather flat. Kinematic parameters reported in Table~\ref{tbl:sample} are those derived from \oiiib. The gas dynamics of this galaxy 
is (barely) dominated by the rotation with a ratio $V/\sigma \approx 1$. 

\begin{figure} \resizebox{\hsize}{!}{\includegraphics{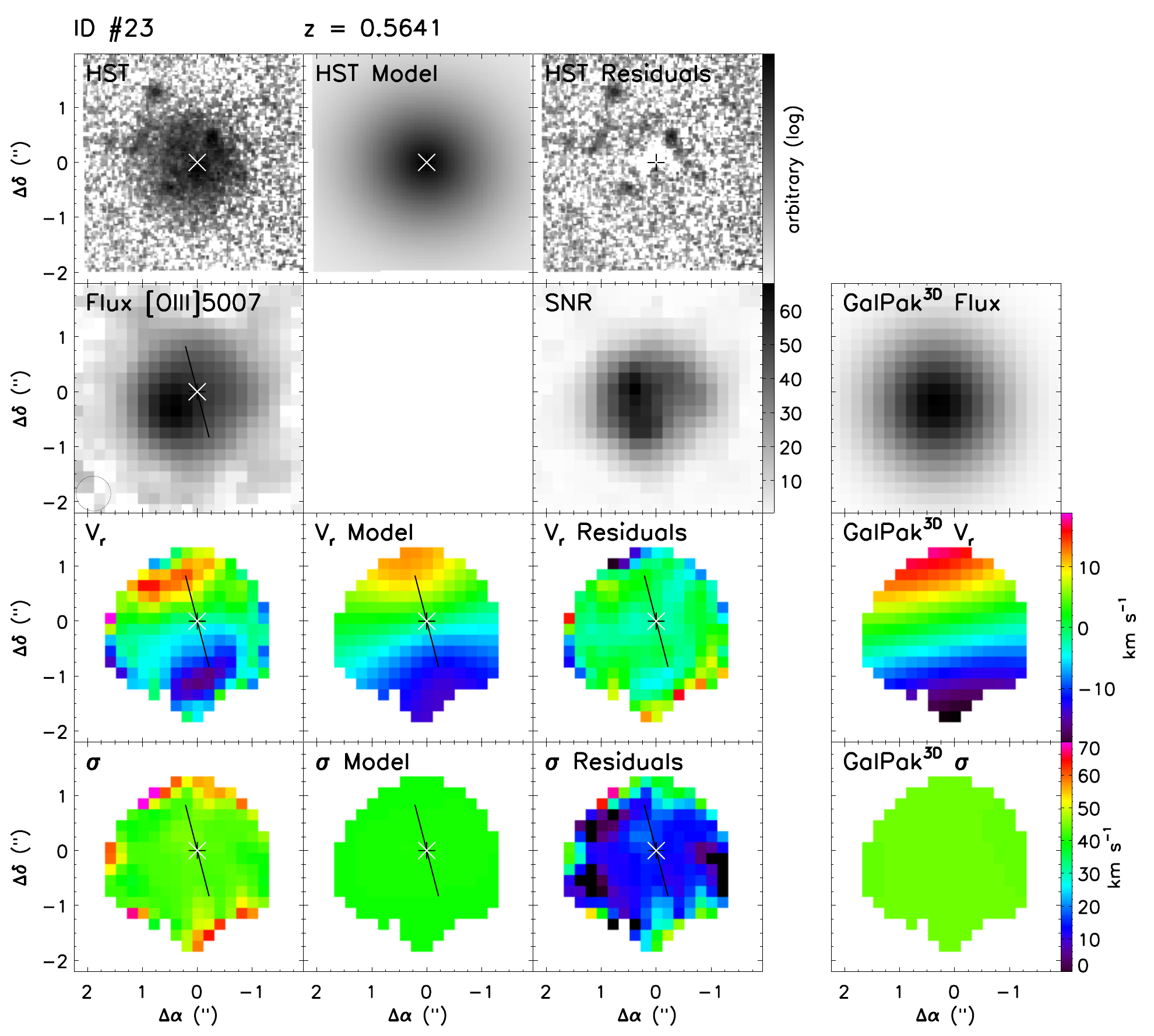}}
\caption{Morpho-kinematics maps for galaxy ID \#23. Same caption as Fig.~\ref{fig:vfexample} but emission-line fitting in the MUSE data cube is based on \oiiib.}
\label{fig:appen_figs_obj23}
\end{figure}

\paragraph{\bf ID \#24}
This galaxy at $z\approx 0.97$ is a nearly face-on disk (around $1$\arcsec\ in diameter) with a bright \oii\ emission together with \mgii\ and \feii\ absorption lines. It displays a surprisingly flat but slightly asymmetric broadband light distribution that is not accurately described by an exponential disk. The velocity field derived from \oii\ shows a regular pattern within $\pm 30$ \kms. Disk modeling (both in 2D and with \gpk) reproduces well the observed velocity fields. The velocity dispersion map is rather flat. The gas dynamics of this galaxy is clearly dominated by the rotation with a ratio $V/\sigma \approx 3.7$.

\begin{figure} \resizebox{\hsize}{!}{\includegraphics{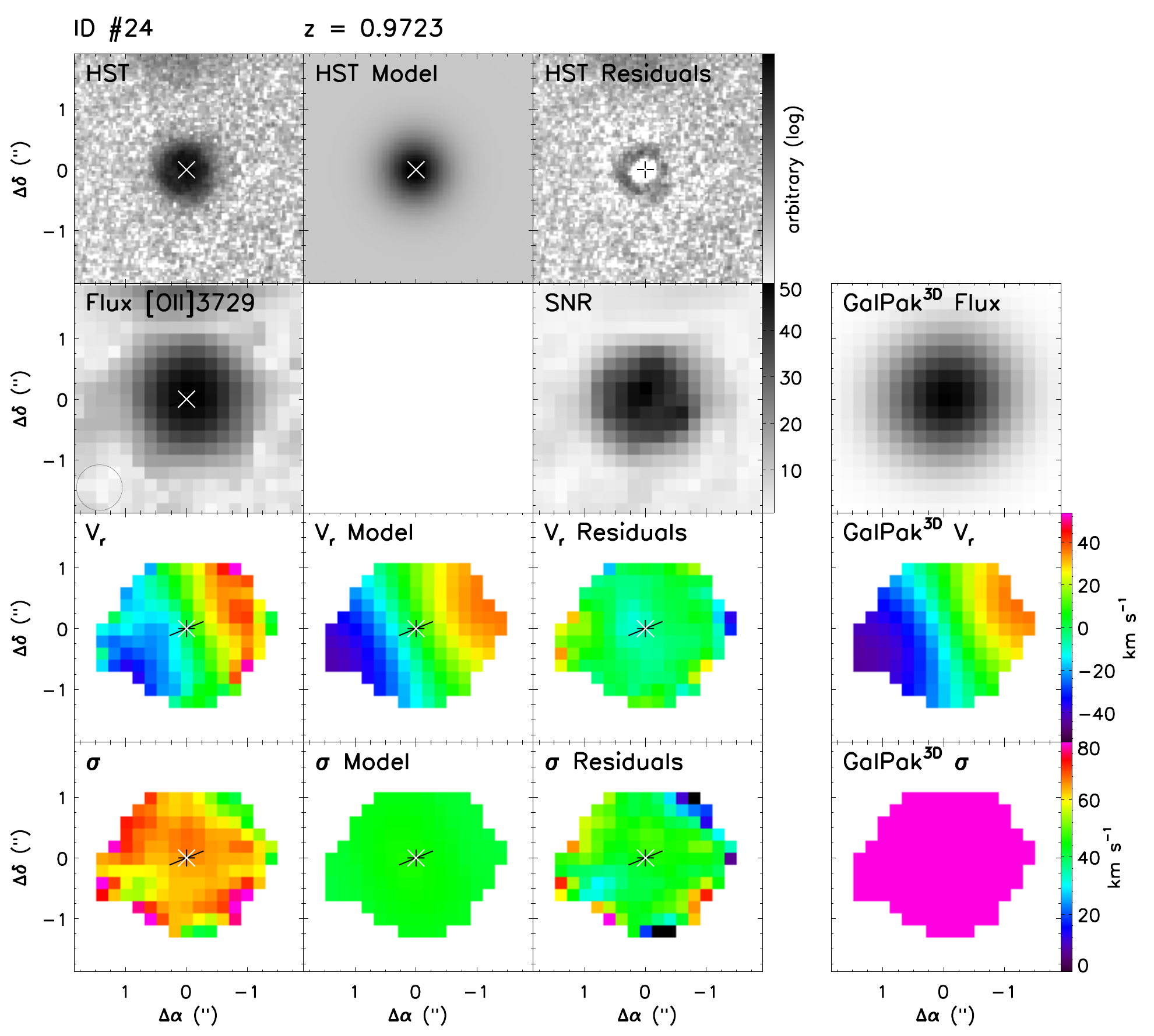}}
\caption{Morpho-kinematics maps for galaxy ID \#24. Same caption as Fig.~\ref{fig:vfexample}}
\label{fig:appen_figs_obj24}
\end{figure}

\paragraph{\bf ID \#26}
This nearly edge-on galaxy at $z\approx 0.22$, of $\approx 2$\arcsec\ size in diameter, is the less massive among our sample ($M\approx 10^{8}$\msun). Velocity fields derived from \halpha\ and \oiiib\ emission lines are consistent and show a regular pattern within $\pm 40$ \kms. Disk modeling (both in 2D and with \gpk) reproduces well the observed velocity fields, but the \gpk\ analysis returns an underestimated disk inclination. This value has thus been further constrained in \gpk\ to be within $70\pm 5\degr$. The velocity dispersion map is quite flat and with low values. Kinematic parameters reported in Table~\ref{tbl:sample} are those derived from \halpha. The gas dynamics of this galaxy is clearly dominated by the rotation with a ratio $V/\sigma \approx 4.8$. 

\begin{figure} \resizebox{\hsize}{!}{\includegraphics{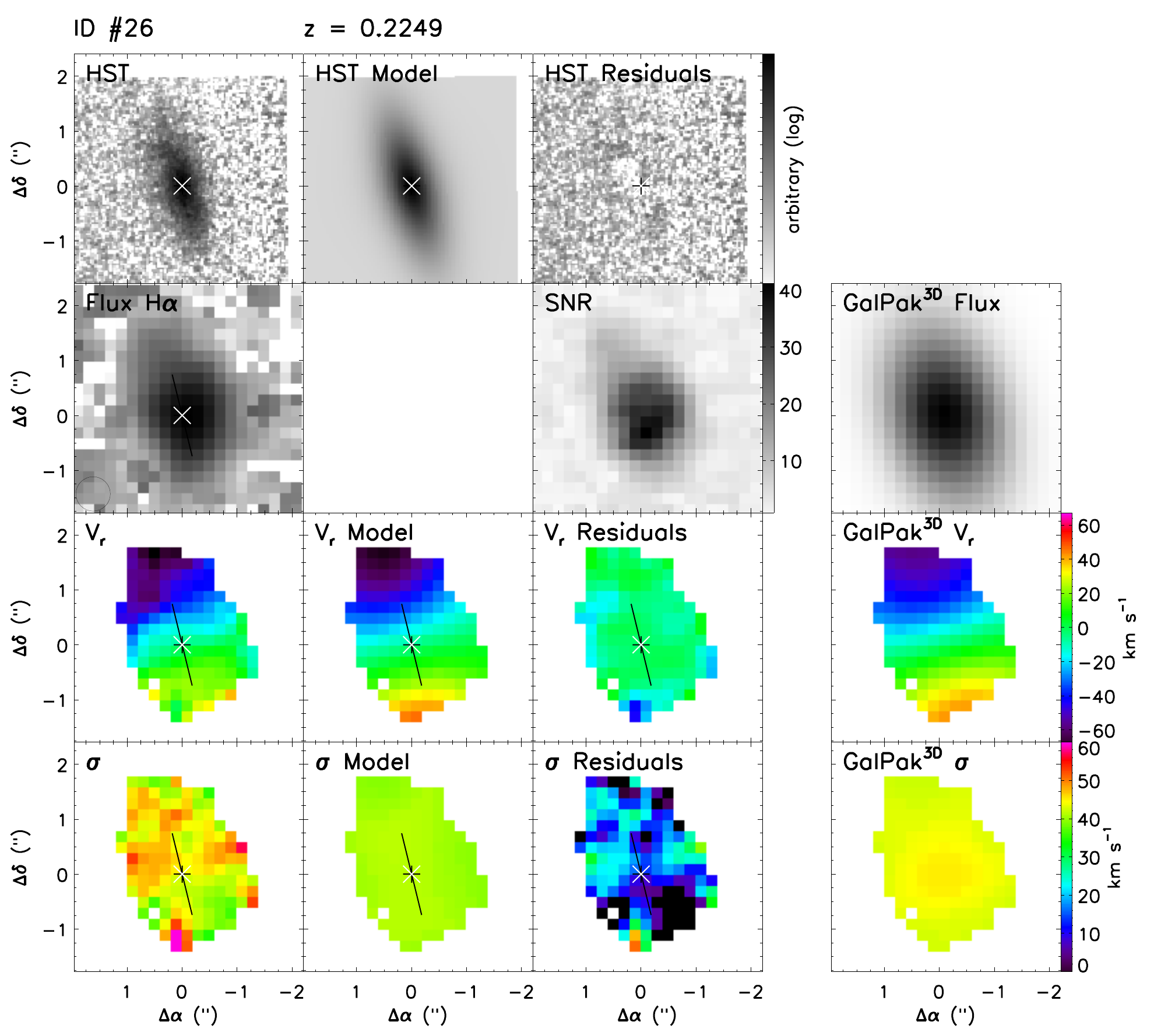}}
\caption{Morpho-kinematics maps for galaxy ID \#26. Same caption as Fig.~\ref{fig:vfexample} but emission-line fitting in the MUSE data cube is based on \halpha.}
\label{fig:appen_figs_obj26}
\end{figure}

\paragraph{\bf ID \#27}
This large ($> 2$\arcsec\ in diameter) galaxy is a member of a group at $z\approx 1.28$, has a very peculiar morphology, and is probably interacting with the galaxy ID \#10 as revealed by low-surface brightness tidal tails in between the two objects. Its broadband morphology is very disturbed and cannot be reproduced by a simple bulge/disk decomposition. Indeed, it shows a main distorted elongation on top of a low-surface brightness filamentary structure, which may be a signature of a merger event with one tidal arm. Two additional small components at about $2$\arcsec\ (northward and southward) are also seen on HST images. One of these, the southern object, is indeed a moderate-mass ($\approx 10^{9.7}$\msun) satellite, ID \#124, located at a projected distance of 14 kpc and with a systemic velocity difference of $\approx 100$ \kms. With a stellar mass ratio of  $\sim 1/10$, it can be considered as a possible future minor merger (Ventou et al., in prep.). The velocity field derived from \oii\ is also 
quite complex. A main gradient ($\approx \pm 70$ \kms) is observed (reproduced by the 2D disk modeling but not with \gpk) from south to north, a direction that is in agreement with the main elongation. Two other components are observed (in the northeast and west directions) with different velocities. In particular the western component has a velocity offset of $\approx 120$ \kms with respect to the systemic velocity. Surprisingly, these two components are not related to any broadband counterpart in the HST images. It could be two low-surface brightness tidal dwarfs or outflowing material in a jet-like structure. The velocity dispersion map shows large values ($\approx 100$ \kms). The feature in the northwest, with a very large velocity dispersion, could be due to sky line residuals, however, it could also be a signature of the interaction with the galaxy ID \#10, where the filament starts. This galaxy is clearly interacting but is still dominated by circular motions with a ratio $V/\sigma \approx 1.7$.  

\begin{figure} \resizebox{0.74\hsize}{!}{\includegraphics{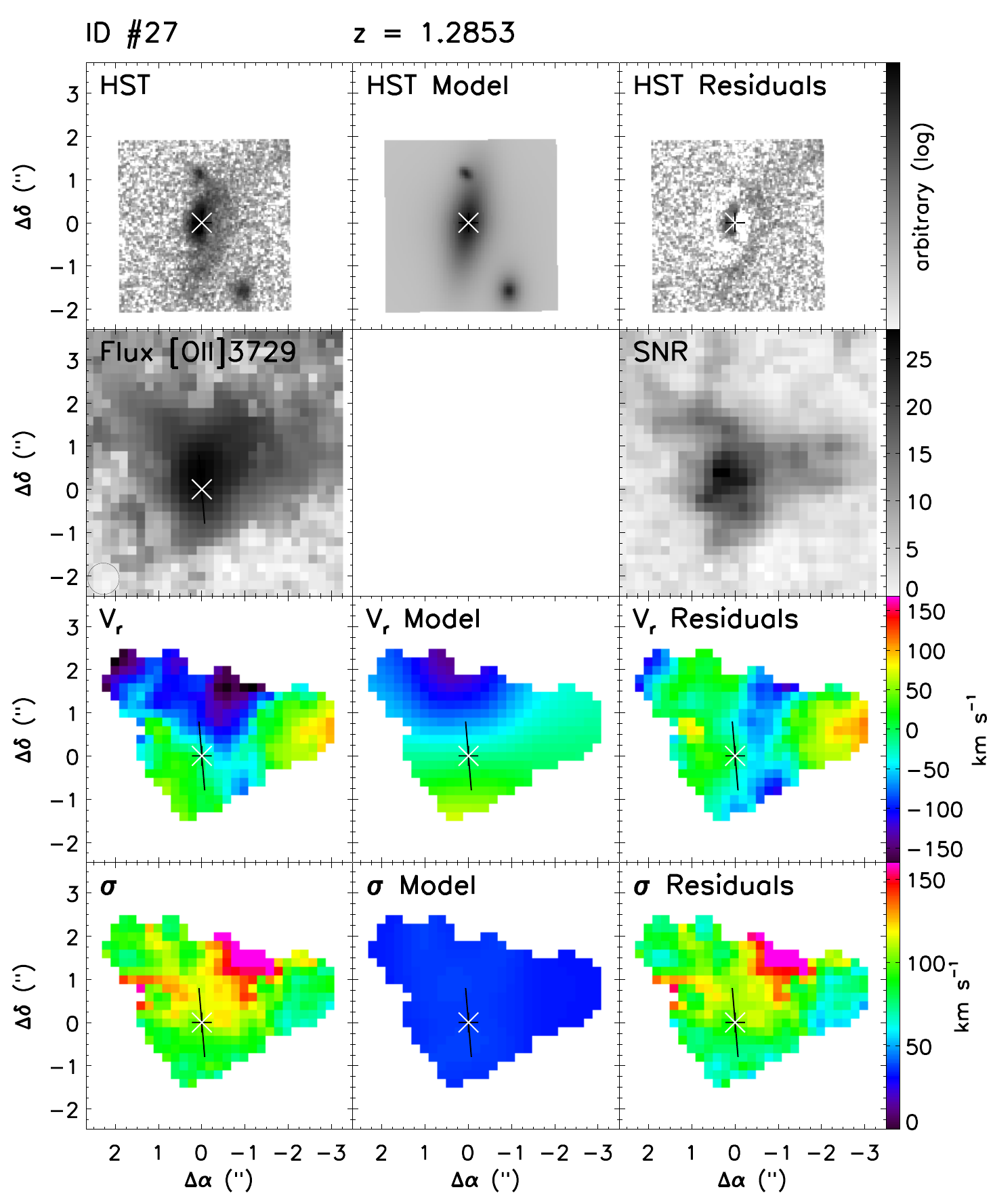}}
\caption{Morpho-kinematics maps for galaxy ID \#27. Same caption as Fig.~\ref{fig:appen_figs_obj17} but emission-line fitting in the MUSE data cube is based on \oiib.}
\label{fig:appen_figs_obj27}
\end{figure}

\paragraph{\bf ID \#28}
This galaxy with a size of $\approx 1.5$\arcsec\ in diameter has a peculiar morphology in HST images.
It has an asymmetric broadband morphology that is surprisingly flat in the center with one bright region offset northward and an arm-like structure in the south. It is therefore not accurately described by an exponential disk model.
With a redshift $z\approx 0.32$, the full set of bright emission lines, from \oii\ to \halpha, is accessible in the MUSE spectral range. Velocity fields derived from \halpha, \oiiib, and \oii\ emission lines are consistent and show a surprisingly low velocity gradients within $\pm 10-20$ \kms. This could indicate that this galaxy is seen nearly face-on, as also indicated by the low inclination derived form {\sc Galfit} and \gpk. The orientation of the velocity gradient differs, depending on the emission line, but is more or less in agreement with the main elongation direction. Disk modeling (both in 2D and with \gpk) reproduces well the observed velocity fields. The velocity dispersion map is rather flat with values quite different depending on the lines used (from $\approx 30$ \kms\ for \oiiib+\hbeta\ and \halpha\ to $\approx 50$ \kms\ for the \oii\ doublet), probably due to some uncertainties in the LSF estimate. Kinematic parameters reported in Table~\ref{tbl:sample} are those derived from \oiiib. The gas dynamics of this galaxy 
is dominated by the rotation with a ratio $V/\sigma \approx 1.6$. 

\begin{figure} \resizebox{\hsize}{!}{\includegraphics{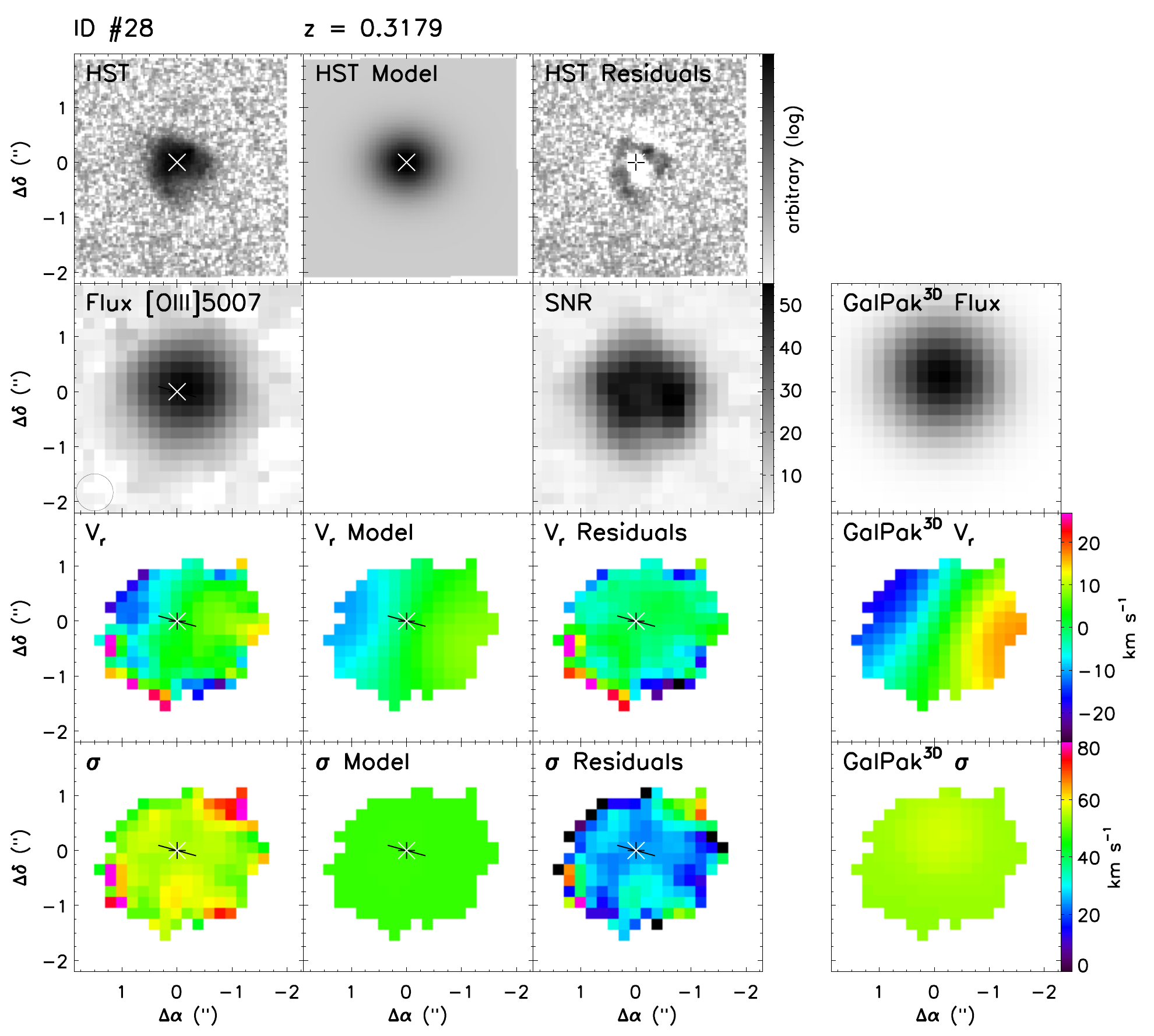}}
\caption{Morpho-kinematics maps for galaxy ID \#28. Same caption as Fig.~\ref{fig:vfexample} but emission-line fitting in the MUSE data cube is based on \oiiib.}
\label{fig:appen_figs_obj28}
\end{figure}

\paragraph{\bf ID \#32}
This galaxy is a highly inclined and rather small ($\approx1$\arcsec\ in diameter) disk at $z\approx 0.56$. Velocity fields derived from \oii\ and \oiiib\ emission lines are consistent and show a regular pattern within $\pm 60$ \kms. Disk modeling (both in 2D and with \gpk) reproduces well the observed velocity fields. The velocity dispersion map shows an increase along the minor axis, which is more pronounced for the \oiiib$+$\hbeta\ lines, maybe due to a better spatial resolution at these wavelengths. This peak is compatible with the unresolved velocity gradients despite it seems offset toward one side. Kinematic parameters reported in Table~\ref{tbl:sample} are those derived from \oii. The gas dynamics of this galaxy is (barely) dominated by the rotation with a ratio $V/\sigma \approx 1.4$. 

\begin{figure} \resizebox{\hsize}{!}{\includegraphics{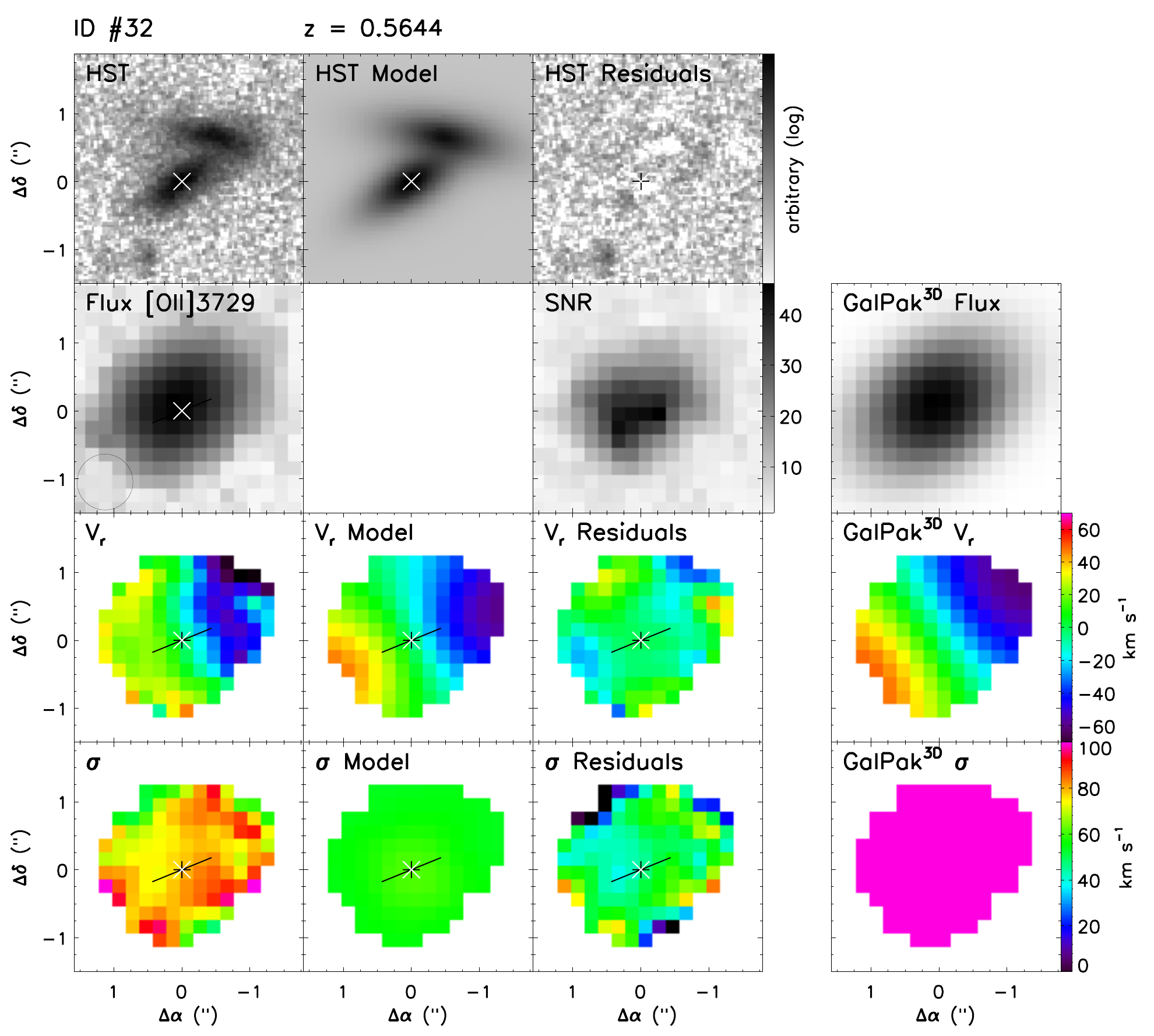}}
\caption{Morpho-kinematics maps for galaxy ID \#32. Same caption as Fig.~\ref{fig:vfexample}}
\label{fig:appen_figs_obj32}
\end{figure}

\paragraph{\bf ID \#35}
This moderately inclined galaxy is also a member of the group at $z\approx 1.28$. It is $\approx 1$\arcsec\ in diameter, shows a main elongation but also displays an asymmetric low-surface brightness component in the HST images. The \oii\ velocity field displays a gradient along the same direction as the morphological elongation which seems to be an indication of a rotating disk ($\approx \pm 80$ \kms). It also shows perturbations at the edges where the S/N is lower, in particular because of strong sky line residuals. Disk modeling (both in 2D and with \gpk) reproduces well the observed velocity fields. The velocity dispersion map shows a strong peak parallel to the minor axis but offset southward by about $0.2$\arcsec. This peak can be partly explained by beam smearing effect. This would indicate that the kinematic center does not match the peak of the \oii\ emission-line flux map. However, the peak may be larger than what is expected from the velocity gradient, suggesting the presence of an inner unresolved gradient. 

\begin{figure} \resizebox{\hsize}{!}{\includegraphics{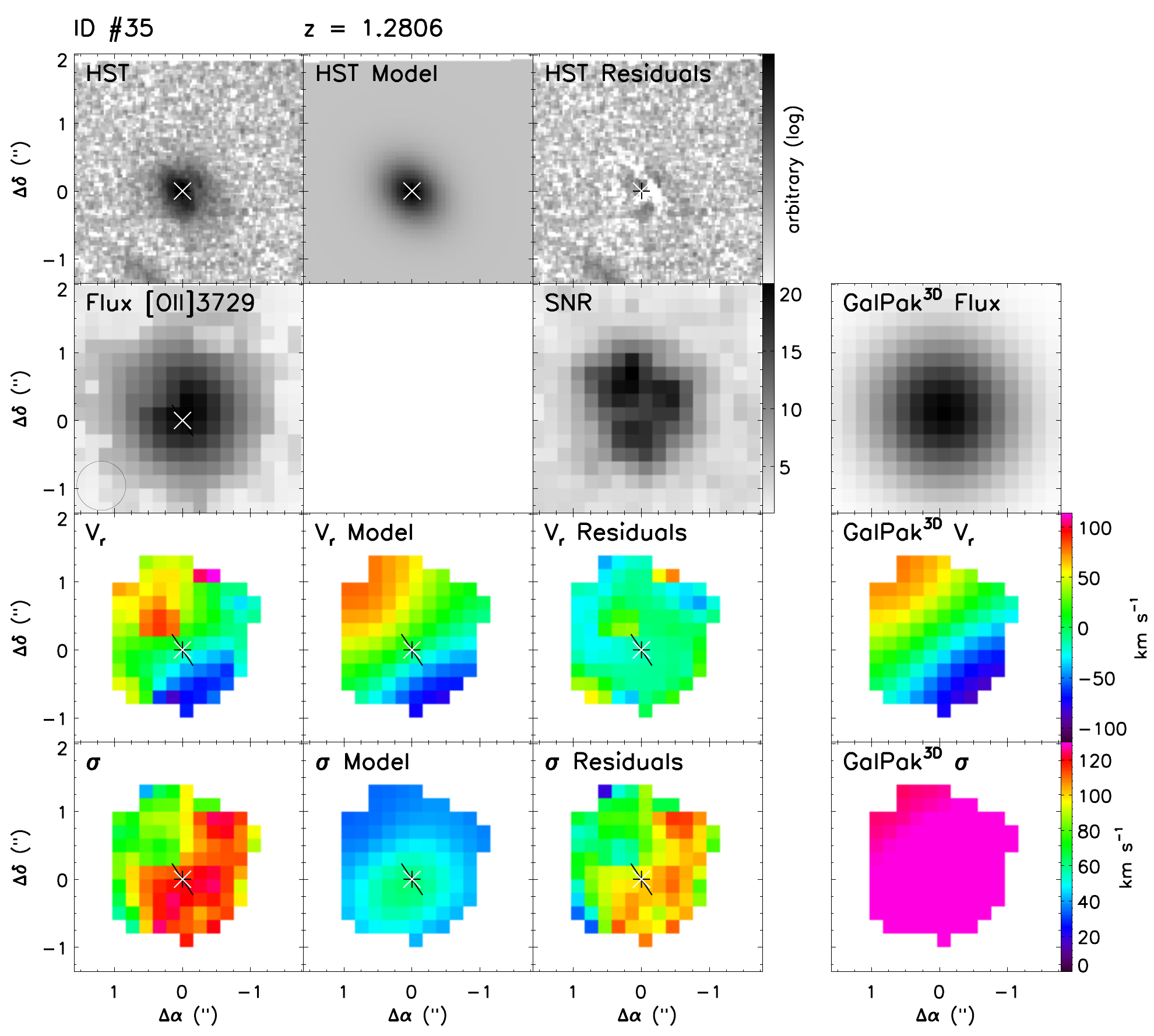}}
\caption{Morpho-kinematics maps for galaxy ID \#35. Same caption as Fig.~\ref{fig:vfexample}}
\label{fig:appen_figs_obj35}
\end{figure}

\paragraph{\bf ID \#37}
This galaxy at $z\approx 1.1$ is rather small ($\approx 1$\arcsec\ in diameter) and seems to be rather elongated in HST images with a derived inclination of $\approx70\degr$ from {\sc Galfit}. However most of the \oii\ emission probed with MUSE spectroscopy comes from the bright central regions. The velocity field derived from \oii\ shows an almost flat pattern within $\pm 10$ \kms. Disk modeling in 2D reproduces well this shallow observed velocity fields but, as a result of convergence problems with \gpk, the inclination has been constrained to be within $70\pm5\degr$ and the kinematical position angle to be within $45\pm 10\degr$. The velocity dispersion map displays rather high values that may indicate that this system is perturbed or has an unresolved velocity gradient which would be however surprising for its size and shape. The gas dynamics of this galaxy, as probed with MUSE, is clearly dominated by non-circular motions in its central parts with a ratio $V/\sigma \approx 0.3$. 

\begin{figure} \resizebox{\hsize}{!}{\includegraphics{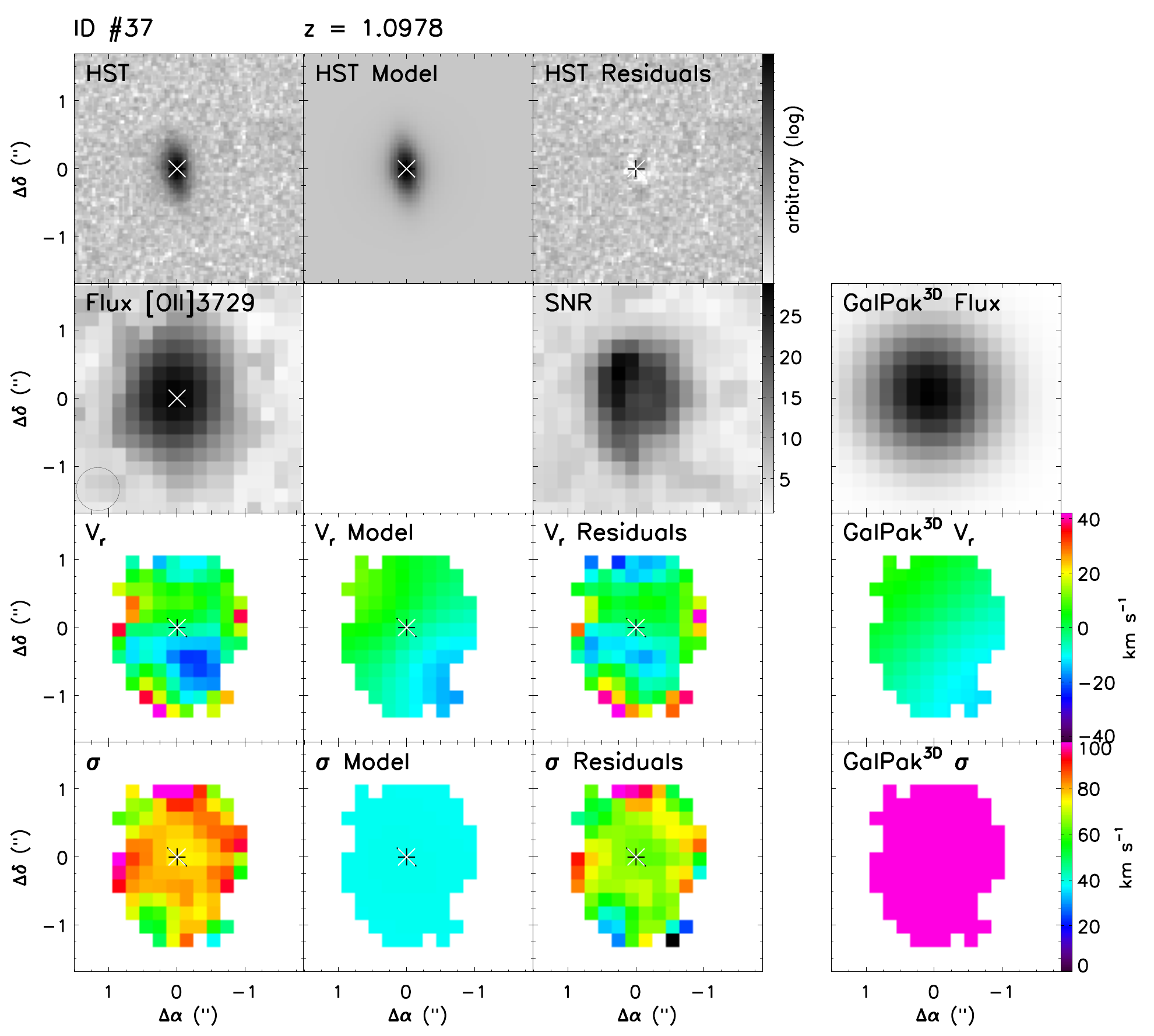}}
\caption{Morpho-kinematics maps for galaxy ID \#37. Same caption as Fig.~\ref{fig:vfexample}}
\label{fig:appen_figs_obj37}
\end{figure}

\paragraph{\bf ID \#38}
This galaxy at $z\approx 1$ has an elongated morphology. It could be a bar-like feature since it also has a lower surface brightness component that seems to have a slightly different orientation. However, the \oii\ velocity field exhibits a regular pattern within $\pm 60$ \kms\ in the same direction of the main elongated structure. Disk modeling (both in 2D and with \gpk) reproduces well this observed velocity fields. The velocity dispersion map shows an increase around the minor axis that may be due to an unresolved velocity gradient. The level of this increase is relatively large compared to the observed velocity gradient, which may indicate that the galaxy rotates faster, at least in the central part. The gas dynamics of this galaxy is dominated by the rotation with a ratio $V/\sigma \approx 2.1$. 

\begin{figure} \resizebox{\hsize}{!}{\includegraphics{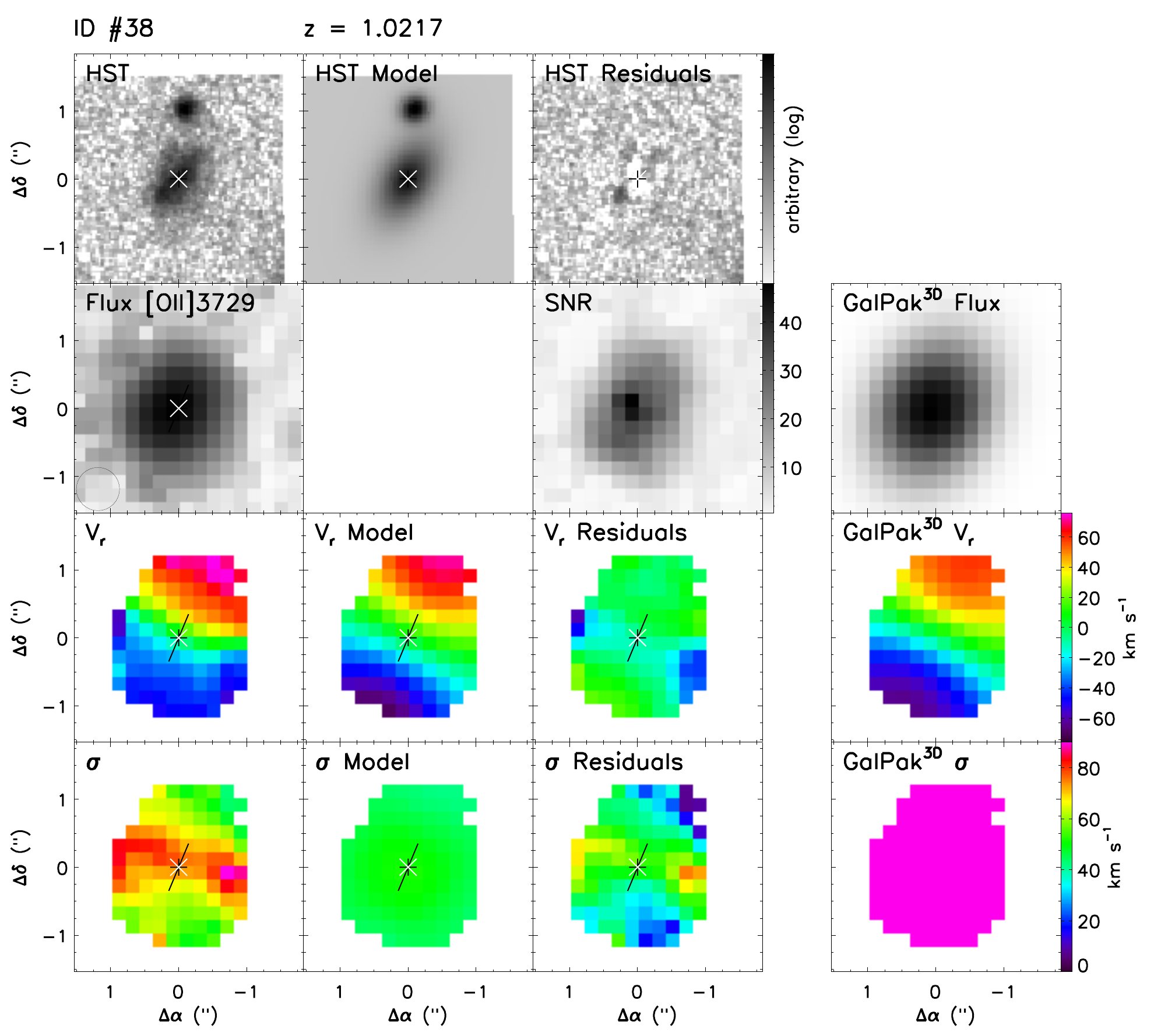}}
\caption{Morpho-kinematics maps for galaxy ID \#38. Same caption as Fig.~\ref{fig:vfexample}}
\label{fig:appen_figs_obj38}
\end{figure}

\paragraph{\bf ID \#49}
This galaxy is a nearly edge-on disk at $z\approx 1$ with a diameter of about $1$\arcsec. This galaxy has a low-mass ($\approx 10^{7.8}$\msun) satellite, ID \#149 in the redshift catalog of \cite{Bacon+2015}, located southward at a projected distance of 14 kpc and with a systemic velocity difference of $60$ \kms. With a stellar mass ratio of  $\sim 1/25$, it can be considered a possible future minor merger (Ventou et al., in prep.). The velocity field derived from \oii\ shows a regular X-shape pattern within $\pm 80$ \kms. Disk modeling (both in 2D and with \gpk) reproduces well this observed velocity field. The velocity dispersion map shows an increase around the minor axis, which is compatible with the unresolved velocity gradient despite it seems slightly offset ($\approx 0.2$\arcsec) toward the northwest side. The gas dynamics of this galaxy is dominated by the rotation with a ratio $V/\sigma \approx 1.7$.

\begin{figure} \resizebox{\hsize}{!}{\includegraphics{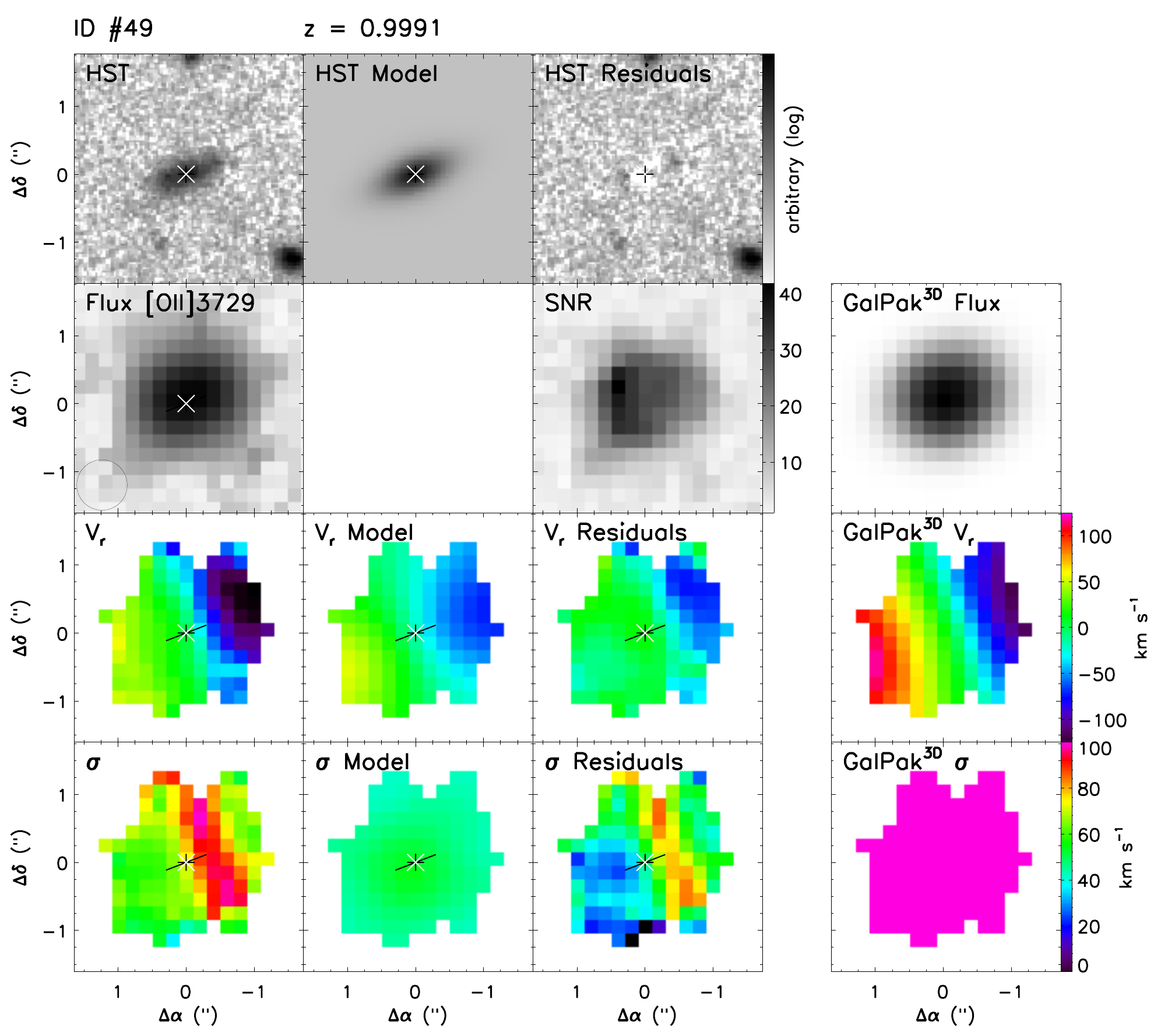}}
\caption{Morpho-kinematics maps for galaxy ID \#49. Same caption as Fig.~\ref{fig:vfexample}}
\label{fig:appen_figs_obj49}
\end{figure}

\paragraph{\bf ID \#68}
This galaxy is a nearly edge-on and rather small ($\approx 1$\arcsec\ in diameter) disk at $z\approx 0.97$. Its broadband HST morphology appears slightly asymmetric. The \oii\ line flux morphology is quite circular, probably due to beam smearing effect. The velocity field derived from \oii\ shows a regular pattern within $\pm 30$ \kms\ compatible with its morphology. Disk modeling (both in 2D and with \gpk) reproduces well this observed velocity field but inclination has been fixed to $74\pm5 \degr$ with \gpk\ becaue of convergence problems. The velocity dispersion map is quite flat. The gas dynamics of this galaxy is dominated by the rotation with a ratio $V/\sigma \approx 1.8$.

\begin{figure} \resizebox{\hsize}{!}{\includegraphics{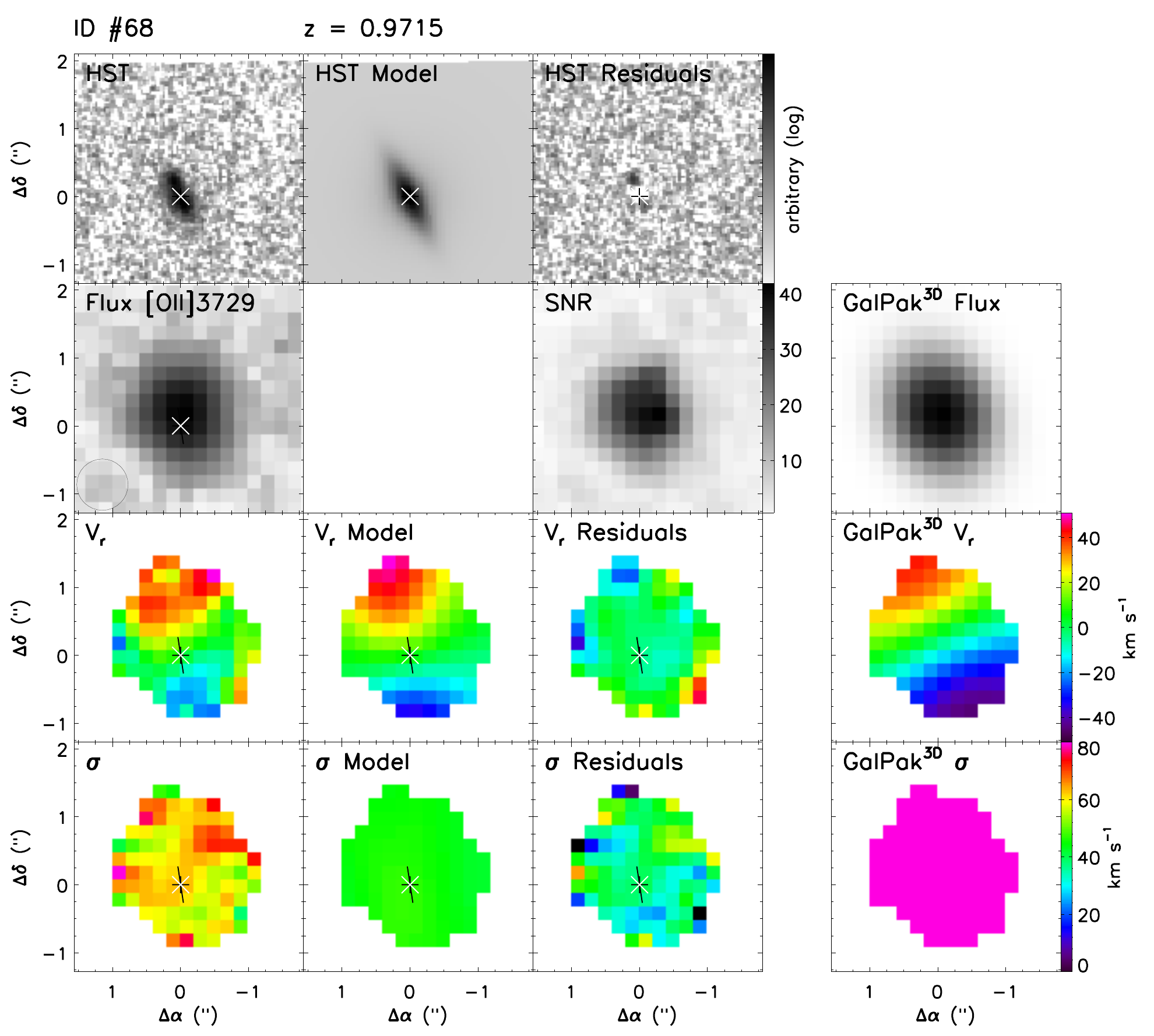}}
\caption{Morpho-kinematics maps for galaxy ID \#68. Same caption as Fig.~\ref{fig:vfexample}}
\label{fig:appen_figs_obj68}
\end{figure}

\paragraph{\bf ID \#72}
This galaxy at $z\approx 0.84$ is rather small ($<1$\arcsec\ in diameter). HST images show a small feature about $0.2$\arcsec\ southward, but this feature may not be gravitationally linked as it is not seen (or resolved) in MUSE data. The velocity field derived from \oii\ shows a regular pattern within $\pm 60$ \kms, which is compatible (both in 2D and with \gpk\ modeling) with a rotating disk almost seen edge-on. We note, however, that inclination has been fixed to $74\pm5 \degr$ with \gpk\ as a result of convergence problems. The gas dynamics of this galaxy is clearly dominated by the rotation with a ratio $V/\sigma \approx 5.7$.

\begin{figure} \resizebox{\hsize}{!}{\includegraphics{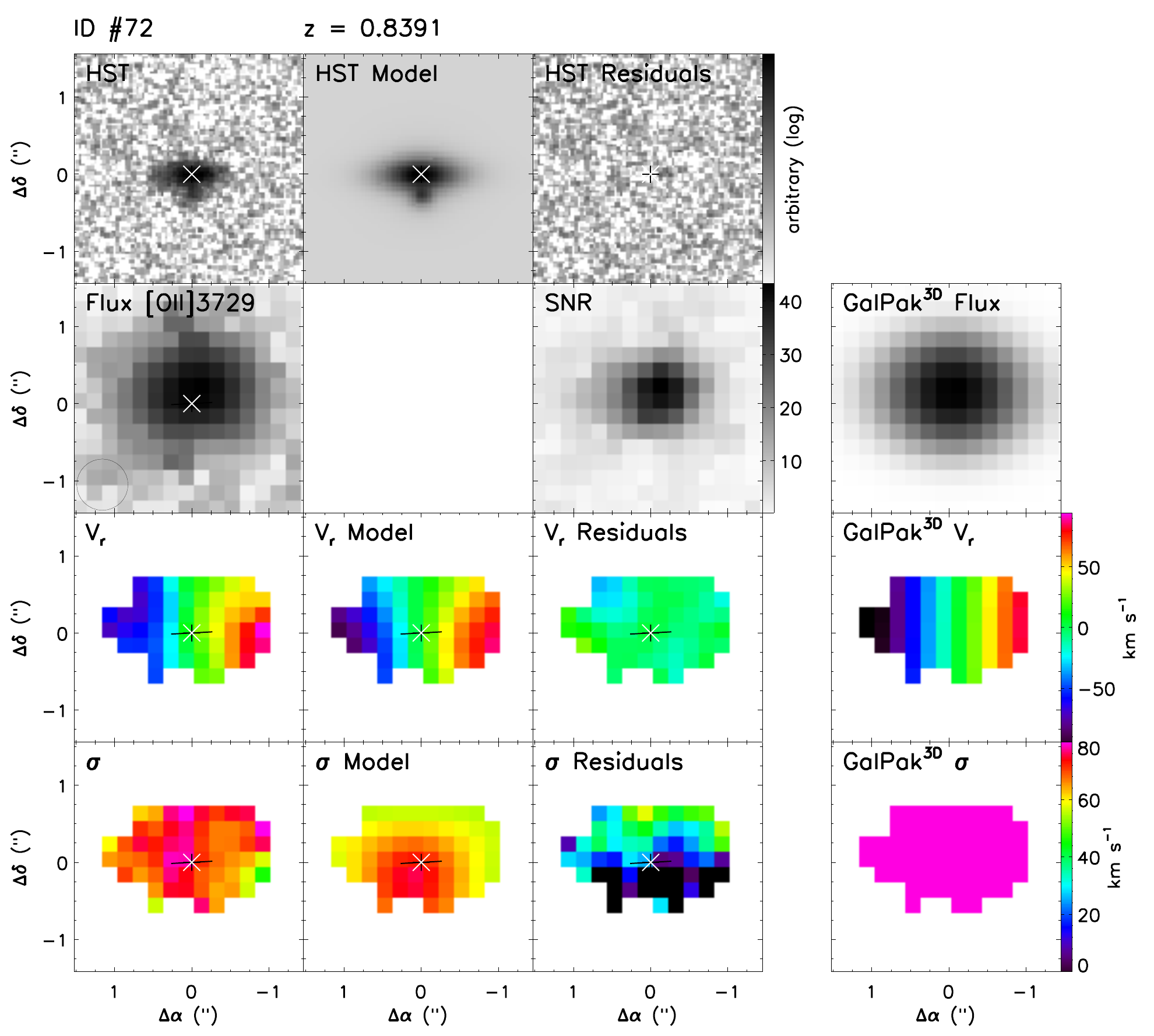}}
\caption{Morpho-kinematics maps for galaxy ID \#72. Same caption as Fig.~\ref{fig:vfexample}}
\label{fig:appen_figs_obj72}
\end{figure}

\paragraph{\bf ID \#88}
With a redshift $z\approx 1.36$ and a size of $\approx 1$\arcsec\ in diameter this galaxy is the farthest of our sample. It has an elongated broadband morphology identified as a merging system. This galaxy has indeed a similar mass ($\approx 10^8$\msun) companion, not identified in the redshift catalog of \cite{Bacon+2015}, located southward at a projected distance of 5 kpc and with a systemic velocity difference of $\approx 20$ \kms. With a stellar mass ratio of  $\sim 1/4$, it can be considered as a possible future major merger (Ventou et al., in prep.). The \oii\ flux distribution is quite circular and small ($<2$\arcsec\ in diameter). The velocity field derived from \oii\ shows a regular pattern within $\pm 20$ \kms, but surprisingly the kinematic axis is perpendicular to the morphological main axis, which is in agreement with the hypothesis of recent interactions. This may also indicate that this galaxy is barely resolved within the MUSE data. Disk modeling in 2D reproduces well this shallow observed velocity fields but, as a result of convergence problems with \gpk, the inclination has been constrained to be within $77\pm5\degr$ and the kinematical position angle to be within $-65\pm10\degr$. The velocity dispersion map is rather flat with no clear feature. The gas dynamics of this galaxy is dominated by the rotation with a ratio $V/\sigma \approx 1.5$.

\begin{figure} \resizebox{\hsize}{!}{\includegraphics{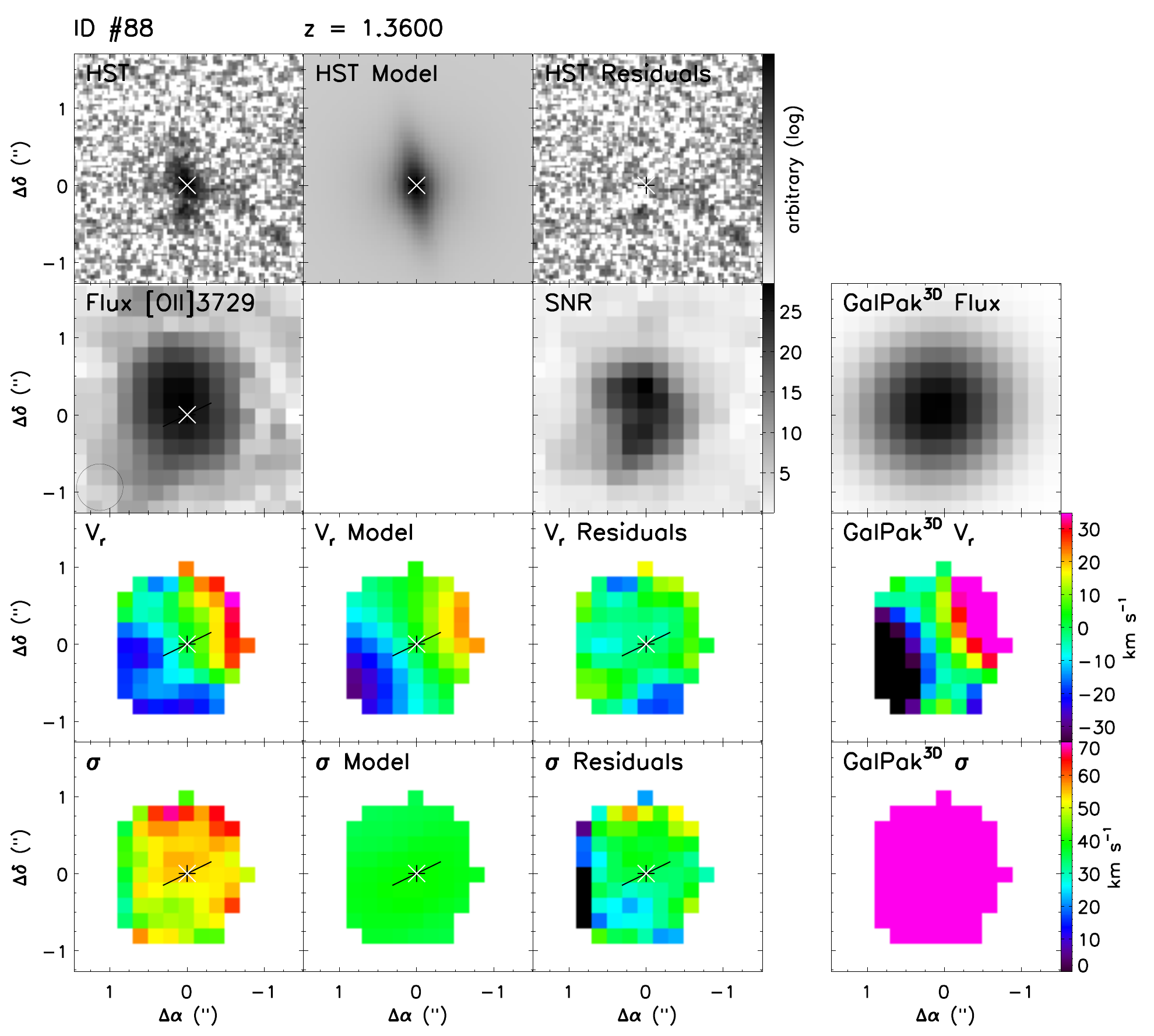}}
\caption{Morpho-kinematics maps for galaxy ID \#88. Same caption as Fig.~\ref{fig:vfexample}}
\label{fig:appen_figs_obj88}
\end{figure}


\end{document}